\documentclass[11pt]{article}
\title{\vspace{25mm} \bf{$A_{\infty } / L_{\infty }$ structure and alternative action for\\ WZW-like superstring field theory} \vspace{15mm} }
\author{\Large{Keiyu Goto\footnote{keiyu.goto.9@gmail.com, [Present address: Komazawa, Setagaya-ku, Tokyo 154-0012, Japan]} , \hspace{2mm} 
Hiroaki Matsunaga}\footnote{matsunaga@fzu.cz} \vspace{5mm}}
\date{${}^{\ast }$Institute of Physics, University of Tokyo, \\ Komaba, Meguro-ku, Tokyo 153-8902, Japan \\ \vspace{3mm}
${}^{\dagger }$Institute of Physics, Academy of Sciences of the Czech Republic, \\ Na Slovance 2, Prague 8, Czech Republic \\ \vspace{2mm}
${}^{\dagger }$Yukawa Institute of Theoretical Physics, Kyoto University, \\ Kyoto 606-8502, Japan \vspace{5mm}}

\usepackage[dvips]{graphicx}
\usepackage{amsmath,amscd, amssymb}
\usepackage[hypertex]{hyperref}
\usepackage{cite}
\usepackage{mathrsfs} 
\usepackage{amsfonts}
\usepackage[dvipdfmx, svgnames]{xcolor}%before tikz
\usepackage{tikz}

\usepackage{upgreek}

\makeatletter
  
  \@addtoreset{equation}{section}
\makeatother
\newcommand{\ld}{ [ \hspace{-0.6mm} [ }
\newcommand{\rd}{ ] \hspace{-0.6mm} ] }
\newcommand{\Ld}{ \big[ \hspace{-1.1mm} \big[ }
\newcommand{\Rd}{ \big] \hspace{-1.1mm} \big] }
\newcommand{\LD}{ \Big[ \hspace{-1.3mm} \Big[ }
\newcommand{\RD}{ \Big] \hspace{-1.3mm} \Big] }
\newcommand{\no}{\nonumber\\}
\newcommand{\niu}[1]{\pagebreak[2]\vspace{7pt}\noindent\underline{{\sf\hspace{4pt}#1\hspace{4pt}}}\vspace{4pt}}
\newcommand{\1}{\mbox{1}\hspace{-0.25em}\mbox{l}}

\newcommand{\teta}{{\widetilde{\eta }}}

\allowdisplaybreaks[3]

\begin{document}

\maketitle 
{\vspace{-128mm}
\rightline{\tt YITP/15-108}
\rightline{\tt UT-Komaba/15-9}
\vspace{128mm}}

\begin{abstract}
We propose new gauge invariant actions for open NS, heterotic NS, and closed NS-NS superstring field theories.
They are based on the large Hilbert space, and have Wess-Zumino-Witten-like expressions which are the $\mathbb{Z}_{2}$-reversed versions of the conventional WZW-like actions. 
On the basis of the procedure proposed in arXiv:1505.01659, we show that our new WZW-like actions are completely equivalent to $A_{\infty }/L_{\infty }$ actions proposed in arXiv:1403.0940 respectively.
\end{abstract}

\thispagestyle{empty}
\clearpage

\tableofcontents
\setcounter{page}{1}

%H.Matsunaga 
\section{Introduction and Summary} 

Recently, the formulation of superstring field theory has progressed gradually: Actions for superstring field theories were constructed and their properties have been clarified \cite{Sen:2015hha, Kunitomo:2015usa, Sen:2015uaa, Matsunaga:2015kra}. 
In addition to these, 1PI effective field theory approach has provided a good insight into self-dual gauge theory and supergravity \cite{Sen:2015nph}. 
One trigger of these developments was given by the establishment of the $A_\infty/L_\infty$ formulation: 
the gauge invariant actions for open NS string \cite{Erler:2013xta}, heterotic NS string, and closed NS-NS string \cite{Erler:2014eba} were constructed by giving the prescription to get the string products satisfying $A_{\infty } / L_{\infty }$ relations. 
This prescription was extended to the cases including the Ramond sector and the equations of motion were provided in the work of \cite{Erler:2015lya}. 
At least at the tree level, these theories reproduce S-matrices of perturbative superstring theory with insertions of picture-changing operators at external lines \cite{Konopka:2015tta}. 
One of our aims in this paper is to develop the understanding of the relation between these $A_\infty / L_{\infty }$ theories and other theories such as WZW-like theories\footnote{For other approaches, see also \cite{Witten:1986qs, Wendt:1987zh, Arefeva:1989cp, Preitschopf:1989fc, Saroja:1992vw, Jurco:2013qra}.}. 

\vspace{2mm} 

While these $A_{\infty } / L_{\infty}$ theories work well, we know that WZW-like theories \cite{Berkovits:1995ab, Berkovits:1998bt, Okawa:2004ii, Berkovits:2004xh, Matsunaga:2013mba, Matsunaga:2014wpa} would also give the same results from \cite{Berkovits:1999bs, Iimori:2013kha}\footnote{For the R sector, see \cite{Berkovits:2001im, Michishita:2004by, Kunitomo:2013mqa, Kunitomo:2014hba, Kunitomo:2014qla}. }. 
Actually, we have clear understandings for open superstrings: for the NS sector, the relation between the $A_\infty$ action given in \cite{Erler:2013xta} and the Berkovits WZW-like action \cite{Berkovits:1995ab} is clarified by the works of \cite{Erler:2015rra, Erler:2015uba, Erler:2015uoa}, 
and for the NS and R sectors, the equivalence of the complete action from which the equations of motion given in \cite{Erler:2015lya} are derived and the complete action proposed in \cite{Kunitomo:2015usa} is provided by the work of \cite{Matsunaga:2015kra}\footnote{See also a new result given by T.Erler, Y.Okawa, and T.Takezaki, JHEP 08 (2016) 012 [arXiv:1602.02582].}. 
It would be important to extend these understandings of open strings to the case of closed strings and to construct complete actions for heterotic and type II theories. 
However, it seems to be difficult to discuss them on the basis of the same procedure\footnote{By taking another approach, namely by constructing an $L_{\infty }$-morphism connecting their equations of motion, one can discuss the {\it on-shell equivalence} of $L_{\infty }$ actions \cite{Erler:2014eba} and conventional WZW-like actions \cite{Berkovits:2004xh, Matsunaga:2014wpa}. 
The existence of the $L_{\infty }$-morphism provides the equivalence of the solution spaces of two equations of motion up to $Q$-exact terms, which implies the on-shell states match each other. 
See \cite{Goto:2015hpa} for more details.} as \cite{Erler:2015rra} and known WZW-like actions \cite{Okawa:2004ii,Berkovits:2004xh, Matsunaga:2013mba,Matsunaga:2014wpa}: 
It necessitates other insights because of {\it the skew} between $A_{\infty } / L_{\infty }$ actions and these conventional WZW-like actions, which we explain. 
It would be helpful to consider their dual versions.

\vspace{2mm} 

Recall that the Berkovits theory is formulated on the large Hilbert space, which is the state space whose superconformal ghost sector is spanned by $\xi (z)$, $\eta (z)$, and $\phi (z)$ \cite{Friedan:1985ge}. 
In this paper, we write $\eta $ for the zero mode of $\eta (z)$ for brevity. 
An NS string field $\Phi $ of the Berkovits theory is a Grassmann even, ghost number $0$, and picture number $0$ state in the large Hilbert space: $\eta \, \Phi \not= 0$, and all string products are defined by Witten's star product\cite{Witten:1985cc}. 
Using a real parameter $t \in [0,1]$ and a path $\Phi (t)$ connecting $\Phi (0 ) = 0$ and $\Phi (1)=\Phi $, the action of the Berkovits theory is given by {\it the Wess-Zumino-Witten-type form} 
\begin{align*}
\nonumber 
S_{\rm Berkovits} [\Phi ] & = \langle e^{- \Phi } \eta e^{\Phi }  , \, e^{-\Phi }  Q e^{\Phi }  \rangle + \int _{0}^{1} dt \, \langle e^{- \Phi (t) } \partial _{t} e^{\Phi (t)} , \, \Ld e^{-\Phi (t)} \eta e^{\Phi (t)} , e^{- \Phi (t)} Q e^{\Phi (t)} \Rd ^{\ast }\rangle ,  
\end{align*}
where $Q$ is the BRST operator, $\langle A , B \rangle$ is the BPZ inner product of $A$ and $B$, and $\ld A , B \rd ^{\ast }$ is the graded commutator for the star products of $A$ and $B$. 
One can find that this action has nonlinear gauge invariances given by 
\begin{align*}
\nonumber 
\delta e^{\Phi } = e^{\Phi } ( \eta \, \Omega _{\sf B}) + (Q \, \Lambda _{\sf B}) e^{\Phi } , 
\end{align*}
where $\Lambda _{\sf B}$ and $\Omega _{\sf B}$ are gauge parameters. 
Like wise $\ld A, B \rd ^{\ast }$, in this paper, we write $\ld d_{1} , d_{2} \rd $ for the graded commutator of operators $d_{1}$ and $d_{2}$, 
\begin{align*}
\Ld d_{1} , d_{2} \Rd = d_{1} \, d_{2} - (-)^{d_{1} d_{2} } d_{2} \, d_{1} . 
\end{align*}
The upper index of $(-)^{d}$ denotes the grading of the operator $d$, namely, its ghost number. 

\niu{Conventional WZW-like form}

It was shown in \cite{Berkovits:2004xh} that Berkovits action can be written as {\it a WZW-like form} 
\begin{align*}
\nonumber 
S_{\rm Berkovits} [\Phi ] = \int_{0}^{1} \, \langle \widetilde{A}_{t} [ \Phi (t) ] , \, \eta \,\widetilde{A}_{Q} [ \Phi (t) ] \big)  \rangle  
\end{align*}
with {\it WZW-like functionals} $\widetilde{A}_{Q} = \widetilde{A}_{Q} [\Phi ]$ of the dynamical string field $\Phi $ defined by  
\begin{align*}
\nonumber 
\widetilde{A}_{Q} [ \Phi (t) ] \equiv e^{- \Phi (t)} \big( Q e^{\Phi (t)} \big)  , \hspace{5mm} 
\widetilde{A}_{t} [\Phi (t) ] \equiv  e^{- \Phi (t)} \big( \partial _{t} e^{\Phi (t)} \big) . 
\end{align*}
The equation of motion is given by the $t$-independent form
\begin{align*}
\eta \, \widetilde{A}_{Q} [ \Phi ] = 0, 
\end{align*}
and we can represent gauge transformations using nilpotent operators $Q_{\widetilde{A}_{Q}[\Phi ]}$ and $\eta$\,, 
\begin{align*}
\nonumber 
\widetilde{A}_{\delta } [\Phi ] = Q_{\widetilde{A}_{Q}[\Phi ]}  \widetilde{\Lambda } + \eta \, \widetilde{\Omega } , 
\end{align*}
where $\widetilde{\Lambda } \equiv e^{\Phi } \Lambda _{\sf B} e^{-\Phi }$ and $\widetilde{\Omega } \equiv \Omega _{\sf B}$ are redefined gauge parameters, and 
\begin{align*}
\nonumber 
\widetilde{A}_{\delta } [ \Phi ] \equiv e^{- \Phi } \big( \delta e^{\Phi } \big) , \hspace{5mm} 
Q_{\widetilde{A}_{Q}[\Phi ] } \widetilde{\Lambda} \equiv Q \, \widetilde{\Lambda } + m_{2} \big( \widetilde{A}_{Q} , \widetilde{\Lambda } \big) + m_{2} \big( \widetilde{\Lambda } , \widetilde{A}_{Q} \big)  .
\end{align*}
We write $m_{2} (A,B)$ for the star product of $A$ and $B$. 
A significant feature of this WZW-like form of the action is that one can obtain all properties of the action by using not explicit forms of functionals $\widetilde{A}_{Q}[ \Phi ] , \widetilde{A}_{t}[\Phi ]$ but only specific algebraical relations of these functionals, which we call {\it WZW-like relations}: 
\begin{align*}
& Q \, \widetilde{A}_{Q} [\Phi ] + m_{2} \big( \widetilde{A}_{Q} [\Phi ] , \widetilde{A}_{Q} [\Phi ] \big) = 0 , 
\no[2mm] \nonumber 
& \hspace{5mm} (-)^{d} d \, \widetilde{A}_{Q} [ \Phi ] = Q_{\widetilde{A}_{Q} [\Phi ]} \, \widetilde{A}_{d} [\Phi ] , 
\end{align*}
where $d = \partial _{t} , \delta , \eta $ and the upper index of $(-)^{A}$ denotes the Grassmann parity of $A$. 
While the first relation provides {\it the constraint} for $\widetilde{A}_{Q}[\Phi ]$, the second relation specifies its derivatives and the properties of {the equation of motion}. 
We call $\widetilde{A}_{d} [\Phi ]$ {\it an associated field}.

\vspace{2mm}

The first relation implies that the WZW-like functional $\widetilde{A}_{Q} [ \Phi ]$, which we call {\it a pure-gauge-like field}\footnote{The Maurer-Cartan equation for $M^{\sf B}$ and the explicit form of $\widetilde{A}_{Q}[\Phi ]$ have the same forms as the equation of motion and pure gauge field in bosonic open string field theory respectively. 
Actually, one can construct $\widetilde{A}_{Q}[\Phi ] \equiv \widetilde{A}_{Q} [ 1 ; \Phi ]$ as the $\tau =1$ value solution $\widetilde{A}_{Q} [ \tau ; \Phi ]$ of the differential equation $\partial _{\tau } \widetilde{A}_{Q} [ \tau ; \Phi ] = Q_{\widetilde{A}_{Q}[\tau ; \Phi ]} \Phi $ with the initial condition $\widetilde{A}_{Q} [ 0 ; \Phi ] = 0$, which is just the defining equation of the pure gauge in bosonic theory. }, gives a solution of the Maurer-Cartan equation for the bosonic open string products 
\begin{align*}
\nonumber 
{\boldsymbol M}^{\sf B} = {\boldsymbol Q} + {\boldsymbol m}_{2} , 
\end{align*} 
which consists of the BRST operator $Q$ and the star product $m_{2}$. 
Recall that $M^{\sf B}$ satisfies $A_{\infty }$-relations: $Q^{2} = 0$\,, $Q \, m_{2} ( A, B ) - m_{2} ( Q A , B ) - (-)^{A} m_{2} ( A , Q B ) = 0$\,, and $m_{2} \big( m_{2} ( A , B ) , C \big) - m_{2} \big( A , m_{2} ( B , C ) \big) = 0$\,. 
The derivation properties for $d = \partial _{t}, \, \delta , \, \eta \,$ also hold: $\ld d , Q \rd = 0$ and $d \, m_{2} (A,B) - m_{2} ( d A , B ) - (-)^{d A} m_{2} ( A , d B ) = 0$. 
In this paper, we symbolically\footnote{We use bold fonts for coalgebraic operations or notions. See appendix A.} write 
\begin{align*}
\nonumber 
\Ld {\boldsymbol M}^{\sf B} , \, {\boldsymbol M}^{\sf B} \Rd = 0 , \hspace{5mm}  \ld {\boldsymbol d} , {\boldsymbol M}^{\sf B} \rd = 0 
\end{align*}
for the $A_{\infty }$-relations of ${\boldsymbol M}^{\sf B}$ and the $d$-derivation properties of ${\boldsymbol M}^{\sf B}$. 
These properties yield the WZW-like relations, and then, the constraint equation and the equation of motion are given by 
\begin{align*}
{\rm Constraint}\,(\textrm{C-WZW}) \, : & \hspace{5mm} Q \, \widetilde{A}_{Q} [\Phi ] + m_{2} \big( \widetilde{A}_{Q} [\Phi ] , \widetilde{A}_{Q} [ \Phi ] \big) = 0 , 
\no[2mm] \nonumber 
{\rm E. O. M.} \,(\textrm{C-WZW}) \, : & \hspace{15mm} \eta \, \widetilde{A}_{Q} [ \Phi ]= 0 . 
\end{align*}

\vspace{2mm}

This construction is extended to more generic case: 
Open NS superstrings with stubs, heterotic NS strings\cite{Berkovits:2004xh}, and closed NS-NS strings\cite{Matsunaga:2014wpa}. 
As an example, we consider the Berkovits theory with stubs. 
The starting point is a set of generic bosonic open string products\cite{Gaberdiel:1997ia} 
\begin{align*}
\nonumber 
{\boldsymbol M}^{\sf B}_{\sf stub} = {\boldsymbol Q} + {\boldsymbol m}^{\sf st}_{2} +  {\boldsymbol m}^{\sf st}_{3} +  {\boldsymbol m}^{\sf st}_{4} + \dots  
\end{align*} 
which are nonassociateive but satisfy $A_{\infty }$-relations and derivation properties for $d= \partial _{t} , \, \delta ,\, \eta$: 
\begin{align*}
\nonumber 
\Ld {\boldsymbol M}^{\sf B}_{\sf stub} , \, {\boldsymbol M}^{\sf B}_{\sf stub} \Rd = 0 , \hspace{5mm}  \ld {\boldsymbol d} , {\boldsymbol M}^{\sf B}_{\sf stub} \rd = 0 . 
\end{align*}
Using these ${\boldsymbol M}^{\sf B}_{\sf stub}$, $d$, and the NS open string field $\varphi $, one can construct a pure-gauge-like field $\widetilde{A}_{Q}^{\sf st} [ \varphi ]$ and an associated field $\widetilde{A}_{d}^{\sf st} [ \varphi ]$ via the same type of the defining differential equations as those of the theory without stubs. 
The resultant theory is given by the following action 
\begin{align*}
S_{\sf stub} [\varphi ] = \int_{0}^{1} dt \, \langle \widetilde{A}_{t}^{\sf st} [ \varphi ] , \, \eta \, \widetilde{A}_{Q}^{\sf st} [ \varphi ] \rangle ,  
\end{align*}
and characterized by the pair of equations: 
\begin{align*}
{\rm Constraint}\,(\textrm{C-WZW}) \, : & \hspace{5mm} Q \, \widetilde{A}^{\sf st}_{Q} [\varphi ] + m^{\sf st}_{2} \big( \widetilde{A}^{\sf st}_{Q} [\varphi ] , \widetilde{A}^{\sf st}_{Q} [ \varphi ] \big) 
+ \sum_{n= 3}^{\infty } m^{\sf st}_{n} \big( \overbrace{\widetilde{A}^{\sf st}_{Q} [ \varphi ] , \dots , \widetilde{A}^{\sf st}_{Q} [ \varphi ] }^{n}  \big) = 0 , 
\no[2mm] \nonumber 
{\rm E. O. M.} \,(\textrm{C-WZW})\, : & \hspace{35mm} \eta \, \widetilde{A}^{\sf st}_{Q} [ \varphi ]= 0 . 
\end{align*}

However, when we compare this conventional WZW-like action and the $A_{\infty }$ action given in \cite{Erler:2014eba}, we notice that there exists a skew: $A_{\infty }$ theory is characterised by the pair of equations 
\begin{align}
\label{A_infty pair}
\begin{split}
{\rm Constraint} \,(A_\infty) \, : & \hspace{30mm} \eta \, \Psi = 0 , 
\\[2mm] 
{\rm E.O.M.} \,(A_\infty) \, : & \hspace{5mm} Q \, \Psi + M_{2} (\Psi , \Psi ) 
+ \sum_{n= 3}^{\infty } M_{n} \big( \overbrace{\Psi , \dots , \Psi }^{n}  \big) = 0 ,  
\end{split}
\end{align}
where $\Psi $ is the NS open string field of $A_{\infty }$ theory and $\{ M_{n} \} _{n=2}^{\infty }$ are NS superstring products. 
While the BRST operator $Q$ and the string products $\{ m^{\sf st}_{n} \} _{n=2}^{\infty }$ determine the constraint equation for the ingredients of the conventional WZW-like theory, that of $A_{\infty }$ theory is supplied by $\eta $. 
This skew is just the obstacle to obtain the off-shell equivalence of two theory by the naive way. 

\vspace{3mm} 

\niu{Alternative WZW-like form}

It is known that Berkovits action can be also written as {\it alternative WZW-like form} 
\begin{align}
\label{Berkovits action}
S_{\rm Berkovits} [\Phi ] = - \int _{0}^{1} dt \, \langle A_{t} [ \Phi (t) ] , \, Q \, A_{\eta } [ \Phi (t) ] \rangle  
\end{align}
with {\it alternative WZW-like functionals} $A_{\eta } = A_{\eta } [\Phi ]$ of the dynamical string field $\Phi $ defined by 
\begin{align}
\label{WZW-like functionals Berkovits}
A_{\eta } [ \Phi ] \equiv ( \eta e^{\Phi (t)} ) e^{- \Phi (t)} , \hspace{5mm} A_{t} [ \Phi ] \equiv ( \partial _{t} e^{\Phi (t)} ) e^{- \Phi (t)} . 
\end{align}
Then, the equation of motion is given by the $t$-independent form
\begin{align*}
Q \, A_{\eta } [ \Phi ] = 0, 
\end{align*}
and we can represent gauge transformations using nilpotent operators $D^{\ast }_{\eta }$ and $Q$\,, 
\begin{align*}
\nonumber 
A_{\delta } [\Phi ] = D_{\eta }^{\ast } \, \Omega + Q \, \Lambda , 
\end{align*}
where $\Lambda  \equiv \Lambda _{\sf B}$ and $\Omega \equiv e^{- \Phi } \Omega _{\sf B} e^{\Phi }$ are redefined gauge parameters, and 
\begin{align*}
\nonumber 
A_{\delta } [ \Phi ] \equiv \big( \delta e^{\Phi } \big) e^{- \Phi } , \hspace{5mm} 
D_{\eta }^{\ast } \, \Omega \equiv \eta \, \Omega - m_{2} \big( A_{\eta } [\Phi ] , \Omega \big) - (-)^{\Omega } m_{2} \big( \Omega , A_{\eta } [\Phi ] \big)  .
\end{align*}
A significant feature of this alternative WZW-like form is that, as the conventional case, one can obtain all properties of the action by using only specific algebraical relations of alternative WZW-like functionals, which we also call {\it WZW-like relations}\,: 
\begin{align}
\label{WZW-like relations Berkovits}
\begin{split}
& \eta \, A_{\eta } [\Phi ] - m_{2} \big( A_{\eta } [\Phi ] , A_{\eta } [\Phi ] \big) = 0 , 
\\[2mm] 
& \hspace{5mm} (-)^{d} d \, A_{\eta } [ \Phi ] = D_{\eta }^{\ast } \, A_{d} [\Phi ] , 
\end{split}
\end{align}
where $d = \partial _{t} , \delta , Q$. 
Therefore, we can say that this type of WZW-like theory also belongs to the category of so-called the WZW-like formulation. 
While the first relation provides the constraint for $A_{\eta }[\Phi ]$, the second relation specifies the properties of the equation of motion. 

\vspace{2mm}

Note that this type of WZW-like theory, a dual version of the conventional WZW-like theory, is characterized by the pair of equations
\begin{align*}
{\rm Constraint} \, (\textrm{A-WZW}) \, : & \hspace{5mm} \eta \, A_{\eta } [\Phi ] - m_{2} \big( A_{\eta } [\Phi ] , A_{\eta } [ \Phi ] \big) = 0 , 
\no[2mm] \nonumber 
{\rm E.O.M.} \, (\textrm{A-WZW}) \, : & \hspace{15mm} Q \, A_{\eta } [ \Phi ]= 0 . 
\end{align*}
One can find that there is no skew between this type of WZW-like theory and $A_{\infty }$ theory. 
Actually, as demonstrated in \cite{Erler:2015rra}, by decomposing the NS superstring product ${\boldsymbol M} = {\boldsymbol Q} + {\boldsymbol M}_{2} + {\boldsymbol M}_{3} + \dots $ given in \cite{Erler:2013xta} as ${\boldsymbol M} = \widehat{\bf G}^{-1} {\boldsymbol Q} \, \widehat{\bf G}$ and by using a redefined string field $A_{\eta } [\Psi ] \equiv \pi _{1} \widehat{\bf G} (1-\Psi )^{-1}$, we can transform the pair of equations (\ref{A_infty pair}) characterizing the $A_{\infty }$ action into 
\begin{align*}
\begin{split}
{\rm Constraint}  \,(A_\infty)\, : & \hspace{5mm} \eta \, A_{\eta } [ \Psi ]  - m_{2} \big( A_{\eta } [ \Psi ] , A_{\eta } [ \Psi ] \big) = 0 , 
\\[2mm] 
{\rm E.O.M.} \,(A_\infty) \, : & \hspace{15mm} Q \, A_{\eta } [\Psi ] = 0 . 
\end{split}
\end{align*}
Therefore, we expect that one can construct WZW-like actions which are off-shell equivalent to $A_{\infty }$ and $L_{\infty }$ actions proposed in \cite{Erler:2014eba} as this type of WZW-like theories.

\niu{Main results of this paper}

What is the starting point of this type of WZW-like theory? 
We can read it from the constraint for $A_{\eta } [\Phi ]$ or $A_{\eta } [\Psi ]$. 
For the Berkovits theory, it is given by the following $A_{\infty }$-product: 
\begin{align*}
\nonumber 
{\boldsymbol D}^{\boldsymbol \eta }_{\ast } \equiv {\boldsymbol \eta } - {\boldsymbol m}_{2} , 
\end{align*}
which we call {\it the dual products} of\, ${\bf M}$\, of \cite{Erler:2013xta}. 
We write $D^{\eta }_{2}$ for $- m_{2}$. 
Then, one can check that $\eta + D^{\eta }_{2}$ satisfies the $A_{\infty }$-relations of $\eta $: $\eta ^{2} = 0$, $\eta \, D^{\eta }_{2} (A , B) - D^{\eta }_{2}( \eta A , B ) - (-)^{A} D^{\eta }_{2} ( A , \eta B) = 0$, and $D^{\eta }_{2} \big( D^{\eta }_{2} ( A , B ) , C \big) - D^{\eta }_{2} \big( A , D^{\eta }_{2} ( B , C ) \big) = 0$. 
The derivation properties for $d = \partial _{t} , \, \delta , \, Q$\, also hold: $\ld d , \eta \rd = 0$ and $d \, D^{\eta }_{2} (A,B) - D^{\eta }_{2} ( d A , B) - (-)^{A} D^{\eta }_{2} (A, d B ) = 0$. 
We write them as follows: 
\begin{align*} 
\ld {\boldsymbol D}^{\boldsymbol \eta }_{\ast } , \, {\boldsymbol D}^{\eta }_{\ast } \rd = 0 , \hspace{5mm} \ld {\boldsymbol d} , {\boldsymbol D}^{\boldsymbol \eta }_{\ast } \rd = 0, 
\end{align*}
which give the starting point of our alternative WZW-like theory. 
Actually, one can construct the pure-gauge-like field $A_{\eta } [\Phi ] = (\eta e^{\Phi } ) e^{-\Phi }$ of (\ref{WZW-like functionals Berkovits}) and the action (\ref{Berkovits action}) using this $D^{\eta }_{\ast } = \eta - m_{2}$: the pure-gauge-like field $A_{\eta } [\Phi ]$ in (\ref{WZW-like functionals Berkovits}) is given by the $\tau =1$ value $A_{\eta } [ \Phi ] \equiv A_{\eta } [ \tau =1 ; \Phi ]$ of the solution $A_{\eta } [\tau ; \Phi ]$ of the differential equation 
\begin{align*}
\frac{\partial }{\partial \tau } A_{\eta } [ \tau ; \Phi ] = \eta \, \Phi - m_{2} \big( A_{\eta } [ \tau ; \Phi ] , \Phi \big) + m_{2} \big( \Phi , A_{\eta } [\tau ; \Phi ] \big) 
\end{align*}
with the initial condition $A_{\eta } [ \tau = 0 ; \Phi ] = 0$, where $\tau \in [0,1]$ is a real parameter. 

\vspace{2mm} 

In this paper, we show that a nonassociative extended version of this construction gives our new WZW-like theory, which provides the equivalence of $A_{\infty }/L_{\infty }$ formulation and WZW-like formulation. 
The starting point is the following $A_{\infty }$ products 
\begin{align*} 
\nonumber 
{\boldsymbol D}^{\boldsymbol \eta } = {\boldsymbol \eta } + {\boldsymbol D}^{\boldsymbol \eta }_{2} + {\boldsymbol D}^{\boldsymbol \eta }_{3} + {\boldsymbol D}^{\boldsymbol \eta }_{4} + \dots  
\end{align*} 
satisfying the $A_{\infty }$-relations of $\eta $ and derivation properties for $d = \partial _{t} , \, \delta , \, Q \,$: 
\begin{align*} 
\nonumber 
\Ld {\boldsymbol D}^{\boldsymbol \eta } , \, {\boldsymbol D}^{\eta } \Rd = 0 , \hspace{5mm} 
\Ld {\boldsymbol d} , {\boldsymbol D}^{\boldsymbol \eta } \Rd = 0  . 
\end{align*} 
One can construct these products by the dual products of ${\boldsymbol M} = \widehat{\bf G}^{-1} {\boldsymbol Q} \, \widehat{\bf G}$ given in \cite{Erler:2014eba}, namely ${\boldsymbol D}^{\boldsymbol \eta } \equiv \widehat{\bf G} \, {\boldsymbol \eta } \, \widehat{\bf G}^{-1}$. 
Then, using these dual products ${\boldsymbol D}^{\boldsymbol \eta }$, we propose alternative WZW-like actions which are equivalent to $A_{\infty }$ actions proposed in \cite{Erler:2014eba} as this type of WZW-like theories 
\begin{align}
\label{Alternative WZW-like action}
S_{\eta } [\varphi ]= - \int_{0}^{1} dt \, \langle A_{t} [ \varphi (t)] , \, Q \, A_{\eta } [ \varphi (t)] \rangle , 
\end{align}
where $\varphi $ is a dynamical NS string field, $t \in [0 ,1]$ is a real parameter. 
The WZW-like functionals $A_{\eta } [ \varphi ]$ and $A_{t} [ \varphi ]$ satisfy the nonassociative extended versions of (\ref{WZW-like relations Berkovits}): 
\begin{align}
\label{relation}
\begin{split}
& \eta A_{\eta } [ \varphi ] - m_{2} ( A_{\eta } [\varphi ] , A_{\eta } [\varphi ] ) + \sum_{n=3}^{\infty } D^{\eta }_{n} \big( \overbrace{A_{\eta } [ \varphi ] , \dots , A_{\eta } [\varphi ] }^{n} \big) = 0 ,  
\\[2mm] 
& \hspace{25mm} (-)^{d} d \, A_{\eta } [ \varphi ] = D_{\eta } A_{d} [ \varphi ] , 
\end{split} 
\end{align}
where now the $A_{\eta }$-shifted dual product $D_{\eta }$ is given by 
\begin{align*}
& D_{\eta } \Omega \equiv \eta \, \Omega  - m_{2} \big( A_{\eta } [ \varphi ] , \Omega \big) - (-)^{\Omega } m_{2} ( \Omega , A_{\eta } [\varphi ] ) + \sum_{\rm cyclic} \sum_{n=2}^{\infty } (-)^{\rm sign} D^{\eta }_{n+1} \big( \overbrace{A_{\eta } [\varphi ] , \dots , A_{\eta } [ \varphi ] }^{n} , \Omega \big)  . 
\end{align*} 
Note that this WZW-like action for generic open NS strings just reduces to the alternative WZW-like form (\ref{Berkovits action}) of the Berkovits theory when we take the star product. 
We would like to emphasize that we do not need a specific form of $A_{\eta }[\varphi ]$ or $A_{t} [\varphi ]$ as a functional of given dynamical string field $\varphi $ but only their properties (\ref{relation}) to show the properties of the action: 
Its variation, equations of motion, gauge invariance, and so on. 
In appendix D, we will give explicit forms of two realizations of these functionals $A_{\eta }$ and $A_{d}$ using two different dynamical string fields: $\Psi $ in the small Hilbert space and $\Phi $ in the large Hilbert space. 
Namely, the equivalence of $A_{\infty }$ and WZW-like actions for open superstring field theory with stubs. 

\vspace{2mm}

In the above, we take open NS theory as an example to grab a feature of our alternative WZW-like approach and its necessity. 
The details of the above open NS theory are discussed in appendix D. 
In the following sections, our main topic is ``heterotic NS theory'': 
We consider the $L_{\infty }$ action and explain how its WZW-like properties arise. 
On the basis of the WZW-like structure naturally arising from $L_{\infty }$ actions, we propose new gauge invariant actions for heterotic NS (and NS-NS strings in appendix E), as well as that for open NS strings with stubs which we introduced in (\ref{Alternative WZW-like action}). 
These actions are $\mathbb Z_2$-reversed versions of the conventional ones \cite{Berkovits:2004xh, Matsunaga:2014wpa}, 
and we show that they are completely equivalent to the $A_{\infty }/L_{\infty }$ actions proposed in \cite{Erler:2014eba}. 

\vspace{2mm} 

We expect that our new WZW-like actions would provide a first step to construct complete actions for heterotic and type II string field theories\footnote{See a new result given by K.Goto and H.Kunitomo, arXiv:1606.07194.}. 
Actually, an action for open superstring field theory including the R sector was constructed \cite{Kunitomo:2015usa} by starting with this type of WZW-like action: 
The R string field couples to the Berkovits theory for the NS sector gauge-invariantly on the basis of (not the conventional but) this type of WZW-like gauge structure. 
We would like to mention that although we expect that our new WZW-like actions are also equivalent to the conventional WZW-like actions, the all order equivalence of these has not been proven: 
We will show lower-order equivalence to the conventional WZW-like actions only. 

\vspace{2mm} 

This paper is organized as follows.
% K.Goto 
In section 2, after a brief review of the $L_\infty$ formulation, we clarify a WZW-like structure naturally arising from it. 
We show that the $L_\infty$ action can be written in our (alternative) WZW-like form: 
The functionals appearing in the action satisfy alternative WZW-like relations, the $\mathbb Z_2$-reversed versions of the conventional WZW-like relations given in \cite{Berkovits:2004xh}, which guarantees the gauge invariance of the action.
The on-shell condition and the gauge transformation of the (alternative) WZW-like action are derived by using only the (alternative) WZW-like relations. 
We also see how the gauge parameters appearing in the WZW-like form are parameterised by that in the $L_\infty$ form.
Then, we conclude the $L_\infty$ action gives one realization of the alternative WZW-like action $S_{\eta } [\Phi ]$ parametrized by the dynamical string field $\Phi$ in the small Hilbert space. 
% K.Goto 
In section 3, we provide another realisation of the alternative WZW-like action using the string field $V$ in the large Hilbert space. 
The functionals satisfying our WZW-like relations can be defined by the differential equations which are the $\mathbb Z_2$-reversed versions of those given in \cite{Berkovits:2004xh}. 
Utilizing them, we construct a new gauge invariant action $S_{\eta }[V]$ for heterotic NS string field theory.
We derive the condition for the equivalence of the new action $S_{\eta }[V]$ and the $L_\infty$ action on the basis of the procedure demonstrated in \cite{Erler:2015rra}: 
These are different parameterizations of the same WZW-like structure and action. 
Then we also derive the relation between two dynamical string fields $\Phi$ and $V$ from the equivalence condition.
We end with some conclusions and discussions. 

% H.Matsunaga 
Basic facts and definitions of the coalgebraic notation of $A_{\infty } / L_{\infty }$ algebras are summarized in appendix A. 
In appendix B, we derive a formula which is used in section 2. 
In appendix C, we consider the trivially embedding of the string field of the $L_\infty$ action belonging to the small Hilbert space
into the string field in the large Hilbert space, and show the embedded action can also be written in the WZW-like form.
Appendices D and E are devoted to the open NS and the closed NS-NS theories, respectively. 

%%%%%%%%%%%%%%%%%%%%%%%%%%%%%%%%%%%%%%%%%%%%%%%%%%%%%%%%%%%%%%%%%%%%%%%%%%%%%%%%%%%%%%%%

%Keiyu Goto 
\section{WZW-like structure from the $L_{\infty }$ formulation} 

In this section, we clarify a WZW-like structure naturally arising from the $L_\infty$ formulation for NS heterotic string field theory. 
After a brief review of the $L_\infty$ formulation \cite{Erler:2014eba},
we show that the $L_{\infty }$ action can be written in our (alternative) WZW-like form.
We see that the functionals appearing in the action satisfy alternative WZW-like relations, the $\mathbb Z_2$-reversed version of the conventional WZW-like relations in \cite{Berkovits:2004xh}, which guarantees the gauge invariance of the action.

\niu{Preliminaries}

The product of $n$ closed strings is described by 
a multilinear map $b_n:\mathcal H^{\wedge n}\to \mathcal H$,
where $\wedge$ is the {\it symmetrized tensor product} satisfying
$\Phi_1\wedge\Phi_2
=(-)^{{\rm deg}(\Phi_1){\rm deg}(\Phi_2)}\Phi_2\wedge\Phi_1$.
A map $b_n:\mathcal H^{\wedge n}\to \mathcal H$ with degree $1$
naturally induces a map from 
the {\it symmetrized tensor algebra} 
$\mathcal {S(H)}=\mathcal H^{\wedge 0}\oplus\mathcal H^{\wedge 1}\oplus\mathcal H^{\wedge 2}\oplus\cdots$
to $\mathcal S(\mathcal H)$ itself,
called a {\it coderivation}.
A map $b_n:\mathcal H^{\wedge n}\to \mathcal H'$ with degree $0$
also naturally induces a map from 
$\mathcal {S(H)}$ to $\mathcal S(\mathcal H')$,
called a {\it cohomomorphism}.
Since it is convenient to write the functionals of the string field and the action in terms of them,
we briefly introduce the rules of their actions here.

The coderivation ${\bf d}_n:\mathcal {S(H)}\to\mathcal {S(H)}$ is naturally derived from a map $d_n:{\cal H}^{\wedge n}\to\cal H$ with degree one.
It act on $\Phi _{1}\wedge \dots \wedge \Phi _{N} \in \mathcal H^{\wedge N\geq n}\subset  \mathcal{S(H)}$ as
\begin{align}
{\bf d}_n (\Phi _{1}\wedge \dots \wedge \Phi _{N})
&=(d_n\wedge\mathbb I_{N-n})(\Phi _{1}\wedge \dots \wedge \Phi _{N})\no
&= \sum_{\sigma}\frac{(-)^{\sigma}}{n!(N-n)!} d_n(\Phi_{\sigma(1)},\cdots,\Phi_{\sigma(n)})\wedge \Phi_{\sigma(n+1)} \wedge \cdots \wedge\Phi_{\sigma(N)},
\end{align}
and vanishes when acting on $ {\cal H}^{\wedge N\leq n}$.
The graded commutator of two coderivations ${\bf b}_{n}$ and ${\bf c}_{m}$, $[\![{\bf b}_n ,{\bf c}_m]\!]$,
is a coderivation derived from the map $[\![b_n,c_m]\!]:{\cal H}^{\wedge n+m-1}\to \cal H$ which is defined by 
\begin{align}
[\![b_n,c_m]\!]&=b_n(c_m\wedge\mathbb I_{n-1})-(-)^{{\rm deg}(b_n){\rm deg}(c_m)}c_m(b_n\wedge\mathbb I_{m-1}).
\end{align}

A set of degree zero multilinear maps $ \{ \mathsf f_n : \mathcal H^{\wedge n}\to \mathcal H'\}_{n=0}^{\infty}$ 
naturally induces a cohomorphism $\widehat{\mathsf f} : \mathcal{S(H)}\to \mathcal{S(H')}$,
which acts on $\Phi _{1}\wedge \dots \wedge \Phi _{n} \in \mathcal H^{\wedge n}\subset  \mathcal{S(H)}$ as
\begin{align}
\widehat{\mathsf f} (\Phi _{1}\wedge \dots \wedge \Phi _{n})=
\sum_{i \leq n} \sum_{k_{1} < \dots < k_{i}} 
&e^{\wedge \mathsf f_0}\wedge{\sf f}_{k_{1}} ( \Phi _{1} , \dots , \Phi _{k_{1}}) \wedge {\sf f}_{k_{2}-k_{1}} ( \Phi _{k_{1} +1} , \dots , \Phi _{k_{2}} ) \wedge \no
&\hspace{15pt}\dots \wedge  {\sf f}_{k_{i} -k_{i-1}} ( \Phi _{k_{i-1}+1} , \dots , \Phi _{n} ).
\end{align}
For definitions and their more details, see appendix A.

%%%%%%%%%%%%%%%%%%%%%%%%%%%%%%%%%%%%%%%%%%%%%%%%%%%%%%%%%%%%%%%%%%%%%%%%%%%%%%%%%%%%%%%%
\subsection{Construction of $L_\infty$-product and $L_\infty$ action}
In this subsection,
we briefly review the construction of the NS superstring product $\mathbf L$ and the action $S_{\scriptscriptstyle \rm EKS}[\Phi]$
for heterotic string field theory in the $L_{\infty }$ formulation originally given in \cite{Erler:2014eba}.

\niu{NS superstring product ${\bf L}$}

Let us review the construction of the NS string products
$\mathbf L{\scriptstyle[\tau]}= \sum_{p=0}^{\infty} \tau^p \mathbf L_{p+1}$
satisfying $L_\infty$-relation $[\![\mathbf L{\scriptstyle[\tau]},\mathbf L{\scriptstyle[\tau]}]\!]=0,$
the cyclicity $\mathbf L^\dagger =-\mathbf L$,
and the $\eta$-derivation $[\![\eta,\mathbf L{\scriptstyle[\tau]}]\!]=0$.
The $(p+1)$-product $\mathbf L_{p+1}$ carries the ghost number $1-2p$ and the picture number $p$.
The products $\mathbf L{\scriptstyle[\tau]}$ consist of $\eta$, $\xi$, and Zwiebach's bosonic string products $\mathbf L^\mathsf{B}= \mathbf Q + \mathbf L_2^\mathsf B + \mathbf L_3^\mathsf B +\cdots$ satisfying $[\![\mathbf L^\mathsf{B},\mathbf L^\mathsf{B}]\!]=0$ \cite{Zwiebach:1992ie}. 
First, we focus on the condition for $\mathbf L{\scriptstyle[\tau]}$ to satisfy the $L_\infty$ relations, 
\begin{align}
[\![\mathbf L{\scriptstyle[\tau]},\mathbf L{\scriptstyle[\tau]}]\!]=0.
\end{align}
The $L_\infty$ relations holds if we define $\mathbf L{\scriptstyle[\tau]}$ as a solution of the differential equations
\begin{align}
\partial_\tau \mathbf L{\scriptstyle[\tau]} = [\![\mathbf L{\scriptstyle[\tau]},\uplambda^{[0]}{\scriptstyle[\tau]}]\!],
\end{align}
with the initial condition $\mathbf L{\scriptstyle[\tau=0]}=Q$.
Here $\uplambda^{[0]}{\scriptstyle[\tau]}=\sum_{p=0}^\infty \tau^p \boldsymbol \uplambda^{[0]}_{p+2}$
are called gauge products.
We can take any $\uplambda^{[0]}$ as long as they carry correct quantum numbers:
the $(p+2)$-product $\uplambda^{[0]}_{p+2}$ carryies ghost number $-2(p+1)$ and picture number $p+1$.
The solution for the differential equations is given by the similarity transformation of $Q$, 
\begin{align}
{\bf L}{\scriptstyle[\tau]} &= \widehat{\bf G}^{-1}{\scriptstyle[\tau]}{\bf Q} \widehat{\bf G}{\scriptstyle[\tau]},
\end{align}
where $\widehat{\mathbf G}$ is an invertible cohomomorphism defined by the path-ordered exponential of the gauge products $\boldsymbol \uplambda^{[0]}{\scriptstyle[\tau]}$ as follows 
\begin{align}
\widehat{\bf G}{\scriptstyle[\tau]}=\overset{\leftarrow}{\mathcal P} \exp \left( \int_{0}^{\tau} d\tau' {\boldsymbol \uplambda^{[0]}}{\scriptstyle[\tau']}\right),
\label{G}\qquad
\widehat{\bf G}^{-1}{\scriptstyle[\tau]}
=\overset{\rightarrow}{\mathcal P} \exp \left(-\int_{0}^{\tau} d\tau' {\boldsymbol \uplambda^{[0]}}{\scriptstyle[\tau']}\right) . 
\end{align}
Here $\leftarrow$($\rightarrow$) on ${\mathcal P}$ denotes that the operator at later time acts from the right (left).
%the path ordered exponential in sequence of increasing (decreasing) $\tau$,
The differential equations hold since the path-ordered exponentials satisfy 
\begin{align}
\partial_\tau \widehat{\mathbf G}{\scriptstyle[\tau]}= \widehat{\mathbf G} {\scriptstyle[\tau]} \boldsymbol \uplambda^{[0]}{\scriptstyle[\tau]},\qquad
\partial_\tau \widehat{\mathbf G}^{-1}{\scriptstyle[\tau]}=-\boldsymbol \uplambda^{[0]}{\scriptstyle[\tau]} \widehat{\mathbf G}^{-1}{\scriptstyle[\tau]}.
\end{align}
%We may omit the arguments if $\tau_f=1, \tau_i =0$.
We may check directly the $L_\infty$ relations: 
\begin{align}
{\bf L}^2 
=\widehat{\bf G}^{-1} {\bf Q} \widehat{\bf G} \widehat{\bf G}^{-1} {\bf Q} \widehat{\bf G}
=\widehat{\bf G}^{-1} \widehat{\bf Q} {\bf Q} \widehat{\bf G}
=0.
\end{align}
%The $L_\infty$-relations hold as long as the gauge products $\uplambda^{[0]}$.
The cyclicity and the $\eta$-derivation properties follow from the suitable choice of $\uplambda^{[0]}$.

Second, let us consider the cyclicity.
For $\mathbf L$ to be cyclic, it is sufficient to choose $\boldsymbol \uplambda^{[0]}$ to be BPZ-odd, so that $\widehat{\mathbf G}^{-1}=\widehat{\mathbf G}^\dagger$:
\begin{align}
\mathbf L^\dagger 
=(\widehat{\mathbf G}^{-1} \mathbf Q \widehat{\mathbf G} )^\dagger
=-\widehat{\mathbf G}^{-1} \mathbf Q \widehat{\mathbf G} =-\mathbf L.
\end{align}

Third, we check $\eta$ acts as a derivation on $\mathbf L$, namely the $\eta$-derivation properties
\begin{align}
[\![\upeta,\mathbf L{\scriptstyle[\tau]}]\!]=0.
\end{align}
A construction of the suitable gauge product $\uplambda^{[0]}$ is given in \cite{Erler:2014eba}.
It is helpful to consider 
a series of generating functions
\begin{align}
\mathbf L(s,\tau)
%= \mathbf G^{-1}(s;t) \mathbf L^{\rm BOS}(s) \mathbf G(s;t) 
= \sum_{d=0}^{\infty} s^d \mathbf L^{[d]}{\scriptstyle[\tau]}
= \sum_{d=0}^{\infty} \sum_{p=0}^{\infty} s^d\tau^p \mathbf L^{[d]}_{p+d+1},
%=\sum_{d=0}^\infty \sum_{p=0}^\infty s^d \tau^p  \mathbf L^{(p)}_{p+d+1}
%=\sum_{p=0}^\infty \tau^p  \mathbf L^{(p)}(s).
\end{align}
including $\mathbf L{\scriptstyle[\tau]}= \mathbf L^{[0]}{\scriptstyle[\tau]}$
and $\mathbf L^\mathsf{B}_N =\mathbf L^{[N-1]}_N$,
where the superscript $[d]$ denotes the picture deficit relative to what is needed for the NS products.
They satisfy the $L_\infty$-relations and $\eta$-derivation properties, 
\begin{align}
[\![\mathbf L(s,\tau),\mathbf L(s,\tau)]\!]&=0,\\
[\![\upeta,\mathbf L(s,\tau)]\!]&=0,\label{eta_deri_Lstau}
\end{align}
if we define $\mathbf L(s,\tau)$ as a solution for the following differential equations, 
\begin{align}
\partial_\tau \mathbf L(s,\tau)
&=[\![ \mathbf L(s,\tau),\uplambda(s,\tau)]\!] ,\label{DE_Lstau1}\\
\partial_s\mathbf L(s,\tau)
&=[\![\eta, \uplambda(s,\tau)]\!]. \label{DE_Lstau2}
\end{align}
Here $\uplambda(s,\tau)$ is a series of generating functions for the gauge products
\begin{align}
\boldsymbol \uplambda (s,\tau) 
=\sum_{d=0}^\infty s^d \boldsymbol \uplambda^{[d]} {\scriptstyle[\tau]} 
=\sum_{d=0}^\infty \sum_{p=0}^\infty s^d \tau^p \boldsymbol \uplambda^{[d]}_{d+p+2},
%=\sum_{d=0}^\infty \sum_{p=0}^\infty s^d \tau^p \boldsymbol \uplambda^{(p+1)}_{d+p+2}
%=\sum_{p=0}^\infty \tau^p \boldsymbol \uplambda^{(p+1)}(s),
\end{align}
including $\boldsymbol \uplambda {\scriptstyle[\tau]} =\boldsymbol \uplambda^{[0]} {\scriptstyle[\tau]} $.
The expansion of the differential equations (\ref{DE_Lstau1}) and (\ref{DE_Lstau2}) in powers of $s$ and $\tau$
provides the recursive definition of $\mathbf L^{[d]}_N$ and $\boldsymbol\uplambda^{[d]}_N$.
Note that the explicit forms of $\boldsymbol\uplambda^{[d]}_N$ is not determined uniquely even if we require all the desired properties,
which follows from the arbitrariness of the inverse of $\eta$ in (\ref{DE_Lstau2}).
Then we obtain the suitable $\boldsymbol\uplambda^{[0]}{\scriptstyle[\tau]} $ which provides $\eta$-derivation properties of $\mathbf L{\scriptstyle[\tau]}  =\mathbf L^{[0]}{\scriptstyle[\tau]} $,
the $s^0$ part of (\ref{eta_deri_Lstau}).
Hereafter we take $\tau=1$ and omit the argument: 
\begin{align}
\mathbf L:=\mathbf L{\scriptstyle [\tau =1]}.
\end{align}

%%%%%%%%%%%%%%%%%%%%%%%%%%%%%%%%%%%%%%%%%%%%%%%%%%%%%%%%%%%%%%%%%%%%%%%%%%%%%%%%%%%%%%%%
\niu{Action for heterotic string field theory in the $L_{\infty }$ formulation}

A gauge-invariant action in the $L_\infty$ formulation is constructed by \cite{Erler:2014eba},
using the above products $\mathbf L = \{L_k\}_{k\geq 1}$.
The dynamical string field $\Phi$ in the $L_\infty$ formulation carries ghost number $2$ and picture number $-1$
and belongs to the small Hilbert space $\mathcal H_\mathrm{small}$: $\eta\Phi=0$.
The action is written as follows:
\begin{align}
S_{\scriptscriptstyle\rm EKS}[\Phi]
%&=\frac12\langle \xi\Phi, Q\Phi \rangle+\sum_{n=1}^\infty \frac{\kappa^{n}}{(n+2)!}\langle \xi\Phi, L_{n+1}(\overbrace{\Phi,\Phi,...,\Phi}^{n+1}) \rangle\no
&=\sum_{n=0}^\infty \frac{1}{(n+2)!}\langle \xi\Phi, L_{n+1}(\overbrace{\Phi,\Phi,...,\Phi}^{n+1}) \rangle,
\end{align}
where  $L_1=Q$ and $\langle A, B\rangle$ is the BPZ inner product with the $c_0^- =\frac12( c_0-\widetilde{c}_0)$-insertion.

Let us introduce 
the $t$-parametrized string field $\Phi (t)$ with $t \in [ 0 , 1 ]$ satisfying $\Phi ( 0 ) = 0$ and $\Phi (1) = \Phi $,
which is a path connecting $0$ and the string field $\Phi $ in the space of string fields.
Utilizing this $\Phi(t)$, the action can be represented in the following form:
\begin{align}
S_{\scriptscriptstyle \rm EKS } [\Phi]
%&=\sum_{n=0}^\infty \frac{\kappa^{n}}{(n+2)!}\langle \xi\Phi, L_{n+1}(\overbrace{\Phi,\Phi,...,\Phi}^{n+1}) \rangle\no 
& = \int_{0}^{1} dt \, \frac{\partial }{\partial t} \Big( \sum_{n=0}^{\infty } \frac{1 }{(n+2)!} \langle \xi \Phi (t) , L_{n+1} ( \overbrace{\Phi (t) , \dots , \Phi (t)}^{n+1} ) \rangle  \Big) \no
& = \int_{0}^{1} dt \sum_{n=0}^{\infty } \frac{1 }{(n+1)!} \langle \partial_t\xi \Phi (t) , L_{n+1} ( \overbrace{\Phi (t) , \dots , \Phi (t)}^{n+1} ) \rangle.
\end{align}
Here we use $\Phi(t) =\eta\xi\Phi(t)$, $\eta$-derivation property, and the cyclicity of $\mathbf L$ in order to obtain the form in which both $\partial_t$ and $\xi$ act on the first slot of the inner product.
Note that this $t$-dependence is topological and it does not appear in the variation of the action, as we will see later. 
We can represent the action in the coalgebraic notation as follows:
\begin{align}
S_{\scriptscriptstyle \rm EKS } [\Phi]& = \int_{0}^{1} dt \, \langle {\pi_1 } ( {\boldsymbol \xi }_{t} \, e^{\wedge \Phi (t)} ) , \,  {\pi_1 } \big( {\bf L}( e^{\wedge \Phi (t)} ) \big) \rangle,
\end{align}
where $e^{\wedge \Phi}$ is the {\it group-like element} defined by
\begin{align}
e^{\wedge \Phi} = {\bf 1} +\Phi + \frac12 \Phi\wedge\Phi +\frac1{3!}\Phi\wedge\Phi\wedge\Phi +\cdots,
\end{align}
$\pi_1 $ is a projector from the symmetrized tensor algebra to the single-state space 
\begin{align}
\pi _{1} \big( \Phi _{0} + \Phi _{1} \wedge \Phi _{2} + \Phi _{3} \wedge \Phi _{4} \wedge \Phi _{5} + \dots \big) = \Phi _{0} , 
\end{align}
and $\boldsymbol{\xi}_t$ is a coderivation\footnote{We will define $\xi_ {d}$ for the more general class of $ {d}$ later. For $\partial_t $, definitions are equivalent.} derived from the linear map $\xi \partial_t : \Phi  \mapsto \xi \partial _{t} \Phi $. 
(See appendix A.)

The variation of the action can be taken as 
\begin{align}
\delta S_{\scriptscriptstyle \rm EKS } [\Phi]
& =  \langle \xi \delta \Phi , {\pi_1 } \big( {\bf L}( e^{\wedge \Phi } ) \big) \rangle,\label{GT EKS Linf}
\end{align}
and then the equation of motion is given by
\begin{align}
\pi_1{\bf L}( e^{\wedge \Phi} ) =0.\label{EOM EKS}
\end{align}
Since $\mathbf L^2 =0$, the action is invariant under the gauge transformation 
\begin{align}
\delta \Phi =\pi_1\mathbf L(\lambda \wedge e^{\wedge\Phi}),\label{GT EKS}
\end{align}
where the gauge parameter $\lambda$ is in the small Hilbert space and carries ghost number $1$ and picture number $-1$. 

%%%%%%%%%%%%%%%%%%%%%%%%%%%%%%%%%%%%%%%%%%%%%%%%%%%%%%%%%%%%%%%%%%%%%%%%%%%%%%%%%%%%%%%%
\subsection{Alternative WZW-like form of $L_{\infty }$ action}
\label{SUBSEC: WZW in EKS}

The action in the $L_\infty$ formulation can be transformed as 
\begin{align}
S_{\scriptscriptstyle{\rm EKS}} [\Phi]
& = \int_{0}^{1} dt \, \langle {\pi_1 } ( \boldsymbol{\xi }_{t} \, e^{\wedge \Phi (t)} ) , \, {\pi_1 } \big( \widehat{\bf G}^{-1} \, {\bf Q} \, \widehat{\bf G}  ( e^{\wedge \Phi (t)} ) \big) \rangle 
\no 
& = \int_{0}^{1} dt \, \langle {\pi_1 } \big( \widehat{\bf G} ( \boldsymbol{\xi }_{t} e^{\wedge \Phi (t)} ) \big)  ,  \, {\pi_1 }\,  \mathbf Q \big(\widehat{\bf G} ( e^{\wedge \Phi (t)} ) \big) \rangle .
\end{align}
See \cite{Goto:2015hpa} for heterotic strings, and see also \cite{Erler:2015rra} for open strings. 
We find that the functionals $\Psi _{\eta } = \Psi _{\eta } [ \Phi ]$ of the dynamical string field $\Phi$ defined by 
\begin{align}
\Psi _{\eta } [\Phi(t)] &\equiv \pi _{1} \widehat{\bf G} \big( e^{\wedge \Phi (t)} \big) , \label{pure-gauge L-infty sec2}\\[2pt]
\Psi _{ {d}} [\Phi( t)] &\equiv \pi _{1} \widehat{\bf G} \big( {\boldsymbol \xi }_{ {\mathbf d}} e^{\wedge \Phi (t) } \big)\label{associated L-infty sec2},
\end{align}
appear in the action, and we will find that these functionals play important roles.
One can show that, by introducing a certain set of products satisfying $L_\infty$-relations, 
\begin{align}
\eta ,\quad [\:\:\cdot\:\:,\:\:\cdot\:\:]^\eta ,\quad [\:\:\cdot\:\:,\:\:\cdot\:\:,\:\:\cdot\:\:]^\eta,\quad\cdots ,
\label{dual products}
\end{align}
the functionals $\Psi_\eta[\Phi(t)]$ and $\Psi_ {d}[\Phi(t)]$ satisfy the (alternative) WZW-like relations:
\begin{align}
0&= \eta \, \Psi _{\eta } + \sum_{n=1}^{\infty } \frac{1 }{(n+1)!} [\overbrace{ \Psi _{\eta } , \dots , \Psi _{\eta }  }^{n+1} ]^{\eta }  , \label{W1}
\\ 
\qquad (-)^{ {d}}  {d} \, \Psi _{\eta } &
= \eta \, \Psi _{ {d}} + \sum_{k=1}^{\infty } \frac{1}{k! } \big[ \overbrace{\Psi _{\eta } , \dots , \Psi _{\eta }  }^{k} , \Psi _{ {d}} \big] ^{\eta }, \label{W2}
\end{align}
which are $\mathbb Z_2$-reversed versions of conventional WZW-like relations in \cite{Berkovits:2004xh}. 
In this subsection, after defining these $L_\infty$ products (\ref{dual products}), which we call the dual $L_{\infty }$ products for EKS's $L_{\infty }$ products, we confirm a pair of fields (\ref{pure-gauge L-infty sec2}) and (\ref{associated L-infty sec2}) satisfy the WZW-like relations (\ref{W1}) and (\ref{W2}). 
We write ${\bf L}^{\boldsymbol \eta }$ for this set of dual $L_{\infty }$ products of (\ref{dual products}). 

\niu{Construction of the dual $L_\infty$ product $\mathbf L^\eta$}

The dual $L_\infty$ products $\mathbf L^\eta$ can be constructed using the cohomomorphism $\widehat{\mathbf G}$ which provides the NS heterotic string products ${\bf L } = \widehat{\bf G}^{-1} \, {\bf Q } \, \widehat{\bf G}$.\footnote{See also appendix D or section 3.2 of \cite{Matsunaga:2015kra} for generic properties of these types of products.} 
The product $\mathbf L^\eta$ is defined as the similarity transformation of $\eta $:
\begin{align}
{\bf L}^{\boldsymbol \eta }
= \widehat{\bf G} \, {\boldsymbol \eta } \, \widehat{\bf G}^{-1} 
=\sum_{n=1}^\infty \mathbf L^\eta _n.
\end{align}
These $\mathbf L^\eta$ carry odd degree, and the $n$-product $\mathbf L^\eta_n$ carries ghost number $3 - 2n$ and picture number $n - 2$.
$\mathbf L^\eta$ satisfiy the $L_{\infty }$-relations, which follow from its definition:
\begin{align}
({{\bf L}^{\boldsymbol \eta } })^2 
= \widehat{\bf G} \, {\boldsymbol \eta } \, \widehat{\bf G}^{-1} \widehat{\bf G} \, {\boldsymbol \eta } \, \widehat{\bf G}^{-1}
= \widehat{\bf G} \, {\boldsymbol \eta } {\boldsymbol \eta } \, \widehat{\bf G}^{-1}
=0.
\end{align}
The $Q$-derivation properties of $\mathbf L^\eta$ follow from $\eta$-derivatioies property of $\mathbf L$:
\begin{align}
{\bf Q} \, {\bf L}^{\boldsymbol \eta } = \widehat{\bf G} \, ( \widehat{\bf G}^{-1} \, {\bf Q} \, \widehat{\bf G} )\, {\boldsymbol \eta } \, \widehat{\bf G}^{-1} = - \widehat{\bf G} \, {\boldsymbol \eta } \, ( \widehat{\bf G}^{-1} \, {\bf Q} \, \widehat{\bf G} ) \, \widehat{\bf G}^{-1} = - {\bf L}^{\boldsymbol \eta } \, {\bf Q}. 
\end{align}
The cyclicity of $L^\eta$ follows from that the gauge products are BPZ-odd:
%$\mathbf G^{-1}=\mathbf G^\dagger$ 
\begin{align}
({\bf L}^{\boldsymbol \eta })^\dagger
= (\widehat{\bf G} \, {\boldsymbol \eta } \, \widehat{\bf G}^{-1})^\dagger
= (\widehat{\bf G}^{-1})^\dagger  \, {\boldsymbol \eta }^\dagger  \, \widehat{\bf G}^\dagger
= -\widehat{\bf G} \, {\boldsymbol \eta }  \, \widehat{\bf G}^{-1}.
\end{align}

The dual $L_\infty$ products $\mathbf L^\eta$ is the generating function for (\ref{dual products}). 
Namely, we define 
\begin{align}
[ B_{1} , \dots , B_{n} ]^{\eta } 
%={\pi } {\bf G}^{\dagger } \, {\boldsymbol \eta } \, {\bf G}  ( B_{1}\wedge \dots \wedge B_{n} ) 
:= {\pi _1} {\mathbf L^\eta_n } (  B_{1}\wedge \dots \wedge B_{n} )  
%= {\pi } {\mathbf L^\eta_n } ( B_{1},\dots , B_{n} ) . 
\end{align} 
for any states $B_{1} , \dots , B_{n} \in \mathcal{H}$. 
In terms of $[B_{1} , \dots , B_{n} ]^{\eta }$, the dual products satisfy the following $L_\infty$ relations, $Q$-derivation properties, and cyclicity:
\begin{align}
{\sum_{\sigma }}\sum_{k=1}^{n} \frac{1}{k!(n-k)!}(-)^{|\sigma |} \big[ [ B_{i_{\sigma (1)}} , \dots , B_{i_{\sigma (k)}} ]^{\eta } , B_{i_{\sigma (k+1)}} , \dots , B_{i_{\sigma (n) }} \big] ^{\eta } = 0 ,\\
Q \big[ B_{1} , \dots , B_{n} \big] ^{\eta } + \sum_{i=1}^{n} (-)^{B_{1} +\dots +B_{k-1}} \big[ B_{1} , \dots, Q B_{k} , \dots , B_{n} \big] ^{\eta } = 0 ,\\
\langle B_1,[B_2,\cdots,B_{n+1}]^\eta\rangle
=(-)^{B_1+B_2+\cdots+B_n}\langle [B_1,\cdots,B_n]^\eta,B_{n+1}\rangle,
\end{align}
where $(-)^\sigma$ is the sign factor of the permutation 
%from $\{A_1,...,A_n\}$ to 
$\{B_{\sigma(1)}, ... ,B_{\sigma(n)}\}$.

For any state $A\in \mathcal{H}$, the $A$-shifted products of $\mathbf L^\eta$ are defined by
\begin{align}
[B_1,B_2, \cdots,B_n \:]^\eta_{A}
=\sum_{m=0}^\infty \frac{1}{m!}[\:\overbrace{A,A,\cdots,A}^m, B_1,B_2, \cdots,B_n \:]^\eta.
\end{align}
They also satisfy $L_\infty$ relations if $A$ is a solution for the Maurer-Cartan equation of $\mathbf L^\eta$:
\begin{align}
0=\pi_1 \mathbf L^\eta (e^{\wedge A}) = \eta \, A + \sum_{n=1}^{\infty } \frac{1}{(n+1)!} [\overbrace{ A , \dots , A  }^{n+1} ]^{\eta } . 
\end{align}
Let $\Psi_\eta$ be a solution for this Maurer-Cartan equation of $\mathbf L^\eta$.
In particular we write $D_\eta$ for the $\Psi_\eta$-shifted 1-product as 
\begin{align}
\label{shifted L-infty}
D_{\eta } B
&:=[B]^\eta_{\Psi_\eta}
%=\sum_{m=0}^\infty \frac{1}{m!}L^\eta_{m+1}(\:\underbrace{\Psi_\eta,\Psi_\eta,\cdots,\Psi_\eta}_m,\:\:\cdot \:\:  \:)
=\sum_{m=0}^\infty \frac{1}{m!}[\:\overbrace{\Psi_\eta,\Psi_\eta,\cdots,\Psi_\eta}^m,B  \:]^\eta.
%= \pi _{1} {\bf L}^{\boldsymbol \eta } \big(  \mathbb{I} \wedge e^{\Psi_\eta} \big)
\end{align}
From the $L_\infty$-relation of the $\Psi_\eta$-shifted $\mathbf L^\eta$, we find that $D_\eta$ is nilpotent, 
\begin{align}
(D_\eta)^2 B= -[\pi_1 \mathbf L^\eta (e^{\wedge \Psi_\eta}) ,B]^\eta_{\Psi_\eta}=0,
\end{align}
and that $D_{\eta }$ acts on the $\Psi_\eta$-shifted 2-product $[B_1,B_2]^{\eta } _{\Psi_\eta}$
%\begin{align}
%[B_1,B_2]^{\eta } _{\Psi_\eta}
%%=\sum_{m=0}^\infty \frac{1}{m!}L^\eta_{m+2}(\:\underbrace{\Psi_\eta,\Psi_\eta,\cdots,\Psi_\eta}_m,\:\:\cdot \:\:,\:\:\cdot \:\:  \:)
%&=\sum_{m=0}^\infty \frac{1}{m!}[\:\overbrace{\Psi_\eta,\Psi_\eta,\cdots,\Psi_\eta}^m,B_1,B_2  \:]^\eta,
%%= \pi _{1} {\bf L}^{\boldsymbol \eta } \big(  \mathbb{I}_2 \wedge e^{\wedge \pi _{1} {\bf G} ( e^{\wedge \Phi  })} \big),
%\end{align}
as a derivation, 
\begin{align}
D_\eta [B_1,B_2]^\eta_{\Psi_\eta}+[D_\eta B_1,B_2]^\eta_{\Psi_\eta}+(-)^{B_1} [B_1,D_\eta B_2]^\eta_{\Psi_\eta}
=-[\pi_1 \mathbf L^\eta (e^{\wedge \Psi_\eta}) ,B_1,B_2]^\eta_{\Psi_\eta}=0 . 
\end{align}
Note that the shifted products are BPS odd, which follows from that of $\mathbf L^\eta$, 
\begin{align}
\langle B_1,[B_2,\cdots,B_{n+1}]^\eta_{\Psi_\eta}\rangle
=(-)^{B_1+B_2+\cdots+B_n}\langle [B_1,\cdots,B_n]^\eta_{\Psi_\eta},B_{n+1}\rangle.
\end{align}

\niu{WZW-like relations}

Let us confirm a pair of fields (\ref{pure-gauge L-infty sec2}) and (\ref{associated L-infty sec2}) 
satisfy the WZW-like relations, which can be represented as follows:
\begin{align}
\label{WZW relation1 sec2}
\pi _{1} {\bf L}^{\boldsymbol \eta } \big( e^{\wedge\Psi _{\eta }} \big) 
& = 0 ,
%= \eta \, \Psi _{\eta } + \sum_{n=1}^{\infty } \frac{\kappa ^{n} }{(n+1)!} [\overbrace{ \Psi _{\eta } , \dots , \Psi _{\eta }  }^{n+1} ]^{\eta }  , 
\\ 
\label{WZW relation2 sec2} 
(-)^{ {d}}  {d} \, \Psi _{\eta } 
& = D_{\eta } \, \Psi _{ {d}}. 
%= \eta \, \Psi _{ {d}} + \sum_{k=1}^{\infty } \frac{\kappa ^{k-1} }{k! } \big[ \overbrace{\Psi _{\eta } , \dots , \Psi _{\eta }  }^{k} , \Psi _{ {d}} \big] ^{\eta }.
\end{align}

The first relation ($\ref{WZW relation1 sec2}$)
directly follows from the fact that $\Phi$ belongs to small space:
\begin{align}
{\bf L}^{\boldsymbol \eta } \big( e^{\wedge\Psi _{\eta }[\Phi(t)]} \big)
={\bf L}^{\boldsymbol \eta } \big(e^{\wedge\pi _{1} \widehat{\bf G} ( e^{\wedge\Phi (t)} ) } \big)
= ( \widehat{\bf G} \, {\boldsymbol \eta } \, \widehat{\bf G}^{-1} ) \, \widehat{\bf G} \big( e^{\wedge\Phi (t)}   \big) 
= \widehat{\bf G} \, {\boldsymbol \eta } \big( e^{\wedge\Phi (t) } \big) = 0.\label{pre 1st wzw}
\end{align}
We call $\Psi _{\eta }[ \Phi (t ) ]$ satisfying ($\ref{WZW relation1 sec2}$)
the {\it pure-gauge-like field}.

The second relation ($\ref{WZW relation2 sec2}$)
can be confirmed similarly.
The operator $ {\boldsymbol d}$ which we focus on is the derivation on $\mathbf L^{\eta}$.
For example, we can take $d = Q$, $\partial _{t}$, or $\delta$. 
Their derivation property on $\mathbf L^{\eta}$ leads to $\ld \widehat{\bf G}^{-1}  {\boldsymbol d} \widehat{\bf G} , \boldsymbol\eta \rd = 0$,
and we can define the coderivation ${\boldsymbol \xi }_{ {\boldsymbol d}}$
such that 
\begin{align}
\widehat{\mathbf G}^{-1}  {\boldsymbol d} \widehat{\mathbf G} = (-)^{ {\boldsymbol d}} \ld {\boldsymbol \eta } , {\boldsymbol \xi }_{ {\boldsymbol d}} \rd .
\label{def xi X}
\end{align}
Note that for the operator $ {\boldsymbol d}$ which commutes with $\widehat{\bf G}$,
such as $\partial_t$ and $\delta$,
${\boldsymbol \xi }_{ {\boldsymbol d}}$ is a coderivation derived form $ {\boldsymbol d} \xi$.
Then, utilizing this $\boldsymbol \xi_ {\boldsymbol d}$, the following relation holds:
\begin{align}
(-)^{ {\boldsymbol d}}  {\boldsymbol d} \widehat{\bf G} \, \big( e^{\wedge \Phi (t)} \big) 
& = (-)^{ {\boldsymbol d}} \widehat{\bf G} \,  ( \widehat{\mathbf G}^{-1}  {\boldsymbol d} \widehat{\mathbf G}) \, \big( e^{\wedge \Phi (t)} \big) 
\no 
&= \widehat{\bf G} \, {\boldsymbol \eta } \,   {\boldsymbol \xi }_{ {\boldsymbol d}} \big( e^{\wedge \Phi (t)} \big) 
\no  
& = {\bf L}^{\boldsymbol \eta } \, \widehat{\bf G} \, {\boldsymbol \xi }_{ {\boldsymbol d}} \, \Big( e^{\wedge \Phi (t) } \Big) 
\no 
& = {\bf L}^{\boldsymbol \eta } \, \Big( \pi _{1} \widehat{\bf G} \, {\boldsymbol \xi }_{ {\boldsymbol d}} \big( e^{\wedge \Phi (t)} \big)  \wedge e^{\wedge\pi _{1} \widehat{\bf G}(e^{\wedge \Phi (t) }) } \Big) . 
\label{pre second WZW}
\end{align} 
Since $D_{\eta } = \pi _{1} {\bf L}^{\boldsymbol \eta } \big(  \mathbb{I} \wedge e^{\wedge\Psi_\eta} \big)$,
we can see 
$\Psi _{\eta }[\Phi(t)]$ and $\Psi _{ {d} }[\Phi(t)]$
satisfy the WZW-like relation ($\ref{WZW relation2 sec2}$)
using (\ref{pure-gauge L-infty sec2}),(\ref{associated L-infty sec2}),(\ref{def xi X}), and (\ref{pre second WZW}).
We call $\Psi _{d }[\Phi (t) ]$ satisfying ($\ref{WZW relation2 sec2}$)
the {\it associated field}.

Then the action in the $L_\infty$ formulation can be written in the alternative WZW-like form:
\begin{align}
S_{\scriptscriptstyle{\rm EKS}} [\Phi]
& = \int_{0}^{1} dt \, \langle \Psi _{t}[\Phi (t)]   , \, Q \Psi _{\eta } [\Phi(t)] \rangle
%=S_{\eta} [\Phi]
.
\label{Z2 WZW action EKS}
\end{align}
The variation of the action can be taken easily using the WZW-like relations ($\ref{WZW relation1 sec2}$) and ($\ref{WZW relation2 sec2}$),
and the gauge invariances also follows from them,
which can be seen in a similar (but $\mathbb Z_2$-reversed) manner to those in \cite{Berkovits:2004xh}.
We will see them in the next subsection.

\subsection{Variation of the action $S_{\eta }$}
\label{SUBSEC VOA EKS}

Let us take the variation of the action. 
Note that in the computation here we use only the WZW-like relations (\ref{WZW relation1 sec2}) and (\ref{WZW relation2 sec2}), 
and we do not necessitate the explicit forms of the functionals $\Psi_\eta$ and $\Psi_{d}$. 
Therefore, even if the functionals $\Psi_\eta$ and $\Psi_{d}$ have different forms,
the variation of the action can be taken in the same manner as long as they satisfy the WZW-like relations.
In this subsection, we write $\Psi _{\eta } (t)$ for $\Psi _{\eta } [\Phi (t) ]$ and so on for brevity. 

First, consider the variation of the integrand of (\ref{Z2 WZW action EKS}), 
\begin{align}
\delta \big\langle \Psi_t(t), Q \Psi_\eta(t) \big\rangle
= 
\big\langle \delta \Psi_t(t), Q \Psi_\eta(t) \big\rangle
+\big\langle \Psi_t(t), Q \delta \Psi_\eta(t) \big\rangle.
\end{align}
Utilizing the following relation following from ($\ref{WZW relation2 sec2}$), 
\begin{align}
0=[\![  {d_1}, {d_2}]\!] \Psi_\eta
=(-)^{{d_1}+ {d_2}}  D_\eta 
\big(  {d_1} \Psi_{d_2} -(-)^{ {d_1}{d_2}}{d_2} \Psi_ {d_1} +(-)^{{d_2}  {d_1}+{d_2}} [ \Psi_{d_2}, \Psi_ {d_1} ]^\eta_{\Psi_\eta}\big),
\end{align}
the first term can be transformed into
\begin{align}
\big\langle \delta \Psi_t(t), Q \Psi_\eta(t) \big\rangle
=\big\langle \partial_t \Psi_\delta(t) +[\Psi_\delta(t), \Psi_t(t)]^\eta_{\Psi_\eta(t)}, Q \Psi_\eta(t) \big\rangle.
\label{VORWZW1}
\end{align}
Utilizing $[\![D_\eta,  {d} \; ]\!] B= -[ {d} \Psi_\eta,B]^\eta_{\Psi_\eta}$, the second term can be transformed into
\begin{align}
\big\langle \Psi_t(t), Q \delta \Psi_\eta(t) \big\rangle
&=\big\langle \Psi_t(t), Q D_\eta \Psi_\delta(t) \big\rangle\no
&=\big\langle D_\eta Q \Psi_t(t), \Psi_\delta(t) \big\rangle\no
&=\big\langle -Q D_\eta \Psi_t(t) -[Q\Psi_\eta(t), \Psi_t(t)]^\eta_{\Psi_\eta(t)}, \Psi_\delta(t) \big\rangle.
\label{VORWZW2}
\end{align}
The second terms of (\ref{VORWZW1}) and (\ref{VORWZW2}) are canceled because of the cyclicity of the $\Psi_\eta$-shifted $\mathbf L^\eta$.
Then we find that the variation of the integrand of (\ref{Z2 WZW action EKS})
becomes a total derivative of $t$:
\begin{align} 
\delta \big\langle \Psi_t(t), Q \Psi_\eta(t) \big\rangle
&=\big\langle \partial_t \Psi_\delta(t), Q \Psi_\eta(t) \big\rangle-\big\langle Q D_\eta \Psi_t(t) , \Psi_\delta(t) \big\rangle\no
&=\big\langle \partial_t \Psi_\delta(t), Q \Psi_\eta(t) \big\rangle + \big\langle \Psi_\delta(t), Q \partial_t \Psi_\eta(t) \big\rangle\no
&=\partial_t \big\langle \Psi_\delta(t), Q \Psi_\eta(t) \big\rangle.
\end{align}

Integrating over $t$, the variation of the action is given by
\begin{align}
\int_{0}^{1} dt \, \delta \langle \Psi _{t} (t)   ,  Q \Psi _{\eta } (t) \rangle
= \int_{0}^{1} dt \, \partial_t \langle \Psi _{\delta} (t)   ,  Q \Psi _{\eta } (t) \rangle
%= \Big[ \langle \Psi _{\delta} (t)   ,  Q \Psi _{\eta } (t) \rangle\Big]^1_0
= \langle \Psi _{\delta}(1)  ,  Q \Psi _{\eta }(1)\rangle,
\end{align}
where the pure-gauge-like field $\Psi_\eta (t) $ and the associated field $\Psi_d (t) $ vanish at $t=0$.
Then the variation of the action becomes
\begin{align}
\delta S_{\scriptscriptstyle{\rm EKS}}  = \langle \Psi _{\delta}[\Phi]  ,  Q \Psi _{\eta }[\Phi]\rangle.\label{VOA WZW}
\end{align}
%Here we omit the argument $t=1$ of the fields.
We find that the variation of the action does not depend on $t$, and therefore $t$-dependence is topological.

\niu{Equations of motion}

One can derive the on-shell condition from the variation of the action in the WZW-like form (\ref{VOA WZW}),
\begin{align}
Q\Psi_\eta[\Phi] =0.\label{OSC WZW}
\end{align}
For completion, let us discuss the equivalence of (\ref{OSC WZW}) and (\ref{EOM EKS}).

The latter (\ref{EOM EKS}) can be transformed into the following form:
\begin{align}
\pi_1{\bf L}( e^{\wedge \Phi} ) 
=\pi_1 \widehat{\mathbf G}^{-1} {\bf Q} \, \widehat{\bf G} \big( e^{\wedge \Phi } \big) 
= \pi_1 \widehat{\mathbf G}^{-1} {\bf Q} \, \big( e^{\wedge\Psi _{\eta}[\Phi] } \big) 
= \pi_1 \widehat{\mathbf G}^{-1} \big( (Q \Psi _{\eta }[\Phi] )\wedge e^{\wedge\Psi _{\eta }[\Phi] }\big).
\end{align}
Since $\pi_1 \widehat{\mathbf G}^{-1}(\:\:\cdot\:\:\wedge e^{\wedge \Psi_\eta})$ is invertible,
(\ref{OSC WZW}) and (\ref{EOM EKS}) are equivalent:
\begin{align}
\pi_1{\bf L}( e^{\wedge \Phi} ) =0 \iff Q\Psi_\eta[\Phi] =0.
\label{on-shell eq EKSsmall Z2WZW}
\end{align}
Note that the overall factor $\pi_1 \widehat{\mathbf G}^{-1} \big( \:\:\cdot\:\:\wedge \:e^{\wedge\Psi _{\eta } }\big)$
comes from the difference of the $\xi \delta \Phi$ and $\Psi_\delta$.

%%%%%%%%%%%%%%%%%%%%%%%%%%%%%%%%%%%%%%%%%%%%%%%%%%%%%%%%%%%%%%%%%%%%%%%%%%%%%%%%%%%%%%%
\subsection{Gauge transformations}\label{SUBSEC:EKSGT}

It follows from the nilpotency of $Q$ and $D_\eta$ that 
the WZW-like action is invariant under the following form of the gauge transformations, 
\begin{align}
\Psi _{\delta } = D_{\eta } \Omega + Q \Lambda ,
\label{GT Z2}
\end{align}
where $\Omega$ and $\Lambda$ are gauge parameters belonging to the large Hilbert space,
which carry ghost numbers $0$ and $0$, and picture numbers $1$ and $0$, respectively.
In this subsection, we see how the gauge transformation (\ref{GT EKS}) can be represented in the WZW-like form (\ref{GT Z2}).

Let us consider the associated field 
$\Psi _{\delta} [\Phi]= \pi _{1} \widehat{\bf G} \big( {\boldsymbol \xi }_{\delta} e^{\wedge \Phi } \big) 
= \pi _{1} \widehat{\bf G} \big( (\xi\delta\Phi)\wedge e^{\wedge \Phi } \big)$,
with $\delta\Phi$ being the gauge transformation (\ref{GT EKS}) in the $L_\infty$-fomulation:
\begin{align}
\delta \Phi 
=\pi_1\mathbf L(\lambda \wedge e^{\wedge\Phi})
=\pi_1\mathbf L\boldsymbol \eta ( \xi\lambda \wedge e^{\wedge\Phi})
=-\eta\pi_1\mathbf L\boldsymbol ( \xi\lambda \wedge e^{\wedge\Phi}).
\end{align}
We find $\Psi_\delta[\Phi]$ can be transformed as follows:
\begin{align}
\Psi _{\delta}[\Phi]
&=
 \pi _{1} \widehat{\bf G} \Big( \big(-\xi\eta \pi_1\mathbf L (\xi\lambda\wedge e^{\wedge\Phi})\big)\wedge e^{\wedge \Phi } \Big)
\no
&=
\pi _{1} \widehat{\bf G} \Big( \big(\eta\xi \pi_1\mathbf L (\xi\lambda\wedge e^{\wedge\Phi})\big)\wedge e^{\wedge \Phi } \Big)
 -\pi _{1} \widehat{\bf G} \Big( \big(\pi_1\mathbf L (\xi\lambda\wedge e^{\wedge\Phi})\big)\wedge e^{\wedge \Phi } \Big)
\no
&=
\pi _{1} \widehat{\bf G} \boldsymbol \eta \Big( \big(\xi \pi_1\mathbf L (\xi\lambda\wedge e^{\wedge\Phi})\big)\wedge e^{\wedge \Phi } \Big)
 -\pi _{1} \widehat{\bf G} \Big(  \mathbf L (\xi\lambda\wedge e^{\wedge\Phi}) + \xi\lambda\wedge \pi_1 \mathbf L (e^{\wedge\Phi})\wedge e^{\wedge\Phi}\Big)
\no
&=
\pi _{1} \mathbf L^\eta  \widehat{\bf G}\Big( \big(\xi \pi_1\mathbf L (\xi\lambda\wedge e^{\wedge\Phi})\big)\wedge e^{\wedge \Phi } \Big)
 -\pi _{1}  \mathbf Q \widehat{\mathbf G} (\xi\lambda\wedge e^{\wedge\Phi})
 -\pi _{1} \widehat{\bf G} \big( \xi\lambda\wedge \pi_1 \mathbf L (e^{\wedge\Phi})\wedge e^{\wedge\Phi}\big)
\no
&=
\pi _{1} \mathbf L^\eta \Big( \pi_1 \widehat{\mathbf G} \big( (\xi \pi_1\mathbf L (\xi\lambda\wedge e^{\wedge\Phi}))\wedge e^{\wedge \Phi }\big) 
 \wedge e^{\wedge \pi_1 \widehat{\mathbf G} (e ^{\wedge \Phi})}\Big)
 - Q \pi _{1} \widehat{\mathbf G} (\xi\lambda\wedge e^{\wedge\Phi})
 -\Delta_T[\lambda,\Phi]\no
%&=
%D_\eta \big(\pi \mathbf G \big( (\xi \pi_1\mathbf L (\xi\lambda\wedge e^{\wedge\Phi}))\wedge e^{\wedge \Phi }\big) \big)
% - Q \big(\pi _{1}\mathbf G (\xi\lambda\wedge e^{\wedge\Phi})\big)
% -\Delta_T[\Phi].\\
&= D_{\eta } \Omega [\lambda, \Phi]+ Q\Lambda [\lambda,\Phi]  -\Delta_T [\lambda,\Phi]  .
\label{GT_EKS_Z2WZW}
\end{align}
%Since $D_\eta =\pi _{1} \mathbf L^\eta ( \:\:\cdot\:\: \wedge e^{\pi \mathbf G(e ^{\wedge \Phi})})$,
While the first two terms of (\ref{GT_EKS_Z2WZW}) correspond to (\ref{GT Z2}) with the gauge parameters $\Omega =\Omega [\lambda, \Phi] $ and $\Lambda =\Lambda [\lambda, \Phi] $ parameterized by $\lambda$, 
\begin{align}
\Omega [\lambda, \Phi] &=\pi_1 \widehat{\mathbf G} \big( (\xi \pi_1\mathbf L (\xi\lambda\wedge e^{\wedge\Phi}))\wedge e^{\wedge \Phi }\big) ,
\label{gauge para omega1}\\[2pt]
\Lambda [\lambda,\Phi] &= -\pi_1 \widehat{\mathbf G} (\xi\lambda\wedge e^{\wedge\Phi}),
\label{gauge para lambda1}
\end{align}
the third term of (\ref{GT_EKS_Z2WZW}),
\begin{align}
\Delta_T[\lambda,\Phi]= \pi _{1} \widehat{\bf G} \big( \xi\lambda\wedge \pi_1 \mathbf L (e^{\wedge\Phi})\wedge e^{\wedge\Phi}\big),
\label{TGT EKS Z2}
\end{align}
corresponds to the trivial gauge transformation of the WZW-like action.
Thus, the gauge transformation in the $L_\infty$-formulation (\ref{GT EKS}) can be written in the WZW-like form (\ref{GT Z2}). 
%While the gauge invariance under (\ref{GT EKS}) follows from the $L_\infty$ relations of $\mathbf L$, that under (\ref{GT Z2}) follows from the WZW-like relations (\ref{WZW relation1 sec2}) and (\ref{WZW relation2 sec2}).

\niu{Trivial gauge transformation}

Trivial gauge transformation
is a transformation proportional to the equations of motion.
Schematically, it is of the following form, 
\begin{align}
\delta_\mu \varphi^i = (\textrm{EOM})_j \mu^{ji},\qquad
\mu^{ji}=-(-)^{ij}\mu^{ij},
\end{align}
and its gauge invariance follows from the symmetric property of $\mu^{ji}$, 
\begin{align}
\delta_\mu S = (\textrm{EOM})_i \delta_\mu \varphi^i =(\textrm{EOM})_i\,(\textrm{EOM})_j \mu^{ji} =0.
\end{align}
They are no physical significance,
but in general they may appear in the algebra of the nontrivial gauge transformations,
and in the context of the Batalin-Vilkovisky quantization\cite{Batalin:1981jr,Batalin:1984jr} it is convenient to consider them\footnote{For more detail, see \cite{Gomis:1994he}.}.

In our case, by the almost same computation with \cite{Erler:2015uoa},  when $A+B+C=even$, cyclicity of the cohomomorphism $\widehat{H}$ is written as follows:
\begin{align}
\langle \pi_1 \widehat{H} (A \wedge B \wedge e^{\wedge\Phi} ),  \pi_1 \widehat{H} ( C\wedge e^{\wedge\Phi} ) \rangle
=-(-)^{B} \langle \pi_1 \widehat{H} ( A\wedge e^{\wedge\Phi} ),  \pi_1 \widehat{H} (B\wedge C \wedge e^{\wedge\Phi} ) \rangle.
\label{CYC H ABC}
\end{align}
The derivation is in appendix \ref{APP cyc H ABC}.
We take $\widehat{H}=\widehat{\bf G}$, $A=\xi\lambda$, which is even, and $B=C=\pi_1\mathbf Le^{\wedge\Phi}$, which are odd.
The invariance under the trivial gauge transformation (\ref{TGT EKS Z2}) \footnote{
Note that, although the expression does not contain explicit WZW-like on-shell condition $Q\Psi_\eta$,
the on-shell equivalence (\ref{on-shell eq EKSsmall Z2WZW}) guarantees it is trivial gauge transformation.
},
\begin{align}
\langle -\Delta_T[\lambda,\Phi], Q\Psi_\eta[\Phi]\rangle 
&=\big\langle -
\pi _{1} \widehat{\bf G} \big( \xi\lambda\wedge \pi_1 \mathbf L (e^{\wedge\Phi})\wedge e^{\wedge\Phi}\big),
\pi_1 \widehat{\mathbf G} \big( \pi_1\mathbf L(e^{\wedge\Phi})\wedge e^{\wedge\Phi}\big)\big\rangle\no
&=- \big\langle 
\pi _{1} \widehat{\bf G} \big( \xi\lambda\wedge e^{\wedge\Phi}\big),
\pi_1 \widehat{\mathbf G} \big( \pi_1\mathbf L(e^{\wedge\Phi})\wedge \pi_1 \mathbf L (e^{\wedge\Phi})\wedge e^{\wedge\Phi}\big)\big\rangle\no
&=0 , 
\end{align}
follows from the symmetric property:
\begin{align}
\pi_1 \mathbf L e^{\wedge\Phi} \wedge  \pi_1 \mathbf L e^{\wedge\Phi}  =0.
\end{align}

%%%%%%%%%%%%%%%%%%%%%%%%%%%%%%%%%%%%%%%%%%%%%%%%%%%%%%%%%%%%%%%%%%%%%%%%%%%%%%%%%%%%%%%%
%%%%%%%%%%%%%%%%%%%%%%%%%%%%%%%%%%%%%%%%%%%%%%%%%%%%%%%%%%%%%%%%%%%%%%%%%%%%%%%%%%%%%%%%
%\clearpage 
\section{Alternative parameterisation for closed NS string field theory} 

Let $\varphi $ be a some dynamical string field and $t \in [0,1]$ be a real parameter. 
We write $\varphi (t)$ for a path satisfying $\varphi (t=0) = 0$ and $\varphi (1) = \varphi $. 
In section \ref{SUBSEC VOA EKS}, we saw that the gauge invariance of the action is provided by only the algebraic relations ($\ref{WZW relation1 sec2}$) and ($\ref{WZW relation2 sec2}$). 
In other words, once the functionals $\Psi_\eta = \Psi _{\eta } [ \varphi ]$ and $\Psi_d = \Psi _{d} [ \varphi ]$ satisfying the WZW-like relations, 
\begin{align}
0=\pi _{1} {\bf L}^{\boldsymbol \eta } \big( e^{\wedge\Psi_{\eta } [\varphi ]} \big) ,\qquad
 (-)^{d} d \, \Psi_{\eta } [\varphi ]
= D_{\eta } \, \Psi_{d} [\varphi ] , 
\end{align}
are obtained, the gauge invariant action can be constructed as 
\begin{align}
S_{\eta} [\varphi ] = \int^1_0 dt \langle \Psi_t [\varphi (t) ], Q \Psi_\eta [\varphi (t) ] \rangle.
\end{align}
This form of the action is the $\mathbb{Z}_2$-reversed version of that in \cite{Berkovits:2004xh}, which we call the alternative WZW-like form.
We found that by taking $\varphi = \Phi $, (\ref{pure-gauge L-infty sec2}), and (\ref{associated L-infty sec2}), the $L_\infty$ action $S_{\scriptscriptstyle {\rm EKS}} [\Phi ]$ gives one realisation of this WZW-like action, 
\begin{align}
S_{\eta}[\Phi]
=S_{\scriptscriptstyle{\rm EKS}}[\Phi] . 
\end{align}

In this section we provides another realisation of these functionals, which is parameterized by the string field $V$ in the large Hilbert space.
We first see the pure-gauge-like and the associated functional fields can be defined by the differential equations,
which are the $\mathbb Z_2$-reversed versions of the construction in \cite{Berkovits:2004xh}.
Then, the WZW-like action parameterized by the string field $V$, 
\begin{align}
S_{\eta}=S_{\eta}[V] , 
\end{align}
is given in terms of them.
The equivalence of these actions in the different parameterizations is shown
by the almost same procedure performed in \cite{Erler:2015rra}.

%%%%%%%%%%%%%%%%%%%%%%%%%%%%%%%%%%%%%%%%%%%%%%%%%%%%%%%%%%%%%%%%%%%%%%%%%%%%%%%%%%%%%%%%
\subsection{Large space parameterisation} 

Let $V$ be a dynamical string field which belongs to the large Hilbert space and carries ghost number $1$ and picture number $0$.
In this subsection, 
we provides the another paramaterization of the pure-gauge-like field $\Psi_\eta =\Psi_\eta[V]$ and the associated fields $\Psi_d =\Psi_d[V]$.
A set of differential equations which are the $\mathbb Z_2$-reversed version of those in \cite{Berkovits:2004xh}
give these parameterizations
so that WZW-like relations (\ref{WZW relation1 sec2}) and (\ref{WZW relation2 sec2}) hold.
Utilizing these functionals, a new gauge invariant action for the string field $V$ is constructed in the (alternative) WZW-like form.

\niu{Pure-gauge-like field $\Psi_\eta =\Psi_\eta[V]$}

A pure-gauge-like (functional) field $\Psi _{\eta } [V] $ satisfying $\pi _{1} {\bf L}^{\boldsymbol \eta } \big( e^{\wedge\Psi _{\eta }[V]} \big) = 0$
is the solution of the the Maurer-Cartan equation for $\mathbf L^\eta$.
Therefore, $\Psi_\eta[V]$ is obtained by mimicking the pure gauge construction of \cite{Berkovits:2004xh}.
First, we introduce a real parameter $\tau \in [0,1]$.
Second, we solve the differential equation
\begin{align}
\frac{\partial }{\partial \tau } \Psi _{\eta }  [\tau; V]& = D_{\eta }(\tau) V 
\no 
& = \eta \, V + \sum_{k=1}^{\infty } \frac{1 }{k! } \big[ \overbrace{\Psi _{\eta } [\tau;V ], \dots , \Psi _{\eta } [\tau;V ] }^{k} , V \big] ^{\eta },
\label{DE PGF}
\end{align}
with the initial condition
\begin{align}
\Psi_\eta[\tau=0; V]=0.
\end{align}
Finally, we set $\tau =1$ and obtain $\Psi_\eta[V]$ as the $\tau =1$ value solution, 
\begin{align}
\Psi_\eta[V] :=\Psi_\eta[\tau =1 ; V].
\end{align}
One can check that the differential equation actually provides a solution for the Maurer-Cartan equation
\begin{align}
\mathcal F(\Psi_\eta[\tau; V])
=\mathbf L^\eta e^{\wedge \Psi_\eta[\tau; V]}=0,
\label{MCEQLeta}
\end{align}
by differentiating it by $\tau$:
\begin{align}
\partial_\tau \mathcal F(\Psi_\eta)
= \pi_1 \mathbf L^\eta \big( \partial_\tau \Psi_\eta \wedge e^{\wedge \Psi_\eta[\tau]}\big)
= D_\eta ( \partial_\tau \Psi_\eta )
= D_\eta D_\eta V
=-[\mathcal F(\Psi_\eta) ,V]^\eta_{\Psi_\eta}.
\end{align}
With the initial condition $\mathcal F (\Psi_\eta[\tau=0;V])=0$,
this differential equation ensures $\mathcal F (\Psi_\eta[\tau;V])=0$ for arbitrary $\tau$,
and then (\ref{MCEQLeta}) holds.

\niu{Associated fields $\Psi_d =\Psi_d[V]$}

The associated fields $\Psi_d[V]$ are the functionals satisfying
\begin{align}
 {d} \Psi_\eta [ V ] =  (-)^ {d} D_\eta \Psi_ {d} [V] .
\label{2nd WZW rela}
\end{align}
To derive the differential equation which defines $\Psi_ {d}[V]$,
let us introduce
\begin{align}
\mathcal I(\tau) \equiv D_\eta(\tau) \Psi_ {d} [\tau; V]-  (-)^ {d} {d} \Psi_\eta[\tau; V]
\end{align}
and consider its differentiation by $\tau$:
\begin{align}
\partial_\tau \mathcal I(\tau) =
[ V,  \mathcal I (\tau)]^\eta_{\Psi_\eta}
+ D_\eta \Big( \partial_\tau \Psi_ {d} -  {d} V -  [ V,  \Psi_ {d}]^\eta_{\Psi_\eta}\Big).
\label{DE ASF PRE}
\end{align}
We define the functional field $\Psi_ {d}[\tau; V]$ by the differential equation
\begin{align}
\partial_\tau  \Psi _{ {d}} [\tau; V] =  {d} V +  \big[ V , \Psi _{ {d}} [\tau; V] \big] _{\Psi _{\eta }[\tau; V]}^{\eta }
\label{DE ASF}
\end{align}
with the initial condition
$\Psi_ {d}[\tau=0;V]=0$.
Then, the equation (\ref{DE ASF PRE}) becomes $\partial_\tau \mathcal I(\tau) =[ V,  \mathcal I (\tau)]^\eta_{\Psi_\eta}$,
and leads to the vanishing of $\mathcal I(\tau)$ at arbitrary $\tau$ since $\mathcal I(\tau=0) =0$,
which means that $\Psi_ {d} [\tau ; V ]$ defined by (\ref{DE ASF}) actually satisfies the WZW-like relation (\ref{2nd WZW rela}). 
We set $\tau =1$ and obtain the associated (functional) field $\Psi _{d}[V]$ as the $\tau =1 $ value solution 
\begin{align}
\Psi _{d} [V] := \Psi _{d} [ \tau =1 ; V]. 
\end{align}

\niu{Action $S_{\eta}=S_{\eta}[V]$}

We write $V(t)$ for a path satisfying $V(0) = 0$ and $V(1)=V$ with a real parameter $t \in [0,1]$. 
Utilizing the pure gauge string field $\Psi_\eta[V]$ and associated fields $\Psi_ {d}[V]$
which are defined by (\ref{DE PGF}) and (\ref{DE ASF}) to satisfy the WZW-like relations (\ref{MCEQLeta}) and (\ref{2nd WZW rela}),
one can construct a new gauge invariant action as follows:
\begin{align}
S_{\eta}[V]
 = \int_{0}^{1} dt \, \langle  \Psi _{t }[V(t)] , \, Q \, \Psi _{\eta }[V(t)] \rangle.
\end{align}
The variation of the action can be taken in the same manner as that in section \ref{SUBSEC VOA EKS}, 
\begin{align}
\delta S_{\eta}[V] = \langle  \Psi _{\delta }[V] , \, Q \, \Psi _{\eta }[V] \rangle ,
\end{align}
and the equation of motion can be read off from it,
\begin{align}
%D_{\eta } \Psi _{Q} [V]=
Q \, \Psi _{\eta }[V] = 0 .
\end{align}
Since $D_\eta$ and $Q$ are nilpotent, 
the action is invariant under the gauge transformations, 
\begin{align}
\Psi _{\delta }[V] = D_{\eta } \Omega + Q \Lambda ,
\end{align}
where $\Omega$ and $\Lambda$ are gauge parameters belonging to the large Hilbert space,
which carry ghost numbers $0$ and $0$ and picture numbers $1$ and $0$, respectively.

%%%%%%%%%%%%%%%%%%%%%%%%%%%%%%%%%%%%%%%%%%%%%%%%%%%%%%%%%%%%%%%%%%%%%%%%%%%%%%%%%%%%%%%%
\subsection{Equivalence of the actions in the different parameterizations}

%\niu{Field redefinition}
%
%Under the partial gauge fixing $\xi V = 0$, or equivalently $V = \xi \Phi _{V}$, 
%\begin{align}
%\pi _{1} {\bf G} \big( e^{\wedge \Phi (t) } \big)  \cong \Psi _{\eta } [V] \equiv \int_{0}^{1}d \tau D_{\eta }[\tau ] V . 
%\end{align}

In this subsection, we show the equivalence of our new action $S_{\eta}[V]$ and $L_{\infty }$ action $S_{\scriptscriptstyle{\rm EKS}}[\Phi]$, by identifying the pure-gauge-like functional fields $\Psi _{\eta }[V]$ and $\Psi _{\eta } [\Phi ]$. 
We also derive the relation of two dynamical string fields $V$ and $\Phi $ from this identification. 
Let us consider the identification of the pure-gauge-like fields in the both parametarisations:
\begin{align}
\label{Id}
\pi _{1} \widehat{\bf G} \big( e^{\wedge \Phi (t) } \big)  = \Psi _{\eta } [\Phi(t)] \equiv  \Psi _{\eta } [V(t)] =  \int_{0}^{1}d \tau D_{\eta }(\tau) V(t) .
\end{align}
Apparently it provides the equivalence of the equations of motions $Q\Psi_\eta[V] =Q\Psi_\eta[\Phi]$.
In addition, one can see that it also provides the equivalence of the two actions $S_{\eta } [V]$ and $S_{\eta }[\Phi ] = S_{\scriptscriptstyle{\rm EKS} } [\Phi ]$, which are two different parametrisations of the same (alternative) WZW-like theory. 

\niu{Approach 1}

Under the identification $\Psi _{\eta } [\Phi] \equiv  \Psi _{\eta } [V]$, the associated fields $\Psi _{d}$ in two parameterizations are equivalent up to $D_\eta$-exact terms
$\Psi _{d} [\Phi] = \Psi _{d } [V] + ( D_\eta\textrm{-exact terms})$,
which is guaranteed by the WZW-like relation $(-)^d d \Psi_\eta = D_\eta \Psi_d$. 
We thus find 
\begin{align}
D_\eta \big( \Psi _{ {d}} [\Phi(t)]  -\Psi _{ {d}} [V(t)] \big)
=(-)^{ {d}} {d} \big( \Psi _{\eta} [\Phi(t)]  -\Psi _{\eta} [V(t)] \big)=0.
\end{align}
The $D_\eta$-exact terms in the associated fields do not affect to the WZW-like relation $(-)^d d \Psi_\eta = D_\eta \Psi_d$.
Recall that there exists the arbitrariness to add $D_\eta$-exact terms in the associated fields.
 %Note that it may be overconstrained to identify the associated fields exactly.
%\begin{align}
%\pi _{1} {\bf G} \boldsymbol \xi_t \big( e^{\wedge \Phi (t) } \big)  \cong \Psi _{t } [V] \equiv,
%\end{align}
Besides, since $Q \Psi _{\eta }$ is $D_\eta$-exact,
the difference between $\Psi _{t } [\Phi]$ and $\Psi _{t } [V]$ does not contribute to the action.
Then, two actions are shown to be equivalent:
\begin{align}
S_{\scriptscriptstyle{\rm EKS}} [\Phi]
= \int_{0}^{1} dt \, \langle \Psi _{t} [\Phi(t)]   ,  Q \Psi _{\eta } [\Phi(t)] \rangle
= \int_{0}^{1} dt \, \langle \Psi _{t} [V(t)]   ,  Q \Psi _{\eta } [V(t)] \rangle
=S_{\eta}[V].
\end{align}
The correspondence of the gauge parameters is given by (\ref{gauge para omega1}) and (\ref{gauge para lambda1}),
with the mixing of the trivial transformation ($\ref{TGT EKS Z2}$).
Note that we only use the WZW-like relations here,
and therefore this identification provides the equivalence of the WZW-like actions in the arbitrary parameterizations 
as long as the WZW-like relations hold.

\niu{Approach 2} 

We write $\xi $ for an operator satisfying $\ld \eta , \xi \rd = 1$. 
Let us consider a linear operator $f$ defined by 
\begin{align}
f \equiv \sum_{n=0}^{\infty } \big( \xi (D_{\eta } - \eta ) \big) ^{n} . 
\end{align}
One can quickly check this $f$ satisfies $\Ld f \xi , D_{\eta } \Rd = 0$. 
Since $Q \Psi _{\eta } = - D_{\eta } \Psi _{Q}$, $(D_{\eta } )^{2} = 0$, and 
\begin{align}
\Psi _{t} & =  (f \xi D_{\eta } + D_{\eta } f \xi ) \Psi _{t} =  f \xi \partial _{t} \Psi _{\eta } + D_{\eta } f \xi \Psi _{t} 
\no 
& =  \sum_{n=0}^{\infty } \big( \xi ( D_{\eta } -\eta ) \big) ^{n } \xi \partial _{t} \Psi _{\eta } +  D_{\eta } f \xi \Psi _{t} , 
\end{align}
we can rewrite the action as the following single functional form: 
\begin{align}
S_{\eta } & = \int_{0}^{1} dt \, \langle \Psi _{t} (t) , Q \Psi _{\eta } (t) \rangle = \sum_{n=0}^{\infty } \int_{0}^{1} \langle \big( \xi ( D_{\eta } (t) -\eta ) \big) ^{n } \xi \partial _{t} \Psi _{\eta }(t) , \, Q \Psi _{\eta } (t) \rangle . 
\end{align}
This form of the WZW-like action consists of the functional field $\Psi _{\eta }$ and operators $Q$, $\xi$, and $\partial _{t}$. 
Hence the identification (\ref{Id}) automatically provides the equivalence of two actions.

\niu{Relation of the fields $\Phi$ and $V$, and Partial-gauge-fixing}

%The identification $\Psi _{\eta } [\Phi] \equiv  \Psi _{\eta } [V]$ can be solved by $\Phi$.
%By expanding $\Phi=\Phi[V] = \Phi_1(V) + \Phi_2(V,V) + \Phi_3(V,V,V)+\cdots$ in powers of $V$,
%one can determine $\Phi_n$ perturbatively,
%and the condition which provides $S_{\eta}[V]$ from the $L_\infty$-action $S_{\scriptscriptstyle EKS}[\Phi]$ is obtained:

The identification $\Psi _{\eta } [\Phi] \equiv  \Psi _{\eta } [V]$ can be solved by $\Phi$.
Exponentiating both hand side, $e^{\wedge\Psi _{\eta } [\Phi]}= e^{\wedge\Psi _{\eta } [V]}$,
and using the property of the cohomomorphism and group-like element, the identification becomes
\begin{align}
e^{\wedge\Psi _{\eta } [\Phi]} 
=e^{\wedge\pi_1 \widehat{\mathbf G} (e^{\wedge\Phi})} 
= \widehat{\mathbf G} e^{\wedge\Phi}
=e^{ \wedge\Psi _{\eta } [V]}.
\end{align}
Since $\widehat{\mathbf G}$ is invertible, by acting $\widehat{\mathbf G}^{-1}$ and by projecting by $\pi_1$, 
the condition of the field corresponding relation which provides $S_{\eta}[V]$ from the $L_\infty$-action $S_{\scriptscriptstyle \rm  EKS}[\Phi]$ is obtained:
\begin{align}
\Phi[V] = \pi_1 \widehat{\mathbf G}^{-1}( e^{\wedge\Psi _{\eta } [V]}).
\end{align}
Expanding it in powers of $V$, it reads
\begin{align}
\Phi[V]&=\eta V
-\frac{1}2 \eta \uplambda^{[0]}_2 (V,\eta V)
+\frac{1}{12} \eta \Big(-\uplambda^{[0]}_3 (V,\eta V,\eta V) 
+2\uplambda^{[0]}_2 (\uplambda^{[0]}_2 (V,\eta V),\eta V) \no
&\hspace{90pt}-\uplambda^{[0]}_2 (V,\uplambda^{[0]}_2(\eta V, \eta V)) 
+2\uplambda^{[0]}_2 (V, \eta \uplambda^{[0]}_2(V, \eta V))\Big) + O(V^4) , 
\end{align}
where $\lambda ^{[0]}_{2}$ and $\lambda ^{[0]}_{3}$ are gauge products. 
(See section 2 or \cite{Erler:2014eba} for their explicit forms of $\lambda ^{[0]}$.) 

The identification $\Psi _{\eta } [\Phi] \equiv  \Psi _{\eta } [V]$ can be solved also by $V$ when the $\eta$-symmetry is fixed.
By expanding $V=V[\Phi]=V_1(\Phi)+V_2(\Phi,\Phi) + V_3(\Phi,\Phi,\Phi)+\cdots$ in powers of $\Phi$,
and by acting $\xi$ on the both hand sides of $\Psi _{\eta } [\Phi] \equiv \Psi _{\eta } [V]$, one can determine $V_n$ perturbatively.
The simplest choice of the partial-gauge-fixing condition is $\xi V=0$, which provides $\xi\eta V= V$.
Then the explicit form of the partially-gauge-fixed string field 
$V(\Phi)$ which provides the $L_\infty$-action $S_{\scriptscriptstyle \rm EKS}[\Phi]$ from $S_{\eta}[V]$ is obtained as follows:
\begin{align}
V[\Phi] &=\xi \Phi
+\frac{1}2 \xi\eta \uplambda^{[0]}_2(\xi \Phi,\Phi)
+\frac{1}{12}\xi\eta \Big( \uplambda^{[0]}_3 (\xi \Phi,\Phi,\Phi) 
-2\uplambda^{[0]}_2 (\uplambda^{[0]}_2 (\xi \Phi,\Phi),\Phi) 
+\uplambda^{[0]}_2 (\xi \Phi,\uplambda^{[0]}_2(\Phi, \Phi)) \no
&\hspace{90pt}
+\uplambda^{[0]}_2 (\xi \Phi, \eta \uplambda^{[0]}_2(\xi \Phi, \Phi))
+3\uplambda^{[0]}_2(\xi\eta \uplambda^{[0]}_2(\xi \Phi,\Phi),\Phi)\Big)
+ O(V ^{4}) .
\end{align}
If we choose the form of $\mathbf G$ as given in \cite{Erler:2014eba}, it reads
\begin{align}
V[\Phi] &=\xi\Phi
+\frac{1}{3!}\xi[\xi\Phi,\Phi]
+\frac{1}{4!}\Big(
\frac 14 \xi[ X\xi\Phi,\Phi,\Phi] +\frac14\xi X[\xi\Phi ,\Phi,\Phi ] +\frac12 \xi [X\Phi,\xi\Phi,\Phi]
\no
&\hspace{60pt}
+\frac13 \xi [\xi\Phi, \xi[\Phi,\Phi]] -\frac23 \xi [\xi\Phi,[\xi\Phi,\Phi]] +\frac43 \xi [\Phi,\xi[\xi\Phi,\Phi]]\Big)
+ O(V ^{4}) . 
\end{align}

\section{Conclusions and Discussions} 
In this paper, we clarified a WZW-like structure naturally arising from the $L_\infty$ formulation: 
The pure-gauge-like field $\Psi _{\eta } [\Phi ] \equiv \pi_{1} \widehat{\bf G} (e^{\wedge \Phi })$ and the associated fields $\Psi_{d} [\Phi ] \equiv \pi _{1} \widehat{\bf G} ( {\boldsymbol \xi }_{d} e^{\wedge \Phi } )$ satisfy the alternative WZW-like relations, which are $\mathbb Z_2$-reversed versions of the conventional WZW-like relations given in \cite{Berkovits:2004xh}.
We found that once WZW-like functionals $\Psi_{\eta }[\varphi ]$ and $\Psi_{d}[\varphi ]$ of some dynamical string field $\varphi $ satisfying these (alternative) WZW-like relations are given, one can construct a gauge invariant WZW-like action $S_{\eta }[ \varphi ]$ in terms of them. 
The $L_\infty$ action $S_{\scriptscriptstyle{\rm EKS}}$ just gives one realisation of this (alternative) WZW-like action $S_{\eta } [ \Phi ]$ parameterised by $\Phi$. 
On the basis of this procedure, we constructed a new WZW-like action(s) as another parameterisation of the WZW-like action $S_{\eta }[V]$ by using the string field $V$ living in the large Hilbert space. 
(For open NS or NS-NS theory, see appendix D or E, respectively.) 
The pure-gauge-like field $\Psi_\eta[V]$ and the associated fields $\Psi_{d} [V]$ can be defined by the differential equations which are the $\mathbb Z_2$-reversed versions of those in \cite{Berkovits:2004xh}.
We showed the equivalence of our new action $S_{\eta }[V]$ and the $L_{\infty }$ action $S_{\eta } [ \Phi ] \equiv S_{\scriptscriptstyle{\rm EKS}} [\Phi ]$ on the basis of the procedure demonstrated in \cite{Erler:2015rra}. 
The direct relation between two dynamical string fields $\Phi $ and $V$, a field redefinition, can be derived from this equivalence with partially gauge-fixing of $V$ or trivial uplift of $\Phi $. 

%\red{NS-NS.}

%%%%%%%%%%%%%%%%%%%%%%%%%%%%%%%%%%%%%%%%%%%%%%%%%%%%%%%%%%%%%%%%%%%%%%%%%%%%%%%%%%%%%%
\niu{Towards the equivalence to the conventional WZW-like action}

Although we expect the equivalence between our (alternative) WZW-like action $S_{\eta }[ V ]$ and Berkovits-Okawa-Zwiebach's (conventional) WZW-like action $S_\mathrm{\scriptscriptstyle BOZ}[ V_{\rm c} ]$, their relation remains to be understood. 
We write $V$ for the dynamical NS heterotic string field of the alternative WZW-like theory, and write $V_{\rm c}$ for that of the conventional WZW-like theory as follows: 
\begin{align*}
S_\eta[V]
&=
\frac12 \langle V, Q \eta V\rangle +\frac{\kappa}{3!}\langle V, Q [V,\eta V]^\eta\rangle 
+\frac{{\kappa}^2}{4!}\langle V, Q\big( [V,\eta V,\eta V]^\eta + \big[ V,[V,\eta V]^\eta \big] ^\eta \big) \rangle + \cdots,
\\
\!\!\!\!\!\!
S_\mathrm{\scriptscriptstyle BOZ}[V_{\rm c}] 
&= \frac12 \langle \eta V_{\rm c}, QV_{\rm c} \rangle
\!+\!\frac{\kappa}{3!} \langle \eta V_{\rm c},[V_{\scriptscriptstyle \rm c},QV_{\rm c}]\rangle
\!+\!\frac{\kappa^2 }{4!} \langle \eta V_{\rm c},
\!\big( [V_{\rm c},QV_{\rm c},QV_{\rm c}]
\!+\! \big[ V_{\rm c},[V_{\rm c},Q V_{\rm c}] \big] \big) \rangle+\cdots . 
\end{align*}
Both string fields $V$ and $V_{\rm c}$ belong to the large Hilbert space. 
At least, perturbatively, one can check their equivalence: 
We set $V_{\scriptscriptstyle \rm c}= V + O(\kappa^2)$ because of $\mathbf L^\eta_2 = -\mathbf L^\mathrm{BOS}_2$, which implies that the first nontrivial order is $\kappa^2$, namely the quartic interaction. 
Let us check how the equivalence can be shown in this order.
It would be crucial that the 3-products $\mathbf L^\eta_3 $ and $\mathbf L^\mathrm B_3$ are made from the same gauge product $\uplambda^{[1]}_3$:
\begin{align*}
\mathbf L^\eta_3 = -\frac{1}2 [\![ \mathbf Q,\uplambda^{[1]}_3]\!], \qquad\mathbf L^{\rm B}_3=\frac12[\![\boldsymbol \eta, \uplambda^{[1]}_3]\!].
\end{align*}
Utilizing them, the quartic interactions in both actions can be written in term of $\uplambda^{[1]}_3$ as
\begin{align}
\langle {V}, Q [{V},\eta {V},\eta {V}]^\eta\rangle 
%&=-\frac{1}2 \langle {V}, Q [\![ \mathbf Q,\uplambda^{[1]}_3]\!] ({V},\eta {V},\eta {V})\rangle \no
%&=-\frac{1}2 \langle Q {V}, \uplambda^{[1]}_3 \mathbf Q ({V},\eta {V},\eta {V})\rangle \no
&=-\frac{1}2 \langle Q{V},  \uplambda^{[1]}_3 (Q{V},\eta {V},\eta {V})\rangle +\langle Q{V},  \uplambda^{[1]}_3 ({V},Q\eta {V},\eta {V})\rangle,\label{S Z2 4}\\
\langle \eta V_{\scriptscriptstyle \rm c},[V_{\scriptscriptstyle \rm c},QV_{\scriptscriptstyle \rm c},QV_{\scriptscriptstyle \rm c}]\rangle
%&=\frac12\langle \eta V_{\scriptscriptstyle \rm c}, [\![\eta, \uplambda^{[1]}_3]\!](V_{\scriptscriptstyle \rm c},QV_{\scriptscriptstyle \rm c},QV_{\scriptscriptstyle \rm c})\rangle\no
%&=-\frac12\langle \eta V_{\scriptscriptstyle \rm c}, \uplambda^{[1]}_3 \eta (V_{\scriptscriptstyle \rm c},QV_{\scriptscriptstyle \rm c},QV_{\scriptscriptstyle \rm c})\rangle\no
&=-\frac12\langle \eta V_{\scriptscriptstyle \rm c}, \uplambda^{[1]}_3 (\eta V_{\scriptscriptstyle \rm c},QV_{\scriptscriptstyle \rm c},QV_{\scriptscriptstyle \rm c})\rangle-\langle \eta V_{\scriptscriptstyle \rm c}, \uplambda^{[1]}_3 (V_{\scriptscriptstyle \rm c},Q\eta V_{\scriptscriptstyle \rm c},QV_{\scriptscriptstyle \rm c})\rangle \label{S 4}.
\end{align}
%Since the cyclicity of the gauge product $\uplambda_3^{[1]}$ is assumed,
% for the gauge invariance of the $\mathbb Z_2$-reversed action,
The difference between the quartic interactions comes from the second terms of (\ref{S Z2 4}) and (\ref{S 4}): 
\begin{align}
S_{\eta,4}[V] - S_{{\scriptscriptstyle\rm WZW},4}[V_{\scriptscriptstyle c}] = \frac{\kappa^2}{4!}\langle Q\eta V, 2 \uplambda^{[1]}_3 (V,\eta V,QV)\rangle + O(\kappa^3),
\end{align}
which can be compensated by the following off-shell field redefinition, 
\begin{align}
V_\mathrm{\scriptscriptstyle c}= V + \frac{2\kappa^2}{4!} \uplambda^{[1]}_3 (V,\eta V,QV) + O(\kappa^3)\label{off shell equiv k2}.
\end{align}
Thus, 
under the identification of the string fields $V_\mathrm{\scriptscriptstyle c}$ and $V$ by (\ref{off shell equiv k2}),
two actions $S_\eta[V]$ and $S_\mathrm{\scriptscriptstyle WZW}[V_{\scriptscriptstyle \rm c}]$ are equivalent at $\kappa^2$.
 
%Utilizing $\uplambda^{[1]}_3 = \frac{2}{4}( \xi L^B_3 -L_3^B \xi)$,
%\begin{align}
%V'=V -\frac{\kappa^2}{4!} \Big( \xi[V,\eta V,QV]-[\xi V,\eta V,QV] +[V,\xi \eta V,QV]+[V,\eta V,\xi QV]\Big).
%\end{align}

To prove their all-order equivalence, it would be helpful to characterize the conventional WZW-like action in terms of the gauge products $\uplambda^{[d]}_N$.
In fact, $Q_\mathcal G$, the pure-gauge shifted BRST operator which plays a key role in conventional WZW-like formulation,
can be written as $Q_\mathcal G = \mathcal E_V^{-1} Q \mathcal E_V$ using a linear map $\mathcal E_V $ \cite{Goto:2015hpa}.
It may provide a solution of this problem to understand the relation between two cohomomorphisms $\mathcal E_V$ and $\widehat{\bf G}$.

%%%%%%%%%%%%%%%%%%%%%%%%%%%%%%%%%%%%%%%%%%%%%%%%%%%%%%%%%%%%%%%%%%%%%%%%%%%%%%%%%%%%%%%%

\niu{Ramond sector of closed superstring fields}

The concept of the WZW-like structure based on the dual product $\mathbf L^\eta$ will play crucial roles
in construction of the complete action of heterotic string field theory including both NS and R sectors.
As \cite{Kunitomo:2015usa}, our WZW-like action $S_{\eta } [V]$ will provide a good starting point of the construction of complete actions.\footnote{See a new result given by K.Goto and H.Kunitomo, arXiv:1606.07194.} 
% including the Ramond sector.
%For the open sting \cite{Kunitomo:2015usa}, the NS sector part of the complete action is the WZW-like action $S_\eta$.
%and the R sector part of the complete action are written using a linear operator $F$ satisfying $[\![D_\eta, F\Xi]\!] =1$.
It is expected that as demonstrated in \cite{Matsunaga:2015kra} for open superstrings, the WZW-like structure including the Ramond sector would be written by the $L_{\infty }$ products ${\bf L}^{\boldsymbol \eta } = \widehat{\bf G} \, {\boldsymbol \eta } \, \widehat{\bf G}^{-1}$ which are dual to the $L_{\infty }$ products of \cite{Erler:2015lya}: $\widetilde{\bf L} = \widehat{\bf G}^{-1} ({\bf Q} + \dots )\, \widehat{\bf G}$ satisfying $\Ld {\boldsymbol \eta } , \widetilde{\bf L} \rd = 0$.
%EKS 3 to action ga icchi suru nara dual product G -1 eta G with EKS 3 's G,

It is also expected that the action for closed NS-R and R-NS strings can be constructed in the same manner as that for heterotic string. 
For closed R-R strings, it is not clear whether or how the kinetic term can be constructed with no constraint yet. 
If the kinetic term can be constructed, it may be possible to construct the complete action of type II string on the basis of the concept of the WZW-like structure arising from the dual product ${\boldsymbol L}^{\boldsymbol \eta }$.

\section*{Acknowledgement} 
The authors would like to thank Hiroshi Kunitomo and Yuji Okawa for discussions and comments. 
H.M. would like to express his gratitude to his doctor for medical treatments. 
%%%%%%%%%%%%%%%%%%%% 
The work of H.M. was supported in part by Research Fellowships of the Japan Society for the Promotion of Science for Young Scientists.

\part*{Appendices}
\appendix

%by GM1 
\section{Basic facts of $A_{\infty}$ and $L_{\infty }$} 

In this section, we give a short review of coalgebraic description for $A_{\infty }/L_{\infty }$ algebras. 
For more details, see \cite{Lada:1992wc, Kajiura:2003ax, Getzler:2007} or some mathematical manuscripts.

\niu{Coalgebras: tensor algebra $\mathcal{T(H)}$ and its symmetrization $\mathcal{S(H)}$}

Let $\cal C$ be a set.
Suppose that a coproduct $\Delta : \cal C \to C \otimes ^{\prime } C$ is defined on $\cal C$ and it is coassociative
\begin{align}
(\Delta\otimes ^{\prime } \1)\Delta = (\1\otimes ^{\prime } \Delta)\Delta . 
\end{align}
Then, the pair $({\cal C},\Delta)$ is called a {\it coalgebra}. 
We write $\cal T(H)$ for a {\it tensor algebra} of a graded vector space $\mathcal{H}$: 
\begin{align}
\mathcal {T(H)}=\mathcal H^{\otimes 0} \oplus \mathcal H^{\otimes 1}\oplus \mathcal H^{\otimes 2} \oplus \cdots .
\end{align} 
One can define a coassociative coprduct $\Delta:{\cal T(H)}\to{\cal T(H)} \otimes ^{\prime } {\cal T(H)}$ by 
\begin{align}
\Delta(\Phi_1\otimes ...\otimes \Phi_{n})
\equiv \sum_{k=0}^{n} (\Phi_{1}\otimes ...\otimes \Phi_{k}) \otimes ^{\prime } (\Phi_{k+1}\otimes ...\otimes \Phi_{n})  
\end{align}
with $\otimes ^{\prime } \equiv \otimes $, where $\Phi_1\otimes...\otimes \Phi_{n}\in {\cal H}^{\otimes n} \subset \mathcal{T(H)}$. 
Then, the pair $({\cal T(H)},\Delta)$ gives a coalgebra. 
As well as $\mathcal{T(H)}$, its symmetrization $\mathcal{S(H)}$ also gives a coalgebra. 
Recall that the {\it symmetrized tensor product} $\wedge$ for $\Phi _{1}, \Phi _{2} \in \mathcal{H}$ is defined by 
\begin{align}
\Phi_1\wedge\Phi_2
\equiv \Phi_1\otimes\Phi_2
+(-)^{{\rm deg}(\Phi_1){\rm deg}(\Phi_2)}\Phi_2\otimes\Phi_1 , 
%\:\:,\:\: \Phi_i\in\cal H.
\end{align}
and it satisfies the following properties for $\Phi _{1} , \dots , \Phi _{n} \in \mathcal{H}$:
\begin{align}
\Phi_1\wedge\Phi_2&=(-)^{{\rm deg}(\Phi_1){\rm deg}(\Phi_2)}\Phi_2\wedge\Phi_1,\\
(\Phi_1\wedge\Phi_2)\wedge\Phi_3&=\Phi_1\wedge(\Phi_2\wedge\Phi_3),\\
\Phi_1\wedge\Phi_2\wedge...\wedge \Phi_n
&=\sum_{\sigma}(-)^{\sigma}\Phi_{\sigma(1)}\otimes\Phi_{\sigma(2)}\otimes...\otimes \Phi_{\sigma(n)}.
\end{align}
We write $\mathcal{H}^{\wedge n}$ for the vector space spanned by $n$-fold symmetrized tensor $\Phi _{1} \wedge \dots \wedge \Phi _{n}$, and $\mathcal{S(H)}$ for the symmetrization of $\mathcal{T(H)}$, which is called the {\it symmetrized tensor algebra} $\cal S(H)$: 
\begin{align}
\mathcal {S(H)}=\mathcal H^{\wedge 0}\oplus\mathcal H^{\wedge 1}\oplus\mathcal H^{\wedge 2}\oplus\cdots.
\end{align}
A coassociative coprduct $\Delta:{\cal S(H)}\to{\cal S(H)} \otimes ^{\prime } {\cal S(H)}$ can be defined by 
\begin{align}
\Delta(\Phi_1\wedge...\wedge \Phi_{n})
\equiv \sum_{k=0}^{n}{\sum_\sigma}'(-)^{\sigma}
(\Phi_{\sigma(1)}\wedge...\wedge \Phi_{\sigma(k)})
\otimes ^{\prime }
(\Phi_{\sigma(k+1)}\wedge...\wedge \Phi_{\sigma(n)})  
\end{align}
with $\otimes ^{\prime } \equiv \otimes$, where $\Phi_1\wedge...\wedge \Phi_{n}\in {\cal H}^{\wedge n}$ and $\sigma$ runs over $(k,n-k)$-unshuffle. 
Then, the pair $({\cal S(H)},\Delta)$ gives coalgebra. 

\vspace{2mm}

In the case of open superstring field theory, $\mathcal{H}$ is the state space of open superstrings, which is a $\mathbb{Z}_{2}$-graded vector space, and its grading which we call degree is given by the Grassmann parity minus one mod $2$. 
Them, the pair $(\mathcal{T(H)} , \Delta )$ is the Fock space of open superstrings. 
On the other hand, in the case of closed superstring field theory, $\mathcal{H}$ is the state space of closed superstrings, which is a $\mathbb{Z}_{2}$-graded vector space, and its grading which we call degree is given by the Grassmann parity mod $2$. 
Then, the pair $(\mathcal{S(H)} , \Delta )$ is the Fock space of closed superstrings.

\niu{Multi-linear maps as a coderivation}

A linear operator $m:{\cal C}\to {\cal C}$ which raise the degree one is called {\it coderivation} if it satisfies
\begin{align}
\Delta m = (m\otimes \1)\Delta +(\1\otimes m)\Delta.
\end{align}
Multilinear maps with degree $1$ and $0$ naturally induce the maps from $\cal T(H)$ to $\cal T(H)$ or from $\mathcal{S(H)}$ to $\mathcal{S(H)}$. 
They are called a coderivation and a cohomomorphism respectively. 

\vspace{2mm} 

Recall that from a $n$-fold multilinear product $b_n$ of $\Phi _{1} , \dots , \Phi _{n} \in \mathcal H$, one can define a linear map on $\mathcal{H}^{\otimes n}$, which we write $b_n:\mathcal H^{\otimes n} \to \mathcal H$, by 
\begin{align}
b_n(\Phi_1\otimes \Phi_2 \otimes ...\otimes \Phi_n) \equiv b_n(\Phi_1,\Phi_2,...,\Phi_n) , 
\end{align}
where $\Phi _{1} \otimes \dots \otimes \Phi _{n} \in \mathcal{H}^{\otimes n}$. 
Let $A:\mathcal H^{\otimes k} \to \mathcal H^{\otimes l}$ and $B:\mathcal H^{\otimes m}\to \mathcal H^{\otimes n}$ be multilinear maps. 
One can naturally define a product of these maps $A \otimes B: \mathcal H^{\otimes k+m}\to \mathcal H^{\otimes l+n}$ by 
\begin{align}
A\otimes B(\Phi_1 \otimes ...\otimes \Phi_{k+m}) =\sum_{k} (-)^{B (\Phi _{1} + \dots + \Phi _{k})} A(\Phi_{1} \otimes ...\otimes \Phi_{k}) \otimes B(\Phi_{k+1} \otimes ...\otimes \Phi_{k+m}).
\end{align}
%%%
Then, the identity operator on $\mathcal{H}^{\otimes n}$ and ${\cal H}^{\wedge n}$ is defined by 
\begin{align}
\mathbb I_n= \overbrace{ \mathbb I\otimes\mathbb I\otimes...\otimes\mathbb I }^{n} 
= \frac1{n!} \overbrace{\mathbb I\wedge\mathbb I\wedge...\wedge\mathbb I }^{n} .
\end{align}
%Note that we need the coefficient $\frac 1{n!}$.
Using a degree one map $b_n:{\cal H}^{\otimes n}\to \cal H$, one can naturally construct a coderivation ${\bf b}_n: \mathcal{T(H)} \to \mathcal{T(H)}$ by
\begin{align} 
{\bf b}_n \Phi = \sum_{k=0}^{N-n} \big( \mathbb{I}_{k} \otimes b_n \otimes \mathbb I_{N-n-k} \big) \Phi , \hspace{5mm} \Phi \in {\cal H}^{\otimes N\geq n}\subset \mathcal {T(H)},
\end{align}
and ${\bf b}_n$ vanishes when acting on $ {\cal H}^{\otimes N\leq n}$.
%We will call $\mathbf b_n$ a {\it $n$-coderivation }.

\vspace{2mm} 

Similarly, a symmetric multilinear product $b_n$ of $\Phi _{1} , \dots \Phi _{n} \in \mathcal{H}$ naturally defines a map $b_n:\mathcal H^{\wedge n}\to \mathcal H$ by
\begin{align}
b_n(\Phi_1\wedge\Phi_2\wedge...\wedge \Phi_n)=b_n(\Phi_1,\Phi_2,...,\Phi_n).
\end{align}
The symmetric tensor product of two multilinear maps $A:\mathcal H^{\wedge k}\to \mathcal H^{\wedge l}$ and $B:\mathcal H^{\wedge m}\to \mathcal H^{\wedge n}$,
$A\wedge B:\mathcal H^{\wedge k+m}\to \mathcal H^{\wedge l+n}$,
can also be defined naturally by
\begin{align}
A\wedge B(\Phi_1\wedge...\wedge \Phi_{k+m})
={\sum_\sigma}'(-)^{\sigma}
A(\Phi_{\sigma(1)}\wedge...\wedge \Phi_{\sigma(k)})
\wedge
B(\Phi_{\sigma(k+1)}\wedge...\wedge \Phi_{\sigma(k+m)}).
\end{align}
A degree map $b_n:{\cal H}^{\wedge n}\to\cal H$ naturally gives a coderivation ${\bf b}_n:\mathcal {S(H)}\to\mathcal {S(H)}$ defined by 
\begin{align}
{\bf b}_n\Phi =(b_n\wedge\mathbb I_{N-n})\Phi\:\:,\:\:\Phi \in {\cal H}^{\wedge N\geq n}\subset \mathcal {S(H)},
\end{align}
and ${\bf b}_n$ vanishes when acting on $ {\cal H}^{\wedge N\leq n}$.
%We will call $\mathbf b_n$ a {\it $n$-coderivation }.
For example, we find that ${\bf b}_1$ acts as follows: 
\begin{align}
\begin{array}{rccl}
{\bf b}_1 : & 1 & \to &0 \\
&\Phi_1&\to &b_1(\Phi_1)\\
&\Phi_1\wedge\Phi_2&\to &b_1(\Phi_1)\wedge \Phi_2 +(-)^{{\rm deg}(\Phi_1){\rm deg}(b_1)}\Phi_1\wedge b_1(\Phi_2).
\end{array}
\end{align}
In particular, the coderivation ${\bf b}_0:\cal S(H)\to S(H)$ derived from a map $b_0:\mathcal H^{0}\to \mathcal H$ is given by 
\begin{align}
{\bf b}_0 \Phi =(b_0\wedge\mathbb I_{N})\Phi =b_0\wedge\Phi\:\:,\:\:\Phi \in {\cal H}^{\wedge N}\subset \mathcal {S(H)} 
\end{align}
and acts as 
\begin{align}
\begin{array}{rccl}
{\bf b}_0:&1&\to&b_0\\
&\Phi_1&\to&b_0\wedge \Phi_1\\
&\Phi_1\wedge\Phi_2&\to &b_0\wedge\Phi_1\wedge \Phi_2 .
\end{array}
\end{align}
%It is useful in the field redefinition and the gauge transformation.

Given two coderivations ${\bf b}_n$ and ${\bf c}_m$ which are derived from $b_n:{\cal H}^{\wedge n}\to \cal H$ and $c_m:{\cal H}^{\wedge m}\to \cal H$ respectively,
the graded commutator $[\![{\bf b}_n ,{\bf c}_m]\!]$ becomes the coderivation derived from 
the map $[\![b_n,c_m]\!]:{\cal H}^{\wedge n+m-1}\to \cal H$ which is defined by
\begin{align}
[\![b_n,c_m]\!]&=b_n(c_m\wedge\mathbb I_{n-1})-(-)^{{\rm deg}(b_n){\rm deg}(c_m)}c_m(b_n\wedge\mathbb I_{m-1}).
\end{align}

\niu{Multilinear maps as a cohomomorphism}

Given two coalgebras $C,C'$, a {\it cohomomorphism} $\widehat{\sf f}: C \to C'$ is a map of degree zero satisfying
\begin{align}
\Delta \widehat{\sf f} = (\widehat{\sf f}\otimes ^{\prime } \widehat{\sf f})\Delta.\label{cohom}
\end{align}

A set of degree zero multilinear maps $ \{ \mathsf f_n : \mathcal H^{\wedge n}\to \mathcal H'\}_{n=0}^{\infty}$ 
naturally induces a cohomorphism $\widehat{\mathsf f} : \mathcal{S(H)}\to \mathcal{S(H')}$, which we denote as $\widehat{\mathsf f}= \{ \mathsf f_n\}_{n=0}^{\infty}$.
Its action on $\Phi _{1}\wedge \dots \wedge \Phi _{n} \in \mathcal H^{\wedge n}\subset  \mathcal{S(H)}$ is defined by
\begin{align}
\widehat{\mathsf f} (\Phi _{1}\wedge \dots \wedge \Phi _{n})=
\sum_{i \leq n} \sum_{k_{1} < \dots < k_{i}} 
&e^{\wedge \mathsf f_0}\wedge{\sf f}_{k_{1}} ( \Phi _{1} , \dots , \Phi _{k_{1}}) \wedge {\sf f}_{k_{2}-k_{1}} ( \Phi _{k_{1} +1} , \dots , \Phi _{k_{2}} ) \wedge \no
&\hspace{15pt}\dots \wedge  {\sf f}_{k_{i} -k_{i-1}} ( \Phi _{k_{i-1}+1} , \dots , \Phi _{n} ).
\end{align}

Its explicit actions are given as follows:
\begin{align}
\begin{array}{rccl}
\widehat{\mathsf f} :&1&\to& e^{\wedge \mathsf f_0}\\
&\Phi&\to&e^{\wedge \mathsf f_0}\wedge \mathsf f_1(\Phi) \\
&\Phi_1\wedge\Phi_2&\to &e^{\wedge \mathsf f_0}\wedge \mathsf f_1(\Phi_1)\wedge  \mathsf f_1(\Phi_2) 
+e^{\wedge \mathsf f_0}\wedge \mathsf f_2(\Phi_1\wedge\Phi_2) .
\end{array}
\end{align}

\niu{Cyclic $A_\infty$ and cyclic $L_{\infty }$}

Let $\cal H$ be a graded vector space and $\cal T(H)$ be its tensor algebra.
A {\it weak $A_\infty$-algebra} $(\mathcal{H},{\bf M})$ is a coalgebra $\cal T(H)$ with a coderivation ${\bf M} ={\bf M}_0+{\bf M}_1+{\bf M}_2 +...$ satisfying
\begin{align}
({\bf M})^2=0. \label{M^2=0}
\end{align}
We denote the collection of the multilinear maps $\{ M_k \}_{k\geq 0}$ also by $\bf M$. 
In particular, if ${\bf M}_0 = 0$, $(\mathcal{H},{\bf M})$ is called an {\it $A_\infty$-algebra}.
Note that for fixed $n$, the equation (\ref{M^2=0}) gives  
\begin{align}
{\bf M}_n\cdot{\bf M}_1+{\bf M}_{n-1}\cdot{\bf M}_2+\cdots+{\bf M}_2\cdot{\bf M}_{n-1}+{\bf M}_1\cdot{\bf M}_n=0.
\end{align}
We can act it on $B_1\otimes B_2\otimes ... \otimes B_n \in\mathcal H^{\otimes n}$ 
to get the $A_\infty$ relations for the maps $\{M_k\}$:
\begin{align}
\sum_{k=0}^{n-i} (-)^{B_{1} + \dots + B_{k}} M_{n-i+1} \big(  B_{1} , \dots , B_{k} ,  M_i ( B_{k+1}, \dots , B_{k+i} ), B_{k+i+1} , \dots , B_{n} \big) = 0 .
\end{align}

Let $\langle \cdot ,\cdot\rangle:\mathcal H^{\otimes 2}\to \mathbb C$ be the BPZ inner product, which gives the graded symplectic form $\langle\omega|:\mathcal H^{\otimes 2}\to \mathbb C$: 
\begin{align}
\langle A,B\rangle
=(-)^A\langle\omega|A\otimes B.
\end{align}
Given the operator $\mathcal O_n$, we can define its BPZ-conjugation $O^\dagger_n$ as follows:
\begin{align}
\langle\omega|\mathbb I \otimes \mathcal O_n =
\langle\omega| \mathcal O^\dagger_n \otimes\mathbb I.
\end{align}
A pair $({\cal T(H)},{\bf M},\omega)$ is called {\it cyclic $A_\infty$-algebra}
if each $M_n$ is BPZ-odd,
\begin{align}
M_n^\dagger = -M_n. 
\end{align}

\vspace{2mm} 

Let $\cal H$ be a graded vector space and $\cal S(H)$ be its symmetrized tensor algebra.
A {\it weak $L_\infty$-algebra} $(\mathcal{H},{\bf L})$ is a coalgebra $\cal S(H)$ with a coderivation ${\bf L} ={\bf L}_0+{\bf L}_1+{\bf L_2}+...$ satisfying
\begin{align}
({\bf L})^2=0.\label{L^2=0}
\end{align}
We denote the collection of the multilinear maps $\{L_k\}_{k\geq 0}$ also by $\bf L$.
In particular,  if ${\bf L}_0=0$, $(\mathcal{H},{\bf L})$ is called an {\it $L_\infty$-algebra}.

In the case of an $L_\infty$-algebra, the part of (\ref{L^2=0}) that correspond to an $n$-fold multilinear map $\mathcal H^{\wedge n}\to\mathcal H$ is given by
\begin{align}
{\bf L}_n\cdot{\bf L}_1+{\bf L}_{n-1}\cdot{\bf L}_2+\cdots+{\bf L}_2\cdot{\bf L}_{n-1}+{\bf L}_1\cdot{\bf L}_n=0.
\end{align}
We can act it on $B_1\wedge B_2\wedge ... \wedge B_n \in\mathcal H^{\wedge n}$ 
to get the $L_\infty$ relations for the multilinear maps $\{L_k\}$:
\begin{align}
0=\sum_{i+j=n+1}{\sum_{\sigma}}'(-)^\sigma
L_j( L_i(B_{\sigma(1)},\dots,B_{\sigma(i)}),B_{\sigma(i+1)},\dots,B_{\sigma(n)}).
\end{align}
A set $({\cal S(H)},{\bf L},\omega)$ is called {\it cyclic $L_\infty$-algebra}
if each $L_n$ is BPZ-odd,
\begin{align}
L_n^\dagger = -L_n.
\end{align}

\niu{Projector and group-like element}

We can naturally define a {\it projector} ${ \pi_1}: {\cal S(H)}\to {\cal H}$
whose action on $\Phi\in\cal S(H)$ is given by
\begin{align}
{ \pi_1} \Phi = \Phi_1, 
\Phi =\sum_{n=1}^\infty \Phi _{1}\wedge \dots \wedge \Phi _{n} \in \mathcal{S(H)}.\label{pi}
\end{align}
Note that $\pi_1$ acts trivially on $\cal H$ 
and commutes with one-coderivations.

Let ${\cal H}_0$ be the degree zero part of $\cal H$.
The following exponential map of $\Phi\in {\cal H}_0$, 
\begin{align}
e^{\wedge \Phi} = {\bf 1} +\Phi + \frac12 \Phi\wedge\Phi +\frac1{3!}\Phi\wedge\Phi\wedge\Phi +\cdots , 
\end{align}
is called a {\it group-like element}. It satisfies
\begin{align}
\Delta e^{\wedge\Phi} =e^{\wedge\Phi}\wedge e^{\wedge\Phi} . 
\end{align}
 
The action of a coderivation on a group-like element is given by
\begin{align}
{\bf b}_n (e^{\wedge\Phi}) = \frac1{n!} b_n(\Phi^{\wedge n})\wedge e^{\wedge\Phi} , 
\end{align}
where we promise $0!=1$.
Note that we can not distinguish a one-coderivation $\mathbf b_1$ derived from a linear map $b_1:\mathcal H \to \mathcal H; \Phi \mapsto b_1(\Phi)$
and a zero-coderivation $\mathbf b_0$ derived from $b_0:{\cal H}^{\wedge 0} \to \mathcal H; 1\mapsto b_0=b_1(\Phi)$ 
when acting on group-like element, 
\begin{align}
{\bf b}_0 (e^{\wedge\Phi})=b_0 \wedge(e^{\wedge\Phi})=b_1(\Phi) \wedge(e^{\wedge\Phi})={\bf b}_1 (e^{\wedge\Phi}).
\end{align}

One of the important property of a cohomomorphism is its action on the group-like element:
\begin{align}
\Delta \widehat{\mathsf f} (e^{\wedge \Phi}) =(\widehat{\mathsf f} \otimes \widehat{\mathsf f})\Delta e^{\wedge \Phi}=(\widehat{\mathsf f} \otimes \widehat{\mathsf f})e^{\wedge \Phi}\wedge e^{\wedge \Phi}
= \widehat{\mathsf f} (e^{\wedge \Phi})\wedge \widehat{\mathsf f} (e^{\wedge \Phi}).
\end{align}
We can see that the cohomomorphisms preserves the group-like element : $\widehat{\mathsf f} (e^{\wedge \Phi})=e^{\wedge \Phi'}$.

Utilizing the projector and the group-like element, 
the {\it Maurer-Cartan element} for an $L_\infty$-algebra $(\mathcal H, \mathbf L)$ is given by
\begin{align}
\mathcal F_\Phi
:= \pi_1{\bf L} (e^{\wedge \Phi}) 
= \mathbf L_1(\Phi) + \frac12 \mathbf L_2(\Phi\wedge\Phi) +\frac1{3!}\mathbf L_3(\Phi\wedge\Phi\wedge\Phi) +\cdots.
\end{align}
The {\it Maurer-Cartan equation} for an $L_\infty$-algebra $(\mathcal H, \mathbf L)$ is given by
$\mathcal F_\Phi =0$,
which correspond to the on-shell condition in string field theory based on the $L_\infty$-algebra$(\mathcal H, \mathbf L)$.

%K.Goto 
\section{Derivation of ($\ref{CYC H ABC}$)}
\label{APP cyc H ABC}
For cyclic cohomomorphism $\widehat{H}$, the following relation holds:
\begin{align}
\langle \omega| \pi_2 
=\langle \omega| \pi_2 \widehat{H} 
=\langle \omega| \nabla (\pi_1 \otimes ' \pi_1) \Delta \widehat{H} 
=\langle \omega| \nabla (\pi_1 \widehat{H} \otimes ' \pi_1 \widehat{H}) \Delta,
\end{align}
where $\pi_2$ is the projector $\mathcal S( \mathcal H) \to \mathcal H \wedge \mathcal H$,
and $\nabla$ is the product of the tensor algebra $(\mathcal T(\mathcal H), \nabla)$.

Let us consider its action
\begin{align}
&\frac{1}{1-\Phi}\otimes A\otimes \frac{1}{1-\Phi}\otimes B\otimes \frac{1}{1-\Phi}\otimes C\otimes \frac{1}{1-\Phi}\no
&\quad+(-)^{AB}\frac{1}{1-\Phi}\otimes B\otimes \frac{1}{1-\Phi}\otimes A\otimes \frac{1}{1-\Phi}\otimes C\otimes \frac{1}{1-\Phi}\no
&\qquad+(-)^{BC}\frac{1}{1-\Phi}\otimes A\otimes \frac{1}{1-\Phi}\otimes C\otimes \frac{1}{1-\Phi}\otimes B\otimes \frac{1}{1-\Phi}.
\label{calc cyclicity cohom 1}
\end{align}
The left hand side vanishes since $\pi_2$ is acting on $\mathcal H^{\wedge n\geq 3}$.
For the right hand side, the first term of (\ref{calc cyclicity cohom 1}) becomes
\begin{align}
&\langle \omega| \nabla (\pi_1 \widehat{H} \otimes ' \pi_1 \widehat{H}) \Delta
\big(\frac{1}{1-\Phi}\otimes A\otimes \frac{1}{1-\Phi}\otimes B\otimes \frac{1}{1-\Phi}\otimes C\otimes \frac{1}{1-\Phi}\big)\no
&=
\langle \omega| \nabla (\pi_1 \widehat{H} \otimes ' \pi_1 \widehat{H}) 
\Big(\frac{1}{1-\Phi}\otimes' \frac{1}{1-\Phi}\otimes A\otimes \frac{1}{1-\Phi}\otimes B\otimes \frac{1}{1-\Phi}\otimes C\otimes \frac{1}{1-\Phi}\no
&\hspace{120pt} +\frac{1}{1-\Phi}\otimes A\otimes \frac{1}{1-\Phi}\otimes'\frac{1}{1-\Phi}\otimes B\otimes \frac{1}{1-\Phi}\otimes C\otimes \frac{1}{1-\Phi}\no
&\hspace{120pt} +\frac{1}{1-\Phi}\otimes A\otimes \frac{1}{1-\Phi}\otimes B\otimes \frac{1}{1-\Phi}\otimes' \frac{1}{1-\Phi}\otimes C\otimes \frac{1}{1-\Phi}\no
&\hspace{120pt} +\frac{1}{1-\Phi}\otimes A\otimes \frac{1}{1-\Phi}\otimes B\otimes\frac{1}{1-\Phi}\otimes C\otimes  \frac{1}{1-\Phi}\otimes' \frac{1}{1-\Phi}\Big).
\label{calc cyclicity cohom 2}
\end{align}
Hereafter we assume $A+B+C = even$. 
The inner product in the large Hilbert space is
\begin{align}
\langle a,b\rangle =(-)^a\langle \omega | a\otimes b,
\end{align}
and the 1st term and 4th term of (\ref{calc cyclicity cohom 2}) cancel since $\langle a,b\rangle =(-)^{(a+1)(b+1)}\langle b,a\rangle$.
Now, gathering the contribution from all terms of (\ref{calc cyclicity cohom 1}),
and utilizing the identities
\begin{align}
a\wedge e^{\wedge\Phi} &=
 \frac{1}{1-\Phi}\otimes a\otimes \frac{1}{1-\Phi},\\
 a\wedge b\wedge e^{\wedge\Phi} &=
 \frac{1}{1-\Phi}\otimes a\otimes \frac{1}{1-\Phi}\otimes b\otimes \frac{1}{1-\Phi} 
+(-)^{ab}\frac{1}{1-\Phi}\otimes b\otimes \frac{1}{1-\Phi}\otimes a\otimes \frac{1}{1-\Phi},
\end{align}
one can obtain the BPZ property for $\widehat{H}$: 
\begin{align}
0 = 
(-)^{B} \langle \pi_1 \widehat{H} (A \wedge B \wedge e^{\wedge\Phi} ),  \pi_1 \widehat{H} ( C\wedge e^{\wedge\Phi} ) \rangle
+ \langle \pi_1 \widehat{H} ( A\wedge e^{\wedge\Phi} ),  \pi_1 \widehat{H} (B\wedge C \wedge e^{\wedge\Phi} ) \rangle.
\end{align}
Note that in the computation some terms cancel by the symmetric property of the inner product.

%Note that in the calculation we used 
%\begin{align}
%0=&+(-)^A\langle\pi_1H(\frac{1}{1-\Phi}\otimes A\otimes \frac{1}{1-\Phi}),\pi_1 H(\frac{1}{1-\Phi}\otimes B\otimes \frac{1}{1-\Phi}\otimes C\otimes \frac{1}{1-\Phi})\rangle\no
%&+(-)^{A(B+C)+B+C}\langle\pi_1H(\frac{1}{1-\Phi}\otimes B\otimes \frac{1}{1-\Phi}\otimes C\otimes \frac{1}{1-\Phi}),\pi_1 H(\frac{1}{1-\Phi}\otimes A\otimes \frac{1}{1-\Phi})\rangle,
%\end{align}
%which again follows from the symmetric property of the inner product.

%By Keiyu Goto 
%%%%%%%%%%%%%%%%%%%%%%%%%%%%%%%%%%%%%%%%%%%%%%%%%%%%%%%%%%%%%%%%%%%%%%%%%%%%%%%
\section{Embedding to the large Hilbert space of NS closed string}

In this appendix, we consider 
the trivial embedding of the NS closed string field $\Phi$ in the $L_\infty$-formulation into the string field $\widetilde{V}$ in the large Hilbert space,
and see that the trivially embedded action is also WZW-like form.
In other words, we provide a parameterization of the pure-gauge-like field and the associated fields by another large space string field $\widetilde{V}$.
In particular, we discuss the relation between the gauge transformations in two representations.

We define the trivial embedding by replacing $\xi \Phi$ to  ${\widetilde{V}}$ which is a string field belonging to the large Hilbert space,
so that under the partial-gauge-ficing condition $\xi {\widetilde{V}}=0$, the trivial solution ${\widetilde{V}} =\xi \Phi$ reproduce the $L_\infty$-formulation 
parameterized by $\Psi$, the string field in the small Hilbert space.
The action for $\widetilde{V}$ is given by 
\begin{align}
S_{\scriptscriptstyle{\rm EKS}}[{\widetilde{V}}] 
&= \int_{0}^{1} dt \, \langle \partial_t \widetilde{V}(t)  , {\pi_1 } \big( {\bf L}( e^{\wedge \eta\widetilde{V} (t)} ) \big) \rangle \no \label{B1}
&= \int_{0}^{1} dt \, \langle {\pi_1 } \widehat{\bf G} \big( (\partial_{t} \widetilde{V}(t)) \wedge  e^{\wedge \eta\widetilde{V}(t)} \big)  , Q \, {\pi_1 }\,  \widehat{\bf G} ( e^{\wedge \eta\widetilde{V} (t)} )  \rangle .
\end{align}
One can confirm the functionals appearing in the action 
\begin{align}
\Psi _{\eta}[{\widetilde{V}}] &= \pi _{1} \widehat{\bf G} \big( e^{\wedge \eta {\widetilde{V}} } \big),\label{PG_TD}\\
\Psi _{ {d}}[{\widetilde{V}}] &= \pi _{1} \widehat{\bf G} \big(  {d} {\widetilde{V}}\wedge e^{\wedge \eta {\widetilde{V}} } \big) \qquad (\textrm{for }{d} = \partial_t , \delta )
\label{ASF_TD}
\end{align}
satisfy the WZW-like relations:
\begin{align}
0=\pi _{1} {\bf L}^{\boldsymbol \eta } \big( e^{\wedge\Psi_{\eta }[{\widetilde{V}}]} \big) ,\qquad
 (-)^{d} d \, \Psi_{\eta }[{\widetilde{V}}] = D_{\eta } \, \Psi_{d}[{\widetilde{V}}].
\end{align}
The first relation follows from the projection invariance $e^{\wedge \eta {\widetilde{V}}}=\eta \xi e^{\wedge \eta {\widetilde{V}}}$ as (\ref{pre 1st wzw}).
To check the second relation, consider
\begin{align}
\Psi _{ {d}}[{\widetilde{V}}] 
%=\pi _{1} {\bf G} \big(  {d} {\widetilde{V}}\wedge e^{\wedge \eta {\widetilde{V}} } \big) 
= \pi _{1} \widehat{\bf G} \big(  {d}(\eta \xi + \xi \eta) {\widetilde{V}}\wedge e^{\wedge \eta {\widetilde{V}} } \big) 
= \pi _{1} \widehat{\bf G} \big(  {d} \xi \eta {\widetilde{V}}\wedge e^{\wedge \eta {\widetilde{V}} } \big)  
+(-)^dD_\eta \pi_1 \widehat{\bf G} \big( d \xi  {\widetilde{V}}\wedge e^{\wedge \eta {\widetilde{V}} } \big),
\end{align}
where we used
\begin{align}
(-)^d\pi _{1} \widehat{\bf G} \big( d \eta \xi   {\widetilde{V}}\wedge e^{\wedge \eta {\widetilde{V}} } \big) 
&=\pi _{1} \widehat{\bf G} \eta \big( d \xi {\widetilde{V}}\wedge e^{\wedge \eta {\widetilde{V}} } \big)
=\pi _{1} \mathbf L^\eta \widehat{\bf G} \big(  d\xi {\widetilde{V}}\wedge e^{\wedge \eta {\widetilde{V}} } \big) \no
&=\pi _{1} \mathbf L^\eta \Big( \pi_1 \widehat{\bf G} \big( d\xi {\widetilde{V}}\wedge e^{\wedge \eta {\widetilde{V}} } \big) \wedge \widehat{\mathbf G} e^{\wedge \eta {\widetilde{V}}}\Big)
=D_\eta \pi_1 \widehat{\bf G} \big( d \xi  {\widetilde{V}}\wedge e^{\wedge \eta {\widetilde{V}} } \big).\label{B6}
\end{align}
Since the second term vanishes when acted by $D_\eta$,
the same computation as (\ref{pre second WZW}) can be applied to show $ (-)^{d} d \, \Psi_{\eta }[{\widetilde{V}}] = D_{\eta } \, \Psi_{d}[{\widetilde{V}}]$.

In fact, the embedding procedure $\xi\Phi \to \widetilde V$ has an ambiguity: 
One can consider the replacing $\xi\Phi = \xi \eta \xi \Phi \to \xi \eta \widetilde V$.
However, the difference of $\Psi_d$ produced by the difference of the procedure
is only $D_\eta$-exact term as seen in (\ref{B6}), which does not affect to the WZW-like relation $(-)^d d \Psi_\eta = D_\eta \Psi_d$.
Besides, since $Q \Psi _{\eta }$ is $D_\eta$-exact, the $D_\eta$-exact part of $\Psi_t$ does not affect the action,
that is, the action does not depend on this ambiguity.
Here we fix this ambiguity by requiring the invertibility of $\Psi_{ {d} = \delta , \partial_t}$ as functions of $ {d} {\widetilde{V}}$ \cite{Erler:2015uoa}.

For derivation $ {d}$ which does not commute with $\widehat{\mathbf G}$, such as $Q$, 
we define $\Psi_d[\widetilde V]$ by
\begin{align}
\Psi _{ d} [{\widetilde{V}}] 
=\pi _{1} \widehat{\bf G} \big( \pi _1 \widehat{\mathbf G}^{-1}  {\boldsymbol d} \widehat{\mathbf G} ( {\widetilde{V}} \wedge {\textstyle \int}  e^{\wedge \eta {\widetilde{V}}} ) \wedge e^{\wedge \eta {\widetilde{V}}}\big),
\end{align}
where 
\begin{align}
{\textstyle \int} e^{\wedge \eta {\widetilde{V}}} = \sum_{n=0}^\infty\frac{1}{(n+1)!}(\eta {\widetilde{V}})^{\wedge n}.
\end{align}
For $ {d}=Q$, it reads
\begin{align}
\Psi _{Q} [{\widetilde{V}}] 
=\pi _{1} \widehat{\bf G} \big( \pi_1 \mathbf L ( {\widetilde{V}} \wedge {\textstyle \int}  e^{\wedge \eta {\widetilde{V}}} ) \wedge e^{\wedge \eta {\widetilde{V}}}\big).
\end{align}
To check the WZW-like relations, 
utilizing $\widehat{\mathbf G}^{-1}  {\boldsymbol d} \widehat{\mathbf G} = (-)^{ {\boldsymbol d}} \ld {\boldsymbol \eta } , {\boldsymbol \xi }_{ {\boldsymbol d}} \rd $, 
we write $\Psi_d[\widetilde V]$ as
\begin{align}
\Psi_d[\widetilde V]&=\pi _{1} \widehat{\bf G} \big( \pi_1 \widehat{\mathbf G}^{-1}  {\boldsymbol d} \widehat{\mathbf G} ( {\widetilde{V}} \wedge {\textstyle \int}  e^{\wedge \eta {\widetilde{V}}} ) \wedge e^{\wedge \eta {\widetilde{V}}}\big)\no
&=\pi _{1} \widehat{\bf G} \big( \pi_1 (-)^{ \mathbf {d}} \ld {\boldsymbol \eta } , {\boldsymbol \xi }_{ {\boldsymbol d}} \rd ( {\widetilde{V}} \wedge {\textstyle \int}  e^{\wedge \eta {\widetilde{V}}} ) \wedge e^{\wedge \eta {\widetilde{V}}}\big)\no
&=
\pi _{1} \widehat{\bf G} \big( \pi_1 \boldsymbol\xi_ {\boldsymbol d} e^{\wedge \eta {\widetilde{V}} }  \wedge e^{\wedge \eta {\widetilde{V}}}\big)
+(-)^{ {\boldsymbol d}} D_\eta \pi _{1} \widehat{\bf G} \big( \pi_1 {\boldsymbol \xi }_{ {\boldsymbol d}}( {\widetilde{V}} \wedge {\textstyle \int}  e^{\wedge \eta {\widetilde{V}}} ) \wedge e^{\wedge \eta {\widetilde{V}}}\big).
\end{align}
Since the second term vanishes when $D_\eta$ acts, 
the same computation as (\ref{pre second WZW}) can be applied to show $ (-)^{d} d \, \Psi_{\eta }[{\widetilde{V}}] = D_{\eta } \, \Psi_{d}[{\widetilde{V}}]$.

Thus, the action is written in the WZW-like form, 
\begin{align}
S_{\scriptscriptstyle{\rm EKS}}[{\widetilde{V}}]
& = \int_{0}^{1} dt \, \big\langle \Psi _{t} [{\widetilde{V}}(t)]   ,  Q \Psi _{\eta }[{\widetilde{V}}(t)] \big\rangle.
\end{align}
Its variation and the on-shell condition can be taken by parallel computations in section \ref{SUBSEC VOA EKS}.

\niu{Gauge transformations}

The variation of the action $S_{\scriptscriptstyle{\rm EKS}}[{\widetilde{V}}]$ in the form of (\ref{B1}) can be taken as
\begin{align}
\delta S_{\scriptscriptstyle \rm EKS } [{\widetilde{V}}]
& =  \langle \delta {\widetilde{V}}, {\pi_1 } \big( {\bf L}( e^{\wedge \eta{\widetilde{V}} } ) \big) \rangle,
\end{align}
and one can find the action $S_{\scriptscriptstyle{\rm EKS}}[{\widetilde{V}}]$ is invariant under the gauge transformations\footnote
{
More generally, the second term can be written by
$-\pi_1 \boldsymbol \xi_Q ( \lambda \wedge e^{\wedge\eta {\widetilde{V}}})$.
The representation above is only a choice of $\xi_ {d}$.
However, the difference of the choices becomes $\eta$-exact term and it can be absorbed into $-\eta\xi\omega$.
}
\begin{align}
\delta {\widetilde{V}} =-\eta \xi \omega -\pi_1\mathbf L\boldsymbol ( \xi\lambda \wedge e^{\wedge\eta {\widetilde{V}}}),\label{B14}
\end{align}
where $\omega$ and $\lambda$ are the gauge parameters belonging to the small Hilbert space,
which carry ghost numbers $1$ and $1$, and picture numbers $0$ and $-1$, respectively.
The minus sign is a convention in which $\eta \delta \widetilde V$ becomes $\delta \Phi$ of (\ref{GT EKS Linf}).
In the following, we will discuss how this gauge symmetry can be written in the WZW-like form:
\begin{align}
\Psi_{\delta }[\widetilde V] = D_{\eta } \Omega [\omega, \widetilde{V}]+ Q \Lambda[\lambda, \widetilde{V}].\label{B15}
\end{align}

Inserting the gauge transformation $\delta {\widetilde{V}}$, $\Psi _{\delta} [\widetilde{V}]$ becomes
\begin{align}
\Psi _{\delta} [\widetilde{V}]
&= \pi _{1} \widehat{\bf G} \big( \delta {\widetilde{V}}\wedge e^{\wedge \eta {\widetilde{V}} } \big)\no
&= -\pi _{1} \widehat{\bf G} \big( (\eta\xi\omega) \wedge e^{\wedge \eta {\widetilde{V}} } \big) 
-\pi _{1} \widehat{\bf G} \Big( \big(\pi_1\mathbf L (\xi\lambda\wedge e^{\wedge\eta {\widetilde{V}}})\big)\wedge e^{\wedge \eta {\widetilde{V}}} \Big)\no
&=-\pi _{1} \mathbf L^\eta \Big( \pi_1 \widehat{\mathbf G} \big( \xi\omega\wedge e^{\wedge \eta {\widetilde{V}} }\big)  \wedge e^{\wedge\pi_1 \widehat{\mathbf G} (e ^{\wedge \eta {\widetilde{V}}})}\Big)
 - Q \pi _{1} \widehat{\mathbf G} (\xi\lambda\wedge e^{\wedge\eta {\widetilde{V}}})-\Delta_T[\lambda,\eta{\widetilde{V}}]\no
&= D_{\eta } \Omega [\omega, \widetilde{V}]+ Q \Lambda[\lambda, \widetilde{V}] -\Delta_T[\lambda,\eta{\widetilde{V}}].
\end{align}
The computation here is parallel to (\ref{GT_EKS_Z2WZW}).
Since $\Delta_T$ is a trivial gauge transformation as in section \ref{SUBSEC:EKSGT},
\begin{align}
\Delta_T[\lambda,\eta\widetilde{V}]= \pi _{1} \widehat{\bf G} \big( \xi\lambda\wedge \pi_1 \mathbf L (e^{\wedge \eta {\widetilde{V}}})\wedge e^{\wedge\eta {\widetilde{V}}}\big),
\end{align}
we conclude that the gauge transformations of the trivial embedded theory (\ref{B14}) are written in the WZW-like form (\ref{B15}),
where the gauge parameters $\Omega$ and $\Lambda$ are parameterized as
\begin{align}
\Omega [\omega, \widetilde{V}]&=-\pi_1 \widehat{\mathbf G} (\xi\omega\wedge e^{\wedge\eta {\widetilde{V}}}), \label{B17}\\
\Lambda [\lambda, \widetilde{V}]&= -\pi_1 \widehat{\mathbf G} (\xi\lambda\wedge e^{\wedge\eta {\widetilde{V}}}).\label{B18}
\end{align}
In particular, one can find that the gauge transformations generated by $\eta$ and $Q$ do not mix:
the gauge parameters $\lambda$ and $\omega$ of (\ref{B14})
are related to $\Lambda$ and $\Omega$ of (\ref{B15}) respectively,
as seen in (\ref{B17}) and (\ref{B18}).

%By Hiroaki Matsunaga 
\section{Open NS superstrings with stubs} 

In this section, we construct a new action for generic open NS string field theory, which we call alternative WZW-like action $S_{\eta }$. 
After defining the dual $A_{\infty }$ products and explaining its WZW-like structure, we give two realizations of this type of WZW-like action using two different dynamical string fields $\Phi $ and $\Psi $ in the large and small Hilbert spaces, respectively. 

\niu{Dual $A_{\infty }$-products and Derivation properties}

Let ${\boldsymbol \eta }$ be the coderivation constructed from $\eta $, which is nilpotent ${\boldsymbol \eta }^{2} = 0$, and let ${\boldsymbol a }$ be a nilpotent coderivation satisfying ${\boldsymbol a} {\boldsymbol \eta } = -(-)^{{\boldsymbol a} {\boldsymbol \eta }} {\boldsymbol \eta } {\boldsymbol a}$ and ${\boldsymbol a}^{2} = 0$. 
Then, we assume that $\widehat{\bf G}^{-1} : (\mathcal{H} , {\boldsymbol a}) \rightarrow (\mathcal{H}_{\rm S} , {\boldsymbol D}_{\boldsymbol a})$ is an $A_{\infty }$-morphism, where $\mathcal{H}$ is the large Hilbert space, $\mathcal{H}_{\rm S}$ is the small Hilbert space, and ${\boldsymbol D}_{\boldsymbol a} \equiv \widehat{\bf G}^{-1} {\boldsymbol a} \, \widehat{\bf G}$. 
Note that ${\boldsymbol D}_{\boldsymbol a}$ is nilpotent: ${\boldsymbol D}_{\boldsymbol a}^{2} = ( \widehat{\bf G}^{-1} {\boldsymbol a} \, \widehat{\bf G} ) ( \widehat{\bf G}^{-1} {\boldsymbol a} \, \widehat{\bf G} ) = \widehat{\bf G}^{-1} {\boldsymbol a}^{2} \, \widehat{\bf G} = 0$. 
For example, one can use ${\boldsymbol Q}$, ${\boldsymbol Q}+{\boldsymbol m}_{2}|_{2}$, and so on for ${\boldsymbol a}$, and various $\widehat{\bf G}$ appearing in \cite{Erler:2013xta, Erler:2014eba, Erler:2015lya} for $\widehat{\bf G}$. 
Suppose that the coderivation ${\boldsymbol D}_{\boldsymbol a}$ also commutes with ${\boldsymbol \eta }$, which means 
\begin{align} 
\label{[D_a,eta]=0}
( {\boldsymbol D}_{\boldsymbol a} )^{2} = 0 , \hspace{5mm} \Ld {\boldsymbol D}_{\boldsymbol a} , {\boldsymbol \eta } \Rd = 0. 
\end{align} 
Then, we can introduce a dual $A_{\infty }$ products ${\boldsymbol D}_{\boldsymbol \eta }$ defined by 
\begin{align} 
{\boldsymbol D}^{\boldsymbol \eta } \equiv \widehat{\bf G} \, {\boldsymbol \eta } \, \widehat{\bf G}^{-1} . 
\end{align}
Note that the pair of nilpotent maps $( {\boldsymbol D}^{\boldsymbol \eta } ,{\boldsymbol a})$ have the same properties as $( {\boldsymbol D}_{\boldsymbol a} , {\boldsymbol \eta })$: 
\begin{align} 
\label{[D^eta,a]=0}
( {\boldsymbol D}^{\boldsymbol \eta } ) ^{2} = 0 , \hspace{5mm} \Ld {\boldsymbol D}^{\boldsymbol \eta } , {\boldsymbol a} \Rd = 0 . 
\end{align} 
We can quickly find when the $A_{\infty }$ products ${\boldsymbol D}_{\boldsymbol a}$ commutes with the coderivation ${\boldsymbol \eta }$ as (\ref{[D_a,eta]=0}), its dual $A_{\infty }$ product ${\boldsymbol D}^{\boldsymbol \eta }$ and coderivation ${\boldsymbol a}$ also satisfies (\ref{[D^eta,a]=0}) as follows 
\begin{align} 
{\boldsymbol a} {\boldsymbol D}^{\boldsymbol \eta } &= \big( \widehat{\bf G} \, \widehat{\bf G}^{-1} \big) \, {\boldsymbol a} \, \big( \widehat{\bf G} \, {\boldsymbol \eta } \, \widehat{\bf G}^{-1} \big) = \widehat{\bf G} \, {\boldsymbol D}_{\boldsymbol a} \,  {\boldsymbol \eta } \, \widehat{\bf G}^{-1} 
\no \nonumber 
& = (-)^{{\boldsymbol a}{\boldsymbol \eta }} \widehat{\bf G} \, {\boldsymbol \eta } \, {\boldsymbol D}_{\boldsymbol a}  \, \widehat{\bf G}^{-1} 
= (-)^{{\boldsymbol a}{\boldsymbol \eta }} \, \widehat{\bf G} \, {\boldsymbol \eta } \, \widehat{\bf G}^{-1} \, {\boldsymbol a} \, \widehat{\bf G} \, \widehat{\bf G}^{-1} 
= (-)^{{\boldsymbol a}{\boldsymbol \eta }} {\boldsymbol D}^{\boldsymbol \eta } \, {\boldsymbol a} .
\end{align}

In this paper, as these coderication ${\boldsymbol a}$ and $A_{\infty }$-morphism $\widehat{\bf G}$, we always use ${\boldsymbol a} \equiv {\boldsymbol Q}$ and a gauge product $\widehat{\bf G}$ given by \cite{Erler:2014eba}. 
Therefore, the original $A_{\infty }$ product is equal to the Neveu-Schwarz products ${\boldsymbol M}$ of open stings with stubs given \cite{Erler:2014eba}, 
\begin{align*}
{\boldsymbol D}_{\boldsymbol a}  = {\boldsymbol D}_{\boldsymbol Q}  = {\boldsymbol M} \equiv \widehat{\bf G}^{-1} \, {\boldsymbol Q} \, \widehat{\bf G} , 
\end{align*}
and the dual $A_{\infty }$ products is always given by  
\begin{align} 
\label{D^eta}
{\boldsymbol D}^{\boldsymbol \eta } \equiv \widehat{\bf G} \, {\boldsymbol \eta } \, \widehat{\bf G}^{-1}  = {\boldsymbol \eta } - {\boldsymbol m}_{2} + \dots . 
\end{align} 
The symbol ${\boldsymbol D}^{\boldsymbol \eta }$ always denotes (\ref{D^eta}) in the rest. 
The dual $A_{\infty }$ products ${\boldsymbol D}^{\boldsymbol \eta }$ consists of a linear map $\eta $ and a set of multilinear products $\{ D^{\eta }_{n} \} _{n=2}^{\infty }$. 
Now the bilinear product $D^{\eta }_{2}$ just equals to the star product $- m_{2}$ because we take $\widehat{\bf G}$ of \cite{Erler:2014eba}. 
We write 
\begin{align}
[ B_{1} , \dots , B_{n} ]^{\eta } \equiv D^{\eta }_{n} ( B_{1} \otimes \dots \otimes B_{n} )  
\end{align} 
for higher products of ${\boldsymbol D}^{\boldsymbol \eta }$. 
Then, the Maurer-Cartan element of ${\boldsymbol D}^{\boldsymbol \eta } = {\boldsymbol \eta } - {\boldsymbol m}_{2} + \dots $ is given by 
\begin{align}  
{\boldsymbol D}^{\boldsymbol \eta } \frac{1}{1-A} & = \frac{1}{1-A} \otimes  \pi _{1} \Big( {\boldsymbol D}^{\boldsymbol \eta } \frac{1}{1-A}  \Big) \otimes \frac{1}{1-A}
\no \nonumber 
& = \frac{1}{1-A} \otimes  \Big( {\boldsymbol \eta } A - {\boldsymbol m}_{2} ( A , A ) + \dots \Big) \otimes \frac{1}{1- A } , 
\end{align} 
where $A$ is a state of the large Hilbert space $\mathcal{H}$ and $\pi _{1}$ is an natural $1$-state projection onto $\mathcal{H}$. 
Hence, the solution of the Maurer-Cartan eqiation ${\boldsymbol D}^{\boldsymbol \eta } (1-A)^{-1} = 0$ is given by a state $\widetilde{A}_{\eta }$ satisfying ${\boldsymbol \eta } A_{\eta } - {\boldsymbol m}_{2} ( A_{\eta } , A_{\eta } ) + \dots = 0 $, or equivalently, 
\begin{align*} 
\eta A_{\eta } - m_{2} ( A_{\eta } , A_{\eta } ) + \sum_{n=3}^{\infty } \big[ \overbrace{A_{\eta } , \dots , A_{\eta } }^{n} \big] ^{\eta } = 0 . 
\end{align*}

\vspace{2mm}

\niu{Shift of the dual $A_{\infty }$ products ${\boldsymbol D}^{\boldsymbol \eta }$} 

We introduce the $A_{\eta }$-shifted products $[ B_{1} , \dots , B_{n} ]^{\eta }_{A_{\eta }}$ defined by 
\begin{align} 
\big[ B_{1} , \dots , B_{n} \big] ^{\eta }_{A_{\eta }} \equiv \pi _{1} {\boldsymbol D}^{\boldsymbol \eta } \Big( \frac{1}{1-A_{\eta } } \otimes B_{1} \otimes \frac{1}{1-A_{\eta } } \otimes \dots \otimes \frac{1}{1-A_{\eta } } \otimes B_{n} \otimes \frac{1}{1-A_{\eta } } \Big) . 
\end{align} 
Note that higher shifted products all vanish $[ B_{1} , \dots , B_{n>2} ]^{\eta }_{A_{\eta }} =0$ when higher products of ${\boldsymbol D}^{\boldsymbol \eta } = \eta - m_{2} + D^{\eta }_{3} + D^{\eta }_{4} \dots $ vanish: $D^{\eta }_{n>2} = 0$. 
In particular, we write $D_{\eta } B$ for $[ B ]_{\eta }$: 
\begin{align*}
D_{\eta } B & \equiv \pi _{1} {\boldsymbol D}^{{\boldsymbol \eta }} \Big( \frac{1}{1-A_{\eta } } \otimes B \otimes \frac{1}{1-A_{\eta } } \Big)  
\no 
& = \eta B - \Ld A_{\eta } , B \Rd + \sum_{\rm cyclic} \sum_{n=2}^{\infty } \big[ \overbrace{A_{\eta } , \dots , A_{\eta }}^{n} , B \big] ^{\eta } . 
\end{align*} 
When $\widetilde{A}_{\eta }$ gives a solution on the Maurer-Cartan equation of ${\boldsymbol D}_{\boldsymbol \eta}$, these $A_{\eta }$-shifted products also satisfy $A_{\infty }$-relations, which implies that the linear operator $D_{\eta }$ becomes nilpotent. 
We find 
\begin{align*}
( D_{\eta } )^{2 } B & = \pi _{1} {\boldsymbol D}^{\boldsymbol \eta } \Big( \frac{1}{1-A_{\eta } }  \otimes \pi _{1} {\boldsymbol D}^{\boldsymbol \eta } \Big( \frac{1}{1-A_{\eta } } \otimes B \otimes \frac{1}{1-A_{\eta } } \Big) \otimes \frac{1}{1-A_{\eta } } \Big) 
\no \nonumber 
& = \pi _{1} ( {\boldsymbol D}^{\boldsymbol \eta } )^{2} \frac{1}{1-A_{\eta } } - \LD \pi _{1} {\boldsymbol D}^{\boldsymbol \eta } \Big( \frac{1}{1-A_{\eta } } \Big) , \,  B \, \RD ^{\eta }_{A_{\eta }}  = 0 , 
\end{align*}
where the $A_{\eta }$-shifted commutator $\ld A , B \rd _{A_{\eta }}$ of $A$ and $B$ is defined by 
\begin{align*}
\Ld A ,B \Rd ^{\eta}_{A_{\eta }} \equiv [ A , B ]^{\eta }_{A_{\eta }} - (-)^{AB} [ B , A ]^{\eta }_{A_{\eta }} . 
\end{align*}

\vspace{2mm}

\subsection{Alternative WZW-like relations and action $S_{\eta } [\varphi ]$}

We call a state $A_{\eta }$ a pure-gauge-like (functional) field when $A_{\eta }$ satisfies 
\begin{align} 
\label{p-eq. A}
\eta \, A_{\eta } - m_{2} ( A_{\eta } , A_{\eta } ) + \sum_{n=2}^{\infty } \big[ \overbrace{A_{\eta } , \dots , A_{\eta } }^{n} \big] ^{\eta }= 0  
\end{align}
and has ghost-and-picture number $(1|-1)$, odd Grassmann parity, and even degree. 
Let $d$ is a derivation operator commuting with $\eta $ and let $(d_{\rm g} | d_{\rm p})$ be its ghost-and-picture number. 
We call a state $A_{d}$ a associated (functional) field when $A_{d}$ satisfies 
\begin{align} 
\label{a-eq. A}
(-)^{d} d A_{\eta } = D_{\eta } A_{d} 
\end{align}
and has ghost-and-picture number $(d_{\rm g}-1|d_{\rm p}+1)$. 

Let $\varphi $ be a dynamical string field, $t \in [ 0 , 1 ]$ be a real parameter, and $\varphi (t) $ be a path satisfying $\varphi (0) = 0$ and $\varphi (1) = \varphi $. 
Once a pure-gauge-like (functional) field $A_{\eta }=A_{\eta } [ \varphi ]$ and an associated (functional) field $A_{d} = A_{d} [ \varphi ]$ are constructed as functionals of given dynamical string field $\varphi $, one can construct a gauge invariant action 
\begin{align}
S_{\eta } [\varphi ] = - \int_{0}^{1} dt \, \langle  A_{t} [\varphi (t)] , \, Q A_{\eta } [\varphi (t)] \rangle . 
\end{align}
We write $A_{t} [ \varphi ]$ for the associated field $A_{d}[\varphi ]$ of the parameter differential $d = \partial _{t}$. 
One can check that the $t$-dependence is ``topological''. 
Namely, the variation of the action is given by 
\begin{align*}
\delta S_{\eta } [ \varphi ] = \langle A_{\delta } [ \varphi ] , \, Q A_{\eta } [\varphi ] \rangle , 
\end{align*}
which is independent of the real parameter $t$. 
Since $QA_{\eta } = - D_{\eta } A_{Q}$ and $(D_{\eta } )^{2} = Q^{2} = 0$, we find that $\delta  S = 0$ with the gauge transformations 
\begin{align}
A_{\delta } [ \varphi ] = D_{\eta } \, \Omega + Q \, \Lambda , 
\end{align}
where gauge parameters $\Omega $ and $\Lambda $ have ghost-and-picture number $(-1|1)$ and $(-1|0)$ respectively, and belong to the large Hilbert space. 
Since $A_{\delta } [ \varphi ]$ is an invertible functional of $\delta \varphi$, the explicit form of the gauge transformation of the dynamical string field $\delta \phi$ is obtained from $A_{\delta } [ \varphi ]$. 

In the rest, we give two realisations of this action by using two different dynamical string fields $\varphi = \Psi $ and $\varphi = \Phi $: Small-space parametrisation $S_{\eta }[\Psi ]$ and large-space parametrisation $S_{\eta }[\Phi ]$. 
By construction, the $A_{\infty }$ action proposed in \cite{Erler:2014eba} is equivalent to the small-space parametrisation $S_{\eta } [\Psi ]$. 
Since the large-space parametrisation $S_{\eta }[\Phi ]$ has the same (alternative) WZW-like structure as $S_{\eta }[\Psi ]$, our new actions $S_{\eta } [\Psi ]$ and $S_{\eta }[\Phi ]$ both are equivalent to that of $A_{\infty }$ formulation.

\subsection{Small-space parametrization: $\varphi = \Psi $}

Let us consider an NS dynamical string field $\Psi$ which is a Grassmann odd, ghost number $1$, and picture number $-1$ state of the small Hilbert space: $\eta \, \Psi = 0$. 
We show that one can construct WZW-like ingredients $A_{\eta } [ \Psi ]$ and $A_{d} [ \Psi ]$ as functionals of this dynamical string field $\Psi$. 
Once we obtain these WZW-like ingredients $A_{\eta } [ \Psi ] , A_{d} [\Psi ]$ parametrized by $\Psi $, one can construct the WZW-like action $S[\Psi ]$ parametrized by $\Psi$,  
\begin{align}
S_{\eta } [\Psi ] = - \int_{0}^{1} ds \, \langle  A_{s} [\Psi (s)] , \, Q A_{\eta } [\Psi (s)] \rangle , 
\end{align}
where $s$ is a real parameter $s \in [ 0 , 1]$ and $\Psi (s) $ is the path satisfying $\Psi (s=0)=0$ and $\Psi (s=1) = \Psi $. 
We wrote $A_{s} [ \Psi ]$ for the associated field $A_{d} [\Psi ]$ of $d = \partial _{s}$. 

\niu{Pure-gauge-like field $A_{\eta } [ \Psi ]$} 

When the ${\boldsymbol \eta }$-complex $( \mathcal{H} , {\boldsymbol \eta } )$ is exact, there exist ${\boldsymbol \xi }$ such that $\ld {\boldsymbol \eta } , {\boldsymbol \xi } \rd  = {\boldsymbol 1}$ and $\mathcal{H}$, the large Hilbert space, is decomposed into the direct sum of $\eta $-exacts and $\xi $-exacts $\mathcal{H} = P_{\eta } \mathcal{H} \oplus P_{\xi }\mathcal{H}$, where $P_{\eta }$ and $P_{\xi }$ are projector onto $\eta $-exact and $\xi $-exact states respectively.\footnote{These satisfy $P_{\eta } ^{2} = P_{\eta }$, $P_{\xi }^{2} = P_{\xi }$, $P_{\eta } P_{\xi } = P_{\xi } P_{\eta }=0$,and $P_{\eta } + P_{\xi } = 1$ on $\mathcal{H}$. } 
Note that since the small Hilbert space $\mathcal{H}_{\rm S}$ is defined by $\mathcal{H}_{\rm S} \equiv P_{\eta } \mathcal{H}$ and satisfies $\mathcal{H}_{\rm S} \subset P_{\eta } \mathcal{H}_{\rm S}$, all the states $\Psi $ belonging to $\mathcal{H}_{\rm S}$ satisfy $P_{\eta } \Psi  = \Psi $ and $P_{\xi } \Psi  = 0$, or simply, 
\begin{align} 
\nonumber 
{\boldsymbol \eta } \, \Psi = 0. 
\end{align}
Using this fact, we can construct a desired pure-gauge-like (functional) fields $A_{\eta } [ \Psi ]$, namely, a solution of the Maurer-Cartan equation of the dual $A_{\infty }$ products ${\boldsymbol D}^{{\boldsymbol \eta }} = {\boldsymbol \eta } - {\boldsymbol m}_{2} + \dots $.

\vspace{3mm} 

A pure-gauge-like (functional) field $A_{\eta } [ \Psi ]$ is given by 
\begin{align}
\nonumber 
A_{\eta } [ \Psi ] \equiv \pi _{1} \widehat{\bf G} \, \frac{1}{1- \Psi } 
\end{align}
because it becomes a trivial solution of the Maurer-Cartan equation as follows 
\begin{align} 
{\boldsymbol D}^{\boldsymbol \eta } \frac{1}{1- A_{\eta } [\Psi ] } & = {\boldsymbol D}^{\boldsymbol \eta } \frac{1}{1- \pi _{1} \widehat{\bf G} \, \frac{1}{1- \Psi  } } = {\boldsymbol D}^{\boldsymbol \eta } \, \widehat{\bf G} \, \frac{1}{1- \Psi } =  \widehat{\bf G} \, {\boldsymbol \eta }  \frac{1}{1- \Psi } 
\no \nonumber 
& = \widehat{\bf G} \, \Big( \frac{1}{1- \Psi } \otimes {\boldsymbol \eta } \Psi \otimes \frac{1}{1-\Psi } \Big) = 0 . 
\end{align}
Recall that $\pi _{1} {\boldsymbol D}^{\boldsymbol \eta } (1- A_{\eta }[\Psi ])^{-1} = 0$ is equal to 
\begin{align} 
\nonumber 
\eta \, A_{\eta } [ \Psi ] - m_{2} \big( A_{\eta } [ \Psi ] , A_{\eta } [ \Psi ] \big) + \sum_{n=2}^{\infty } \big[ \overbrace{A_{\eta } [\Psi ] , \dots , A_{\eta } [ \Psi ] }^{n} \big] ^{\eta }= 0  . 
\end{align}

\vspace{2mm} 

\niu{Associated fields $A_{d} [ \Psi ] $}

Let ${\boldsymbol d}$ be a coderivation constructed from a derivation $d$ of the dual $A_{\infty }$ products ${\boldsymbol D}^{\boldsymbol \eta }$, which implies that the $d$-derivation propery $\ld {\boldsymbol d} , {\boldsymbol D}^{\boldsymbol \eta } \rd = 0$ holds. 
Then, we obtain $\ld {\boldsymbol D}_{\boldsymbol d} , {\boldsymbol \eta } \rd = 0 $ with ${\boldsymbol D}_{\boldsymbol d} \equiv \widehat{\bf G}^{-1} \, {\boldsymbol d} \, \widehat{\bf G}$, which means that ${\boldsymbol D}_{\boldsymbol d}$ is ``${\boldsymbol \eta }$-exact'' and there exists a coderivation ${\boldsymbol \xi }_{\boldsymbol d}$ such that 
\begin{align*}
{\boldsymbol D}_{\boldsymbol d}= \widehat{\bf G}^{-1} \, {\boldsymbol d} \, \widehat{\bf G} = ( - )^{{\boldsymbol d}} \ld {\boldsymbol \eta } , {\boldsymbol \xi }_{\boldsymbol d} \rd . 
\end{align*} 
Using this coderivation ${\boldsymbol \xi}_{\boldsymbol d}$, we can construct an associated (functional) field $A_{d} [\Psi ]$. 
Note that the response of ${\boldsymbol d}$ acting on the group-like element of $A_{\eta } [ \Psi ] = \widehat{\bf G} (1-\Psi )^{-1}$ is given by 
\begin{align*}
(-)^{{\boldsymbol d}} {\boldsymbol d} \, \frac{1}{1- A_{\eta } [ \Psi ] } & = (-)^{{\boldsymbol d}} \widehat{\bf G} \, \widehat{\bf G}^{-1} \, {\boldsymbol d} \, \widehat{\bf G}  \frac{1}{1- \Psi } =  \widehat{\bf G} \, {\boldsymbol \eta } \, {\boldsymbol \xi }_{{\boldsymbol d}} \frac{1}{1- \Phi } = {\boldsymbol D}^{{\boldsymbol \eta }} \, \widehat{\bf G} \Big( {\boldsymbol \xi }_{{\boldsymbol d}} \frac{1}{1- \Psi } \Big) 
\no 
&= {\boldsymbol D}^{{\boldsymbol \eta }} \Big( \frac{1}{1- A_{\eta }[ \Psi ] } \otimes  \pi _{1} \, \widehat{\bf G} \Big( {\boldsymbol \xi }_{{\boldsymbol d}} \frac{1}{1- \Psi } \Big)  \otimes \frac{1}{1- A_{\eta } [\Psi ]} \Big) . 
\end{align*}

\vspace{2mm} 

An associated (functional) field of $d$ is given by 
\begin{align}
\nonumber 
A_{d} [ \Psi ] & \equiv \pi _{1} \widehat{\bf G} \Big( {\boldsymbol \xi }_{{\boldsymbol d}} \frac{1}{1- \Psi } \Big)
\end{align}
because one can directly check 
\begin{align*} 
(-)^{{\boldsymbol d}} {\boldsymbol d} \, \frac{1}{1-A_{\eta }[ \Psi ]} & = {\boldsymbol D}^{{\boldsymbol \eta }} \Big( \frac{1}{1-A_{\eta }[\Psi ]} \otimes  \pi _{1} \widehat{\bf G} \Big( {\boldsymbol \xi }_{{\boldsymbol d}} \frac{1}{1-\Psi } \Big)  \otimes \frac{1}{1- A_{\eta } [\Psi ] } \Big)  
\no 
& = {\boldsymbol D}^{{\boldsymbol \eta }} \Big( \frac{1}{1-A_{\eta }[\Psi ]} \otimes  A_{d} [\Psi ] \otimes \frac{1}{1- A_{\eta } [\Psi ] } \Big)  . 
\end{align*}
Picking up the relation on $\mathcal{H}$, or equivalently acting $\pi _{1}$ on this relation on $T(\mathcal{H})$, we obtain  
\begin{align} 
\nonumber 
(-)^{d} d  \, A_{\eta } [ \Psi ] & =  \eta \, A_{d} [ \Psi ] - \Ld A_{\eta } [ \Psi ] , A_{d} [\Psi ] \Rd ^{\eta }_{A_{\eta } [ \Psi ] } ,
\end{align}
which is the simplest case of $(-)^{d} d A_{\eta } [\Psi ] = \pi _{1} {\boldsymbol D}^{\boldsymbol \eta } \mathchar`- {\rm exact} \,\, {\rm term}$.

\subsection{Large-space parametrization: $\varphi = \Phi $}

Let us consider an NS dynamical string field $\Phi$ which is a Grassmann even, ghost number $0$, and picture number $0$ state of the large Hilbert space: $\eta \, \Phi \not= 0$. 
We show that one can construct WZW-like ingredients $A_{\eta } [ \Phi ]$ and $A_{d} [ \Phi ]$ as functionals of this dynamical string field $\Phi$. 
Using these WZW-like ingredients $A_{\eta } [ \Phi ] , A_{d} [\Phi ]$ parametrized by $\Phi $, we obtain the WZW-like action $S_{\eta } [\Phi ]$ parametrized by $\Phi$, 
\begin{align}
S_{\eta } [\Phi ] = - \int_{0}^{1} dt \, \langle  A_{t} [\Phi (t)] , \, Q A_{\eta } [\Phi (t)] \rangle , 
\end{align}
where $t$ is a real parameter $t \in [ 0 , 1]$ and $\Phi (t) $ is the path satisfying $\Phi (t=0)=0$ and $\Phi (t=1) = \Phi $. 
We wrote $A_{t} [ \Psi ]$ for the associated field $A_{d} [\Psi ]$ of $d = \partial _{t}$.

\niu{Pure-gauge-like field $A_{\eta } [ \Phi ]$}

Let us consider a functional $A_{\eta } [ \tau ; \Phi ]$ defined by the differential equation 
\begin{align}
\frac{\partial }{\partial \tau } A_{\eta } [\tau ; \Phi ] & = \eta \, \Phi + \sum_{k=1}^{\infty } \sum_{\rm cyclic} \big[ \overbrace{A_{\eta } [\tau ; \Phi ], \dots , A_{\eta } [\tau ; \Phi ] }^{k} , \Phi \big] ^{\eta } 
\no 
& \equiv D_{\eta } (\tau ) \Phi  
\end{align}
with the initial condition $A_{\eta } [ \tau = 0 ; \Phi ] = 0$, where $\tau $ is a real parameter. 
A few terms of $A_{\eta } [ \tau ; \Phi ]$ is given by 
\begin{align*}
A_{\eta } [ \tau ; \Phi ] = \tau \eta \, \Phi + \frac{\tau ^{2}}{2} \ld \eta \, \Phi , \Phi \rd + \frac{\tau ^{3}}{3} \Big( \sum_{\rm cyclic} \big[ \eta \, \Phi , \eta \, \Phi , \Phi \big] ^{\eta } + \Ld  \ld \eta \, \Phi , \Phi \rd ^{\eta } , \Phi \Rd ^{\eta } \Big) + \dots . 
\end{align*}
We write $A_{\eta } [ \Phi ]$ for the $\tau = 1$ value of the solution $A_{\eta } [ \tau ; \Phi ]$: 
\begin{align*}
A_{\eta } [\Phi ] \equiv A_{\eta} [ \tau =1 ; \Phi ] . 
\end{align*} 
We quickly find that this $A_{\eta }[\Phi ]$, a functional on the dynamical string field $\Phi $, satisfies (\ref{p-eq. A}) and gives a pure-gauge-like (functional) field. 
For brevity, we introduce 
\begin{align*}
F (\tau ) \equiv 
\eta \, A_{\eta } [ \tau ; \Phi ] - m_{2} \big( A_{\eta } [ \tau ; \Phi ] , A_{\eta } [ \tau ; \Phi ] \big) + \sum_{n=2}^{\infty } \big[ \overbrace{A_{\eta } [ \tau ; \Phi ] , \dots , A_{\eta }[ \tau ; \Phi ] }^{n} \big] ^{\eta } .
\end{align*}
The statement ``$A_{\eta } [ \Phi ]$ is a pure-gauge-like (functional) field'' is equivalent to the equation $F (1) =0$. 
By definition, $F(\tau )$ satisfies $F ( \tau = 0 ) = 0$ and the following relation holds: 
\begin{align}
\label{F-eq. A}
\frac{\partial }{\partial \tau } F(\tau ) & = \eta \, \frac{\partial }{\partial \tau } A_{\eta } [ \tau ; \Phi ] %- m_{2} \big( A_{\eta } [ \tau ; \Phi ] , A_{\eta } [ \tau ; \Phi ] \big) 
+ \sum_{n=1}^{\infty } \sum_{\rm cyclic} \big[ \overbrace{A_{\eta } [ \tau ; \Phi ] , \dots , A_{\eta }[ \tau ; \Phi ] }^{n} , \frac{\partial }{\partial \tau } A_{\eta } [ \tau ; \Phi ] \big] ^{\eta } 
\no 
& = D_{\eta } (\tau ) \, D_{\eta } ( \tau ) \, \Phi  = - \Ld  F ( \tau ) , \, \Phi  \, \rd ^{\eta }_{A_{\eta }[\tau ; \Phi ] } .
\end{align}
The solution of the differential equation (\ref{F-eq. A}) with the initial condition $F(0)=0$ is given by $F(\tau ) = 0$ for any $\tau$. 
Hence, $F(1) = 0$ holds and $A_{\eta } [ \Phi ]$ indeed gives a pure-gauge-like (functional) field. 

\niu{Associated field $A _{d} [ \Phi ]$}

Then, we consider a functional $A_{d} [\tau ; \Phi ]$ defined by the differential equation 
\begin{align}
\label{def. eq. of Ad A}
\frac{\partial }{\partial \tau } A_{d} [\tau ; \Phi ] = d \Phi + \Ld \Phi , A_{d} [\tau ; \Phi ] \Rd ^{\eta }_{A_{\eta }[ \tau ; \Phi] } 
\end{align}
with the initial condition $A_{d} [\tau = 0 ; \Phi ] = 0$. 
A few terms of $A_{d} [ \tau ; \Phi ]$ is given by 
\begin{align*}
A_{d} [ \tau ; \Phi ] = \tau d \, \Phi + \frac{\tau ^{2}}{2} \ld d \, \Phi , \Phi \rd + \frac{\tau ^{3}}{3} \Big( \sum_{\rm cyclic} \big[ \eta \, \Phi , d \, \Phi , \Phi \big] ^{\eta } + \frac{1}{2} \Ld  \ld d \Phi , \Phi \rd ^{\eta } , \Phi \Rd ^{\eta } \Big) + \dots . 
\end{align*}
The $\tau =1$ value of the functional $A_{d} [ \tau ; \Phi ]$ of the dynamical string field $\Phi $, which we write 
\begin{align*}
A_{d}[\Phi ] \equiv A_{d} [ \tau = 1 ; \Phi ] , 
\end{align*}
satisfies (\ref{a-eq. A}) and gives an associated (functional) field. 
To prove this fact, we introduce a function of $\tau $ 
\begin{align*}
\mathcal{I} ( \tau ) \equiv  D_{\eta } (\tau ) A_{d} [ \tau ; \Phi ] - (-)^{d} A_{\eta } [\tau ; \Phi ] , 
\end{align*}
whose zeros would provide the WZW-like relation that associated (functional) field must satisfy. 
By definition, $\mathcal{I}(\tau )$ satisfies $\mathcal{I}(\tau = 0) = 0$ and 
\begin{align*}
\frac{\partial }{\partial \tau } \mathcal{I} (\tau ) & = D_{\eta } (\tau ) \, \big( \partial _{\tau } A_{d} [ \tau ; \Phi ] \big) + \big[ \partial _{\tau } A_{\eta } [ \tau ; \Phi ] , A_{d} [\tau ; \Phi ] \big] ^{\eta }_{A_{\eta }[ \tau ; \Phi ] } - (-)^{d} d \big( \partial _{\tau } A_{\eta } [\tau ; \Phi ] \big)  
% \no & = D_{\eta } ( \tau ) \, \big( \partial _{\tau } A_{d} [ \tau ; \Phi ] \big) + \big[  D_{\eta } (\tau ) \Phi , A_{d} [\tau ; \Phi ] \big] ^{\eta }_{A_{\eta }[ \tau ; \Phi ] } - (-)^{d} d D_{\eta } (\tau ) \Phi  
\no 
& = \big[ \Phi , \, \mathcal{I} (\tau ) \big] ^{\eta }_{A_{\eta }[ \tau ; \Phi ] } + D_{\eta } \, \Big( \frac{\partial }{\partial \tau } A_{d} [ \tau ; \Phi ]   - d \Phi - \Ld \Phi , A_{d} [\tau ; \Phi ] \Rd ^{\eta }_{A_{\eta }[ \tau ; \Phi ] } \Big) . 
\end{align*}
Hence, when $A_{d} [ \tau ; \Phi ]$ is defined by (\ref{def. eq. of Ad A}), the function $\mathcal{I} (\tau )$ gives the solution of the differential equation 
\begin{align*}
\frac{\partial }{\partial \tau } \mathcal{I} (\tau ) = \Ld \Phi , \mathcal{I} (\tau ) \Rd ^{\eta }_{A_{\eta } [ \tau ; \Phi ] }  
\end{align*} 
with the initial condition $\mathcal{I} (0) = 0$, namely, $\mathcal{I} ( \tau ) = 0$ for any $\tau $: we obtain $\mathcal{I} (1) = 0$. 

%%%%%%%%%%%%%%%%%%%%%%%%%%%%%%%%%%%%%%%%%%%%%%%%%%%%%%%%%%%%%%%%%%%%%%%%%%%

%H.M and K.G, added and modified by H.M 
\section{NS-NS sector} 

In this section, starting from the NS-NS superstring $L_{\infty }$-products, we clarify a WZW-like structure and give alternative WZW-like form $S_{\eta \widetilde{\eta }} [\varphi ]$ of the action for NS-NS string field theory. 

\niu{Dual $L_{\infty }$ products}

Using path-ordered exponential map $\widehat{\bf G}$, the NS-NS superstring products ${\bf L}$ of the small-space $L_{\infty }$ action of \cite{Erler:2014eba} is given by ${\bf L } = \widehat{\bf G}^{-1} \, {\bf Q } \, \widehat{\bf G}$. 
By construction, it satisfies $\ld {\boldsymbol \eta } , {\bf L} \rd = 0$ and $\ld \widetilde{\boldsymbol \eta } , {\bf L} \rd = 0$. 
(See \cite{Erler:2014eba} or \cite{Erler:2013xta} for details.) 
We start with its dual $L_{\infty }$ products: 
\begin{subequations}
\begin{align}
\label{dual left}
{\bf L}^{\boldsymbol \eta } & \equiv \widehat{\bf G} \, {\boldsymbol \eta } \, \widehat{\bf G}^{-1} , 
\\ \label{dual right} 
{\bf L}^{\widetilde{\boldsymbol \eta }} & \equiv \widehat{\bf G} \, \widetilde{{\boldsymbol \eta }} \, \widehat{\bf G}^{-1} . 
\end{align}
\end{subequations} 
They satisfy $L_{\infty }$-relations $({\bf L}^{\boldsymbol{\alpha }})^{2} = 0$, commutativity $\ld {\bf L}^{\boldsymbol{\eta }} , {\bf L}^{\boldsymbol{\teta }} \rd = 0$, and $Q$-derivation properties 
\begin{align*}
{\bf Q} \, {\bf L}^{\boldsymbol \alpha } = \widehat{\bf G} \, ( \widehat{\bf G}^{-1} \, {\bf Q} \, \widehat{\bf G} )\, {\boldsymbol \alpha } \, \widehat{\bf G}^{-1} = - \widehat{\bf G} \, {\boldsymbol \alpha } \, ( \widehat{\bf G}^{-1} \, {\bf Q} \, \widehat{\bf G} ) \, \widehat{\bf G}^{-1} = - {\bf L}^{\boldsymbol \alpha } \, {\bf Q}
\end{align*} 
for ${\boldsymbol \alpha } = {\boldsymbol \eta} ,\, \widetilde{\boldsymbol \eta }$, which give extensions of $\eta $-constraints. 
In the rest, we write 
\begin{align*}
[ A_{1} , \dots , A_{n} ]^{\alpha } := {\pi_1 } \widehat{\bf G} \, {\boldsymbol \alpha } \, \widehat{\bf G} ^{-1} ( A_{1}\wedge \dots \wedge A_{n} ) , \hspace{5mm} ( \alpha = \eta , \, \widetilde{\eta } ). 
%[ A_{1} , \dots , A_{n} ]^{\widetilde{\eta }} := {\pi _1} \widehat{\bf G}\, \widetilde{\boldsymbol \eta } \, \widehat{\bf G}^{-1}  ( A_{1}\wedge \dots \wedge A_{n} ) . 
\end{align*} 
By definition, these dual $L_{\infty }$ products satisfy $({\bf L}^{\boldsymbol{\alpha }})^{2} = 0$ for $\alpha = \eta , \widetilde{\eta }$, namely $L_{\infty }$-relations, 
\begin{subequations} 
\begin{align}
\sum_{\sigma }\sum_{k=1}^{n} (-)^{|\sigma |} \big[ [ A_{i_{\sigma (1)}} , \dots , A_{i_{\sigma (k)}} ]^{\alpha } , A_{i_{\sigma (k+1)}} , \dots , A_{i_{\sigma (n) }} \big] ^{\alpha } = 0 . 
\end{align}
Note that the commutation relation $\ld {\bf L}^{\boldsymbol{\eta }} , {\bf L}^{\boldsymbol{\teta }} \rd = 0$ can be written as 
\begin{align}
\label{cross}
\sum_{\alpha_{1} , \alpha _{2} = \eta , \widetilde{\eta} }\sum_{\sigma }\sum_{k=1}^{n} (-)^{|\sigma |} \big[ [ A_{i_{\sigma (1)}} , \dots , A_{i_{\sigma (k)}} ]^{\alpha _{1}} , A_{i_{\sigma (k+1)}} , \dots , A_{i_{\sigma (n) }} \big] ^{\alpha _{2}} = 0 . 
\end{align}
For $d = Q , \partial _{t} , \delta$, these products satisfy $d$-derivation properties: 
\begin{align} 
\label{deri}
d\, \big[ B_{1} , \dots , B_{n} \big] ^{\alpha } + \sum_{i=1}^{n-1} (-)^{d (B_{1} +\dots +B_{k-1})} \big[ B_{1} , \dots, d \, B_{k} , \dots , B_{n} \big] ^{\alpha } = 0 . 
\end{align} 
\end{subequations}

\subsection{Alternative Wess-Zumino-Witten-like relations}

Let $\Psi _{\eta \widetilde{\eta }} = \Psi _{\eta \teta } [\varphi ]$ be a Grassmann even, ghost number $2$, left-moving picture number $-1$, and right-moving picture number $-1$ state in the left-and-right large Hilbert space. 
When this $\Psi _{\eta \widetilde{\eta }}$ satisfies the Maurer-Cartan equations for the both dual products (\ref{dual left}) and (\ref{dual right}),  
\begin{subequations}
\begin{align}
\label{WZW relation1}
&\alpha \, \Psi _{\eta \widetilde{\eta }} + \sum_{n=1}^{\infty } \frac{1 }{(n+1)!} [\overbrace{ \Psi _{\eta \widetilde{\eta }} , \dots , \Psi _{\eta \widetilde{\eta }}  }^{n+1} ]^{\alpha }  = 0 ,  \hspace{5mm}  ( \alpha = \eta , \widetilde{\eta } ) , 
\end{align}
we call $\Psi _{\eta \teta } [\varphi ]$ as {\it a pure-gauge-like (functional) field}. 
As we will see, in addition to this $\Psi _{\eta \teta }[\varphi ]$, if one can obtain a state $\Psi _{d} = \Psi _{d}[\varphi ]$ which satisfy {\it the WZW-like relation}, 
\begin{align}
\label{WZW relation2} 
& \hspace{1mm} (-)^{d} d \, \Psi _{\eta \widetilde{\eta }} =  -D_{\eta } \, D_{\widetilde{\eta }} \, \Psi _{d} , 
\end{align} 
then, one can always find a gauge invariant action.  
We call $\Psi _{d} [ \varphi ]$ as {\it a large associated (functional) field}. 
Here, the linear operator $D_{\alpha}$ for $\alpha = \eta , \widetilde{\eta }$ is given by 
\begin{align*}
D_{\alpha } B \equiv \alpha \, B + \sum_{n=1}^{\infty } \frac{1}{n!} \big[ \overbrace{\Psi _{\eta \widetilde{\eta }} , \dots , \Psi _{\eta \widetilde{\eta }} }^{n} , B \big] ^{\alpha } , \hspace{5mm}  ( \alpha = \eta , \widetilde{\eta }  ) , 
\end{align*}
and $d$ is a derivation operator satisfying (\ref{deri}). 
For example, one can take $d = Q$, $\partial _{t}$, or $\delta $. 
Note that $(D_{\eta })^{2} = (D_{\teta })^{2} = 0$ and $D_{\eta } D_{\widetilde{\eta }}\, B = - D_{\widetilde{\eta } } D_{\eta } \, B$ hold because $\Psi _{\eta \widetilde{\eta }}$ satisfies (\ref{WZW relation1}).

\vspace{2mm}

Although it is sufficient to consider the above (\ref{WZW relation1}) and (\ref{WZW relation2}), it would be helpful to consider their {\it small associated (functional) fields} $\Psi _{\eta d} = \Psi _{\eta d} [\varphi ]$ and $\Psi _{d \widetilde{\eta }} = \Psi _{d \teta } [\varphi ]$ which are defined by 
\begin{align} 
\label{WZW relation3}
& \Psi _{\eta d} \equiv \hspace{0mm} D_{\eta }\, \Psi _{d} , \hspace{5mm} \Psi _{d\widetilde{\eta }} \equiv -D_{\widetilde{\eta }} \, \Psi _{d} .  
\end{align}
\end{subequations}
While $\Psi _{d} [\varphi ]$ has the same ghost, left-moving picture, and right-moving picture numbers as $d$, this $\Psi _{\eta d} [\varphi ]$ or $\Psi _{d \widetilde{\eta }}[\varphi ]$ has the same as ``$d$ plus $\eta $'' or ``$d$ plus $\teta $'' respectively.

\subsection{Alternative WZW-like action $S_{\eta \widetilde{\eta }} [ \varphi ]$} 

Let $\varphi $ be a dynamical NS-NS string field and $\varphi (t)$ be a path satisfying $\varphi (0) = 0$ and $\varphi (1) = \varphi $, where $t \in [0,1]$ is a real parameter. 
Once we construct WZW-like ingredients $\Psi _{\eta \widetilde{\eta }} [ \varphi ]$ and $\Psi _{d} [\varphi ]$ as functionals of given dynamical string field $\varphi $, we can obtain a new gauge invariant action 
\begin{align}
\label{NS-NS WZW}
S_{\eta \widetilde{\eta }} [\varphi ] = \int_{0}^{1} dt \, \big{\langle } \Psi _{t }[ \varphi (t)] , \, Q \, \Psi _{\eta \widetilde{\eta }} [\varphi (t)] \big{\rangle } , 
\end{align}
whose gauge transformations are given by 
\begin{align}
\label{NS-NS inv}
\Psi _{\delta } [\varphi ] = D_{\eta } \, \Omega + D_{\widetilde{\eta }} \, \widetilde{\Omega } + Q \, \Lambda . 
\end{align}
Here, we write $\Psi _{t} [\varphi (t) ]$ for $\Psi _{d} [\varphi (t)]$ with $d= \partial _{t} $, and $\Psi _{d} [\varphi ]$ for $\Psi _{d} [\varphi ]$ with $d= \delta $. 
The equation of motion is given by $t$-independent form 
\begin{align}
\label{NS-NS eom}
 Q \, \Psi _{\eta \widetilde{\eta }} [\varphi ] = - D_{\eta } \, \Psi _{Q\widetilde{\eta }} [\varphi ] = - D_{\widetilde{\eta }} \, \Psi _{\eta Q} [\varphi ] = D_{\eta } \, D_{\widetilde{\eta } } \, \Psi _{Q} [\varphi ] = 0 . 
\end{align}
One can find these facts by using WZW-like relations (\ref{WZW relation1}), (\ref{WZW relation2}), and (\ref{WZW relation3}) only, which we explain. 
Note that computations are almost parallel to the conventional WZW-like case \cite{Matsunaga:2014wpa}. 

\niu{Variation of the action}

To derive (\ref{NS-NS inv}) and (\ref{NS-NS eom}) from (\ref{NS-NS WZW}), we compute the variation of (\ref{NS-NS WZW}):
\begin{align*}
\delta S_{\eta \tilde{\eta }} [\varphi ] = \int _{0}^{1} dt \Big( \big{\langle } \delta \Psi _{t} [\varphi (t) ] , \, Q \, \Psi _{\eta \tilde{\eta } } [\varphi (t) ] \big{\rangle } + \big{\langle } \Psi _{t} [\varphi (t) ] , \, \delta \big( Q \, \Psi _{\eta \tilde{\eta } } [\varphi (t) ] \big) \big{\rangle } \Big) . 
%= \langle \Psi _{\delta } [\varphi ] , Q \Psi _{\eta \teta } [\varphi ] \rangle . 
\end{align*}
For this purpose, we define two bilinear maps, so-called shifted $L_{\infty }$-products, 
\begin{align}
\label{bi}
\big[ \, A \, , \, B \, \big] ^{\alpha }_{\Psi _{\eta \eta }} \equiv \sum_{n=0}^{\infty } \frac{1}{n!} \big[ \overbrace{\Psi _{\eta \teta }, \dots , \Psi _{\eta \teta }}^{n} \,  ,\,  A \, , \, B \, \big] ^{\alpha } \, , \hspace{5mm} ( \alpha = \eta , \teta ) .   
\end{align} 
With $D_{\alpha }$, it satisfies $D_{\alpha } [ A , B ]^{\alpha }_{\Psi _{\eta \teta }} + [ D_{\alpha } A , B ]^{\alpha }_{\Psi _{\eta \teta }} + (-)^{A} [ A , D_{\alpha } B]^{\alpha }_{\Psi _{\eta \teta }} = 0$. 
For $d = \partial _{t}$, $\delta $, or $Q$, because of the derivation properties (\ref{deri}) of ${\bf L}^{\boldsymbol{\alpha }}$, we find 
\begin{subequations} 
\begin{align}
(-)^{d} d \, \big( \, D_{\alpha } \, A \, \big) - D_{\alpha } \big( \, d \, A \, \big) - \big[ \, d \, \Psi _{\eta \teta } \, , A \, \big] ^{\alpha }_{\Psi _{\eta \teta }} & = 0 \, . \label{formula 1}
\end{align}
By considering $\ld d_{1} , d_{2} \rd \Psi _{\eta \teta } = 0$ with this formula (\ref{formula 1}), for example, we quickly find 
\begin{align}  
D_{\teta } \Big( d _{1} \, \Psi _{\eta d_{2} } - (-)^{d_{1}d_{2}} d_{2} \Psi _{\eta d_{1} } - (-)^{d_{1}} [ \Psi _{\eta d_{1} } , \Psi _{\eta d_{2} } ]^{\teta } _{\Psi _{\eta \teta }} \Big) & = 0 . \label{formula 2} 
\end{align}
\end{subequations} 

\vspace{2mm} 

For brevity, we omit $\varphi (t)$-dependence of functionals. 
Note that the inner product $\langle A , B \rangle$ includes the $c_{0}^{-}$-insertion: we have $\langle  d A  , B \rangle  = (-)^{dA} \langle A , d   B \rangle $ for $d = D_{\eta }, D_{\teta }, Q$, and we use $\langle A , [ B , C ]^{\alpha }_{\Psi _{\eta \teta } } \rangle = (-)^{AB} \langle B , [ A , C ] ^{\alpha }_{\Psi _{\eta \teta }} \rangle $ for $\alpha = \eta ,\teta $. 
Using (\ref{formula 1}) with (\ref{WZW relation2}) and (\ref{WZW relation3}), we find that the second term can be rewritten as $\langle \Psi _{\delta } , \partial _{t} (Q \Psi _{\eta \teta }) \rangle $ plus extra terms: 
\begin{align}
 \label{2nd term}
\big{\langle } \Psi _{t} , \, \delta ( Q \, \Psi _{\eta \teta } ) \big{\rangle } 
& = \langle \Psi _{t} ,  Q D_{\teta } \Psi _{\eta \delta }  \rangle 
\\ 
& = - \langle \Psi _{t} , D_{\teta } Q  \Psi _{\eta \delta } \rangle 
- \langle \Psi _{t} , [ Q \Psi _{\eta \teta } , \, \Psi _{\eta \delta } ]^{\teta }_{\Psi _{\eta \teta }} \rangle 
\no 
& = - \langle \Psi _{\eta \delta } ,  Q \Psi _{t \teta } \rangle 
+ \langle \Psi _{\eta \delta } , [ \Psi _{t} , \, Q \Psi _{\eta \teta } ]^{\teta }_{\Psi _{\eta \teta }} \rangle 
\no 
& =  \langle \Psi _{\delta } , Q D_{\eta } \Psi _{t \teta } \big) \rangle 
+ \langle \Psi _{\delta } , [ Q \Psi _{\eta \teta } , \, \Psi _{t \teta } ]^{\eta }_{\Psi _{\eta \teta }} \rangle 
+ \langle \Psi _{\eta \delta } , [ \Psi _{t} , \, Q \Psi _{\eta \teta } ]^{\teta }_{\Psi _{\eta \teta }} \rangle 
\no \nonumber 
& = \big{\langle } \Psi _{\delta } , \, \partial _{t} \big( Q \Psi _{\eta \teta } \big) \big{\rangle } 
+ \langle \Psi _{Q \teta } , D_{\eta } \big( [ \Psi _{t\teta } , \, \Psi _{\delta } ]^{\eta }_{\Psi _{\eta \teta }} 
- [ \Psi _{t} , \Psi _{\eta \delta } ]^{\teta }_{\Psi _{\eta \teta }} \big) \rangle . 
\end{align}
Likewise, we find the first term of the variation becomes $\langle \partial _{t} \Psi _{\delta } , Q \Psi _{\eta \teta } \rangle $ plus extra terms: 
\begin{align}
\label{1st term} 
\big{\langle } \delta \Psi _{t} , \, Q \, \Psi _{\eta \teta } \big{\rangle } 
& = - \langle D_{\eta } \delta \Psi _{t} , \Psi _{Q \teta } \rangle 
\\
& = - \langle \delta \big( D_{\eta } \Psi _{t} \big) , \Psi _{Q \teta } \rangle 
+ \langle [ \delta \Psi _{\eta \teta } , \Psi _{t} ]^{\eta }_{\Psi _{\eta \teta }} , \Psi _{Q\teta } \rangle 
\no
& = - \langle \delta \Psi _{\eta t} , \Psi _{Q\teta } \rangle 
+ \langle [ D_{\teta } \Psi _{\eta \delta } , \Psi _{t} ]^{\eta }_{\Psi _{\eta \teta }} , \Psi _{Q\teta } \rangle 
\no
& = \langle \partial _{t} \Psi _{\eta \delta } , \Psi _{Q\teta } \rangle 
+ \langle [ \Psi _{\eta t} , \Psi _{\eta \delta } ]^{\eta }_{\Psi _{\eta \teta }} , \, \Psi _{Q\teta } \rangle 
+ \langle [ D_{\teta } \Psi _{\eta \delta } , \Psi _{t} ]^{\eta }_{\Psi _{\eta \teta }} , \Psi _{Q\teta } \rangle 
\no
& = \langle D_{\eta } \partial _{t} \Psi _{\delta } % , \Psi _{Q \teta } \rangle 
+ \langle [ \partial _{t} \Psi _{\eta \teta } , \Psi _{\delta } ]^{\teta }_{\Psi _{\eta \teta }} , \Psi _{Q \teta } \rangle  
%\no & \hspace{10mm} 
+ \langle \Psi _{Q\teta }  , \,  [ \Psi _{\eta t} , \Psi _{\eta \delta } ]^{\eta }_{\Psi _{\eta \teta }} 
+ [ \Psi _{t } , D_{\teta } \Psi _{\eta \delta } ]^{\eta }_{\Psi _{\eta \teta }} \rangle 
\no \nonumber 
& = \langle \partial _{t} \Psi _{\delta } , Q \Psi _{\eta \teta } \rangle 
+ \langle \Psi _{Q\teta }  , \,  [ D_{\eta } \Psi _{t \teta } , \Psi _{\delta } ]^{\teta }_{\Psi _{\eta \teta }} 
+[ \Psi _{\eta t} , \Psi _{\eta \delta } ]^{\eta }_{\Psi _{\eta \teta }} 
+ [ \Psi _{t } , D_{\teta } \Psi _{\eta \delta } ]^{\eta }_{\Psi _{\eta \teta }} \rangle  . 
\end{align}
From the third line to the forth line, we used (\ref{formula 2}) with (\ref{WZW relation3}). 
If and only if the sum of these extra terms vanishes, the action has a topological $t$-dependence. 
However, (\ref{cross}) provides the cancellation of these extra terms. 
Therefore, using $\varphi (0) = 0$ and $\varphi (1) = \varphi$, we obtain
\begin{align*}
\delta S_{\eta \tilde{\eta }} [\varphi ] = 
\int_{0}^{1} dt \, \Big[ (\ref{2nd term}) + (\ref{1st term}) \Big] 
= \big{\langle } \Psi _{\delta } [\varphi ] , \, Q \, \Psi _{\eta \tilde{\eta } } [\varphi ] \big{\rangle } . 
\end{align*}
We proved that when we have WZW-like functional fields $\Psi _{\eta \teta } [ \varphi ]$ and $\Psi _{d} [\varphi ]$ which satisfy (\ref{WZW relation2}), our NS-NS action $S_{\eta \teta } [\varphi ]$ has topological $t$-dependence: $\delta S_{\eta \teta } [\varphi ] = \langle \Psi _{\delta } [\varphi ] , Q \Psi _{\eta \teta } [\varphi ] \rangle $. 
As a result, this form of the variation of (\ref{NS-NS WZW}) ensures that it is invariant under (\ref{NS-NS inv}) and the equations of motion is given by (\ref{NS-NS eom}) because of (\ref{WZW relation2}).

\vspace{2mm}

In the rest, we give two realisations of this action: Small-space parametrisation $S_{\eta \teta }[\Phi ]$ and large-space parametrisation $S_{\eta \teta }[\Psi ]$. 
By construction, the $L_{\infty }$ action proposed in \cite{Erler:2014eba} is equivalent to $S_{\eta \teta } [\Phi ]$. 
Since $S_{\eta \teta }[\Psi ]$ has the same (alternative) WZW-like structure as $S_{\eta \teta }[\Phi ]$, our new actions $S_{\eta \teta } [\Phi ]$ and $S_{\eta \teta }[\Psi ]$ both are equivalent to that of $L_{\infty }$ formulation.

\subsection{Small-space parametrisation: $\varphi = \Phi $} 

Let $\Phi $ be a NS-NS dynamical string field belonging to the small Hilbert space, which is a Grassmann even, ghost number $2$, left-moving picture number $-1$, and right-moving picture number $-1$ state. 
We show that the pure-gauge-like field $\Psi _{\eta \widetilde{\eta } }$ is given by  
\begin{align}
\label{pure-gauge L-infty}
\Psi _{\eta \widetilde{\eta }} (t) = \pi _{1} \widehat{\bf G} \big( e^{\wedge \Phi (t)} \big) , 
\end{align} 
and the linear operator $D_{\alpha }$ for $\alpha = \eta , \widetilde{\eta }$ becomes 
\begin{align*}
%\label{shifted L-infty}
D_{\alpha } = \pi _{1} {\bf L}^{\boldsymbol \alpha } \big(  \mathbb{I} \wedge e^{\wedge \pi _{1} \widehat{\bf G} ( e^{\wedge \Phi  })} \big) , \hspace{5mm} ( {\boldsymbol \alpha } = {\boldsymbol \eta } , \widetilde{\boldsymbol \eta } ) . 
\end{align*}
Then, the left associated field $\Psi _{d \widetilde{\eta }}$ and the right associated field $\Psi _{\eta d}$ are given by 
\begin{align*}
%\label{associated L-infty}
\Psi _{d \widetilde{\eta }} ( t) & = \pi _{1} \widehat{\bf G} \big( {\boldsymbol \xi }_{d} e^{\wedge \Phi (t) } \big) ,
\\ 
\Psi _{\eta d} ( t) & = \pi _{1} \widehat{\bf G} \big( \widetilde{\boldsymbol \xi }_{d} e^{\wedge \Phi (t) } \big) , 
\end{align*}
respectively. 
The large associated field $\Psi _{d}$, which we call the large associated field, is given by 
\begin{align} 
\label{c field}
\Psi _{d} & = \pi _{1} \widehat{\bf G} \big( {\xi \widetilde{\xi } d \Phi (t) } \wedge e^{\wedge \Phi (t) } \big) 
%\no & = \pi _{1} \widehat{\bf G} \big( {\boldsymbol \xi }_{ \widetilde{\boldsymbol \xi }_{\boldsymbol d}} e^{\wedge \Phi (t) } \big) = - \pi _{1} \widehat{\bf G} \big( \widetilde{{\boldsymbol \xi }}_{{\boldsymbol \xi }_{\boldsymbol d}} e^{\wedge \Phi (t) } \big) . 
\end{align}

\niu{Proofs of properties}

Since the group-like element given by $(\ref{pure-gauge L-infty})$ satisfies 
\begin{align*}
{\bf L}^{\boldsymbol \eta } \big(e^{\pi _{1} \widehat{\bf G} ( e^{\wedge\Phi (t)} ) } \big)  = ( \widehat{\bf G} \, {\boldsymbol \eta } \, \widehat{\bf G}^{-1} ) \, \widehat{\bf G} \big( e^{\wedge\Phi (t)}   \big) = \widehat{\bf G} \, {\boldsymbol \eta } \big( e^{\wedge\Phi (t) } \big) = 0 , 
\end{align*}
we obtain the desired equation $\pi _{1} {\bf L}^{\boldsymbol \eta } \big( e^{\wedge\Psi _{\eta }(t)} \big)  = 0$, or equivalently, 
\begin{align*}
\eta \, \Psi _{\eta \widetilde{\eta }} (t) + \sum_{n=1}^{\infty } \frac{1}{(n+1)!} \big[ \overbrace{\Psi _{\eta \widetilde{\eta }} (t) , \dots , \Psi _{\eta \widetilde{\eta }} (t) }^{n+1} \big] ^{\eta } = 0. 
\end{align*}
Similarly, provided that $d \widehat{\bf G} = \widehat{\bf G} \, d$ and $\ld X , \eta \rd = 0$, after the following computation 
\begin{align*}
(-)^{X} X \widehat{\bf G} \, \big( e^{\wedge \Phi (t)} \big) & = (-)^{d} \widehat{\bf G} \,  \big( d \Phi (t) \wedge e^{\wedge\Phi (t)} \big) 
%\no &
= \widehat{\bf G} \, {\boldsymbol \eta } \, \big( \xi _{d} \Phi (t) \wedge e^{\wedge \Phi (t)} \big) 
\no 
& = {\bf L}^{\boldsymbol \eta } \, \widehat{\bf G} \, \Big( \xi _{d} \Phi (t) \wedge e^{\wedge\Phi (t) } \Big) 
\no 
& = {\bf L}^{\boldsymbol \eta } \, \Big( \pi _{1} \widehat{\bf G} \, \big( \xi _{d} \Phi (t) \wedge e^{\wedge \Phi (t)} \big)  \wedge e^{\wedge\pi _{1} \widetilde{\bf G}(e^{\wedge \Phi (t) }) } \Big) , 
\end{align*} 
we find that the second Wess-Zumino-Witten-like relation holds: 
\begin{align*}
(-)^{d} d \, \Psi _{\eta \widetilde{\eta }} ( t ) = D_{\eta } (t) \Psi _{d \widetilde{\eta }} (t) . 
\end{align*}
Note that $\ld Q , \eta \rd = 0$ but ${\bf Q} \, \widehat{\bf G} = \widehat{\bf G} \, {\bf L}$. 
Utilizing the coderivation ${\boldsymbol \xi }_{Q}$ such that ${\bf L} = - \ld {\boldsymbol \eta } , {\boldsymbol \xi }_{Q} \rd $, we can check that the field \begin{align*}
\Psi _{Q\widetilde{\eta }} (t) = \pi _{1} \widehat{\bf G} \big( {\boldsymbol \xi }_{Q} e^{\wedge \Phi (t)}  \big) 
\end{align*}
satisfies the second Wess-Zumino-Witten-like relation $-Q \, \Psi _{\eta \widetilde{\eta }} = D_{\eta } \, \Psi _{Q\widetilde{\eta }}$ as follows  
\begin{align*}
- {\bf Q} \, \widehat{\bf G} \, \big( e^{\wedge \Phi (t)} \big) & = - \widehat{\bf G} \,  {\bf L} \, \big( e^{\wedge \Phi (t)} \big) 
%\no &
= \widehat{\bf G} \, {\boldsymbol \eta } \,   {\boldsymbol \xi }_{Q} \big( e^{\wedge \Phi (t)} \big) 
%\no  & 
= {\bf L}^{\boldsymbol \eta } \, \widehat{\bf G} \, {\boldsymbol \xi }_{Q} \, \Big( e^{\wedge \Phi (t) } \Big) 
\no 
& = {\bf L}^{\boldsymbol \eta } \, \Big( \pi _{1} \widehat{\bf G} \, {\boldsymbol \xi }_{Q} \big( e^{\wedge \Phi (t)} \big)  \wedge e^{\wedge\pi _{1} \widehat{\bf G}(e^{\wedge \Phi (t) }) } \Big) . 
\end{align*} 
As the almost same way, one can check $\pi _{1} {\bf L}^{\widetilde{\boldsymbol \eta }} e^{\wedge \Psi _{\eta \widetilde{\eta }}} = 0$ and $(-)^{d} d \Psi _{\eta \widetilde{\eta } } = D_{\widetilde{\eta } } \Psi _{\eta d}$. 
Then, we find 
\begin{align*}
\ld D_{\eta } , D_{\widetilde{\eta } } \rd B & = - \pi _{1} \mathbf L^{\eta } \big( \pi_1 \mathbf L^{\widetilde{\eta } } (  e^{\Psi_{\eta \widetilde{\eta } } } ) \wedge B \wedge e^{\Psi_{\eta \widetilde{\eta } } } \big) 
- \pi_1 \mathbf L^{\widetilde{\eta } } \big( \pi_1 \mathbf L^{\eta } (  e^{\Psi_{\eta \widetilde{\eta } } } ) \wedge B \wedge e^{\Psi_{\eta \widetilde{\eta} } } \big)  
\end{align*}
and thus, (\ref{c field}) gives an appropriate large associated (functional) field: (\ref{WZW relation3}) holds. 

\subsection{Large-space parametrization: $\varphi = \Psi $} 

We write $\Psi $ for a dynamical NS-NS string field which has ghost-and-picture numbers $(0|0,0)$ and belongs to the left-large and right-large Hilbert space: $\eta \Psi \not= 0$, $\teta \Psi  \not= 0$, and $\eta \teta \Psi \not= 0$. 
Let us consider the solution $\Psi _{\eta \teta } [ \tau ; \Psi ]$ of the following differential equation, 
\begin{align}
\frac{\partial }{\partial \tau } \Psi_{\eta \teta } [\tau ; \Psi  ] = D_{\eta } D_{\teta } \, \Psi  \label{NS-NS def}
\end{align}
with the initial condition $\Psi _{\eta \teta } [\tau = 0 ; \Psi ] = 0$, where for any state $A \in \mathcal{H}$ and for $\alpha = \eta , \teta $, 
\begin{align*}
D_{\alpha } A \equiv \alpha \, A + \sum_{n=0}^{\infty } \frac{1}{n!} \big[ \overbrace{\Psi _{\eta \teta } [ \tau ; \Psi ] , \dots , \Psi _{\eta \teta } [ \tau ; \Psi ] }^{n}, A \big] ^{\alpha } . 
\end{align*} 
A pure-gauge-like (functional) field $\Psi _{\eta \teta }[ \Psi  ]$ is obtained as the $\tau = 1$ value solution 
\begin{align*}
\Psi _{\eta \teta } [\Psi ] \equiv \Psi _{\eta \teta } [\tau =1 ; \Psi ] . 
\end{align*}    
One can check that this $\Psi _{\eta \teta } [ \Psi ]$ satisfies (\ref{WZW relation1}) by the same (but double) way as NS theory, and then $\ld D_{\eta } , D_{\teta } \rd A = - \big[ \pi _{1} {\bf L}^{\boldsymbol \eta } e^{\Psi _{\eta \teta }} , A \big] ^{\eta }_{\Psi _{\eta \teta }} - \big[ \pi _{1} {\bf L}^{\boldsymbol \teta } e^{\Psi _{\eta \teta }} , A \big] ^{\teta }_{\Psi _{\eta \teta}} = 0$ holds for any state $A \in \mathcal{H}$. 

We consider the solutions $\Psi _{d \teta } [ \tau ; \Psi ]$ and $\Psi _{\eta d} [ \tau ; \Psi ]$ of the following differential equations 
\begin{subequations}
\begin{align}
\frac{\partial }{\partial \tau } \Psi _{d \teta } [\tau ; \Psi ] &= d \, D_{\teta } \, \Psi  + \big[ D_{\teta } \Psi , \Psi _{d \teta } [ \tau ; \Psi ] \big] ^{\eta }_{\Psi _{\eta \teta }[\tau ; \Psi ] } , 
\\ 
- \frac{\partial }{\partial \tau } \Psi _{\eta d} [\tau ; \Psi ] &= d \, D_{\eta } \, \Psi  + \big[ D_{\eta } \Psi , \Psi _{\eta d} [ \tau ; \Psi ] \big] ^{\teta }_{\Psi _{\eta \teta }[\tau ; \Psi ] } , \label{minus} 
\end{align}
\end{subequations} 
with the initial conditions $\Psi _{d \teta } [ \tau = 0 ; \Psi ] = 0$ and $\Psi _{\eta d} [ \tau = 0 ; \Psi ] = 0$. 
Here, we used (\ref{bi}). 
Associated (functional) fields $\Psi _{d \teta } [\Psi ]$ and $\Psi _{\eta d } [\Psi ]$ are $\tau =1$ values of these solutions, 
\begin{align*}
\Psi _{d \teta } [\Psi ] \equiv \Psi _{d \teta } [ \tau =1 ; \Psi ] , \hspace{5mm} \Psi _{\eta d} [ \Psi ] \equiv \Psi _{\eta d} [ \tau =1 ; \Psi ] .
\end{align*}
One can also check these satisfy (\ref{WZW relation2}) by the same way as NS theory. 
The minus sign of (\ref{minus}) comes from the ordering of $D_{\eta }$ and $D_{\teta }$ in the definition of (\ref{NS-NS def}). 

Note that we can only specify the large associated (functional) field $\Psi ^{\ast }_{d}$ up to $D_{\eta }$- and $D_{\teta }$-exact terms, and these ambiguities do not contribute in the action. 
The situation is parallel with that of the conventional WZW-like NS-NS theory \cite{Matsunaga:2014wpa}. 
Thus one may consider a differential equation defining $\Psi ^{\ast }_{d}$ up to $D_{\eta }$- and $D_{\teta }$-exacts by mimicking that of \cite{Matsunaga:2014wpa}, but now we can take more economical way: 
We have operators $F \xi $ and $\widetilde{F} \widetilde{\xi }$ defined by 
\begin{align*} 
F \xi \equiv \sum_{n=0}^{\infty } \Big[ \xi \big( \eta - D_{\eta } \big) \Big] ^{n} \xi , 
\hspace{5mm} 
\widetilde{F} \widetilde{\xi } \equiv \sum_{n=0}^{\infty } \Big[ \widetilde{\xi } \big( \teta - D_{\teta } \big) \Big] ^{n} \widetilde{\xi } , 
\end{align*}
which satisfy $\ld F \xi , D_{\eta } \rd = 1$ and $\ld \widetilde{F} \widetilde{\xi } , D_{\teta } \rd = 1$, respectively. 
These $F \xi $ and $\widetilde{F} \widetilde{\xi }$ consist of the pure-gauge-like (functional) field $\Psi _{\eta \teta } [ \Psi ]$, which is already constructed, and operators $\xi $ and $\widetilde{\xi }$. 
Hence, using these pieces, one can quickly obtain the desired $\Psi _{d} = \Psi _{d} [\Psi ]$ as follows, 
\begin{align*}
\Psi _{d} [\Psi ] \equiv F \xi \Psi _{d \teta} [\Psi ] = - F \widetilde{\xi } \Psi _{\eta d} [\Psi ] . 
\end{align*}

\end{document}